\documentclass[twocolumn]{aastex631}

\usepackage{multirow}

\usepackage[utf8]{inputenc}
\usepackage{natbib}
\usepackage{graphicx}
\usepackage{longtable}
\usepackage{enumerate}
\usepackage{amsmath}
\usepackage{hyperref}
\usepackage{CJK}
\usepackage[title]{appendix}
\usepackage{rotating}
\usepackage{longtable}

\usepackage{amsmath}

\def\msun{\ifmmode {\rm M}_{\mathord\odot}\else $M_{\mathord\odot}$\fi}
\def\rsun{\ifmmode {\rm R}_{\mathord\odot}\else $R_{\mathord\odot}$\fi}
\def\lsun{\ifmmode {\rm L}_{\mathord\odot}\else $L_{\mathord\odot}$\fi}

\def\co{$^{12}$CO}
\def\c18o{C$^{18}$O}
\def\h2{H$_{2}$}
\def\13co{$^{13}$CO}
\def\n2hp{$_{2}$H$^{+}$}

\def\radmc{{\sc radmc-3d}}
\def\cm2{cm$^{-2}$}
\def\cmc{cm$^{-3}$}

\newcommand{\kms}{km~s$^{-1}$}

\def\deg{$^{\circ}$}

\shorttitle{}
\shortauthors{}

\begin{document}
\begin{CJK*}{UTF8}{gbsn}    

\title{Disk Wind Feedback from High-mass Protostars. III. Synthetic CO Line Emission}

\author[0000-0001-6216-8931]{Duo Xu}
\affiliation{Department of Astronomy, University of Virginia, Charlottesville, VA 22904-4235, USA}

\author[0000-0002-3389-9142]{Jonathan C. Tan}
\affiliation{Department of Astronomy, University of Virginia, Charlottesville, VA 22904-4235, USA}
\affiliation{Department of Space, Earth \& Environment, Chalmers University of Technology, SE-412 96 Gothenburg, Sweden}

\author{Jan E. Staff}
\affiliation{Department of Space, Earth \& Environment, Chalmers University of Technology, SE-412 96 Gothenburg, Sweden}
\affiliation{College of Science and Math, University of the Virgin Islands, St Thomas, 00802, United States Virgin Islands}

\author[0000-0002-3835-3990]{Jon P. Ramsey}
\affiliation{Department of Astronomy, University of Virginia, Charlottesville, VA 22904-4235, USA}

\author[0000-0001-7511-0034]{Yichen Zhang}
\affiliation{Department of Astronomy, University of Virginia, Charlottesville, VA 22904-4235, USA}
\affiliation{Department of Astronomy, Shanghai Jiao Tong University, 800 Dongchuan Rd., Minhang, Shanghai 200240, China}

\author[0000-0002-6907-0926]{Kei E. I. Tanaka}
\affiliation{Department of Earth and Planetary Sciences, Tokyo Institute of Technology, Meguro, Tokyo, 152-8551, Japan}

\email{xuduo117@virginia.edu}

\begin{abstract}
To test theoretical models of massive star formation it is important to compare their predictions with observed systems. To this end, we conduct CO molecular line radiative transfer post-processing of 3D magneto-hydrodynamic (MHD) simulations of various stages in the evolutionary sequence of a massive protostellar core, including its infall envelope and disk wind outflow. Synthetic position-position-velocity (PPV) cubes of various transitions of \co, \13co, and \c18o emission are generated. We also carry out simulated Atacama Large Millimeter/submillimeter Array (ALMA) observations of this emission. We compare the mass, momentum and kinetic energy estimates obtained from molecular lines to the true values, finding that the mass and momentum estimates can have uncertainties of up to a factor of four. However, the kinetic energy estimated from molecular lines is more significantly underestimated. Additionally, we compare the mass outflow rate and momentum outflow rate obtained from the synthetic spectra with the true values. Finally, we compare the synthetic spectra with real examples of ALMA-observed protostars and determine the best fitting protostellar masses and outflow inclination angles. We then calculate the mass outflow rate and momentum outflow rate for these sources, finding that both rates {agree with} theoretical protostellar evolutionary tracks.

\end{abstract}

\section{Introduction}
\label{Introduction}

During the process of star formation, protostars launch energetic, collimated bipolar outflows that expel high-velocity gas into the surrounding molecular cloud, injecting significant amounts of mass, momentum and energy into their environment \citep[e.g.,][]{2014prpl.conf..451F,2016ARA&A..54..491B}. Several theoretical models have been proposed to explain the mechanism behind these outflows, including the X-wind model \citep{1994ApJ...429..781S}, pure stellar winds \citep{1968MNRAS.138..359M}, the magnetic tower model \citep{1996MNRAS.279..389L}, and the magnetocentrifugal disk wind model, which is perhaps the most popular \citep{1982MNRAS.199..883B,1983ApJ...274..677P,1986ApJ...301..571P,1997A&A...319..340F}. According to this mechanism, gas in the surface layers of a Keplerian accretion disk is threaded by open magnetic field lines, and the centrifugal force flings material outwards along these lines to form a bipolar outflow that includes a highly collimated fast jet and a slower, wider-angle component \citep[][]{1999ApJ...526L.109M}. In addition to its impact on the surrounding environment, the outflow also extracts angular momentum from the disk, which is a crucial part of the accretion process.
A large number of magnetohydrodynamic (MHD) simulations of disk wind outflows, with a variety of included physics, have been presented \citep[e.g.,][]{2003ApJ...582..292O,2010ApJ...722.1325S,2015MNRAS.446.3975S,2015ApJ...801...84G,2016ApJ...823...28K,2019ApJ...882..123S,2023ApJ...947...40S,2020ApJ...900...59M,2022ApJ...941..202R}.

Observational studies \citep[e.g.,][]{2009A&A...494..147L,2016Natur.540..406B,2017NatAs...1E.152L, 2018ApJ...864...76Z,2020A&A...634L..12D,2022A&A...668A..78D,2023arXiv230107877L,2023A&A...678A.135L} have provided evidence that supports the disk wind model in low- and intermediate-mass protostellar systems. One of the key pieces of evidence supporting this model is the observation that outflowing gas is expelled from a range of locations along the disk and displaced from the central star by up to $\sim$25 AU in some cases, indicating that disk winds are the probable cause, rather than stellar or X-winds. {\citet{2009A&A...494..147L} initially observed CB 26, an edge-on T Tauri star–disk system, and subsequent high-angular-resolution observations indicated a predominantly magnetohydrodynamic disk wind \citep{2023A&A...678A.135L}. Likewise, recent findings by \citet{2023arXiv230107877L} suggest that the outflow from CB26 shares the same rotational direction as the edge-on disk, providing additional support for the magnetocentrifugal disk wind model. The study by \citet{2022A&A...668A..78D} used ALMA CO observations to examine the DG Tau B outflow and disk, concluding that wind-driven shell models fall short in explaining the observed characteristics, while a steady MHD disk wind model successfully accounts for the conical flow's morphology and kinematics, suggesting that molecular outflows trace matter directly ejected from the disk.}

The formation process of high-mass stars and their associated feedback mechanisms remain less well understood than those of their low- and intermediate-mass counterparts, owing to both the relative dearth of nearby high-mass protostellar systems and the limitations of theoretical models. Observationally, bipolar jets and outflows have been detected emanating from massive protostars {\citep[e.g.,][]{1996ApJ...472..225S,1996ApJ...457..267S,2001A&A...378..495R,2002A&A...383..892B,2004ApJ...608..330B,2013A&A...558A.125D,2017NatAs...1E.146H,2019NatCo..10.3630F,2019ApJ...873...73Z},} and are believed to be scaled-up versions of those observed in low- and intermediate-mass stars {\citep[e.g.,][]{2015MNRAS.453..645M,2015A&A...573A..82C,2018Natur.554..334M}.} In terms of theoretical models, three main competing models are currently being debated for the formation of massive stars: core accretion, competitive accretion and protostellar collisions. Core accretion, an extension of the low-mass star formation theory, proposes that dense gas cores formed from clump fragmentation undergo gravitational collapse to form a single star (or small $N$ multiple system via disk fragmentation) \citep{2003ApJ...585..850M}. In competitive accretion, a massive protostar accretes material more chaotically from a surrounding, infalling clump, without forming a massive coherent core, resulting in the contemporaneous formation of a massive star and a cluster of low-mass protostars \citep{2006MNRAS.370..488B,2022MNRAS.512..216G}. Finally, protostellar collisions occur in very dense stellar systems, where the most massive stars are formed through a combination of gas accretion and stellar mergers \citep{2002MNRAS.336..659B}.

In previous papers of this series, \citet{2019ApJ...882..123S,2023ApJ...947...40S} (Papers I and II) conducted 3D MHD simulations of disk wind outflows originating from a massive protostar forming via the Turbulent Core Accretion (TCA) model \citep{2003ApJ...585..850M}. In particular, these simulations consider star formation from an initial core of mass $M_c=60\:M_\odot$ with its structure set assuming it is embedded in a surrounding clump with a mass surface density $\Sigma_{\rm cl}=1\, {\rm g\, cm^{-2}}$, which sets its initial radius to be about 12,000~au. The simulations of Paper II trace the protostellar evolutionary sequence continuously, wherein the mass of the central star, $m_{*}$ grows from 1~\msun\ to more than 24 \msun via accretion of core material. The focus of the simulation is on the evolution of the disk wind and its interaction with the envelope material over a period of 94,000 yr, providing a continuous model with many snapshots in time that can be compared with observations. While this is a single evolutionary track of the Turbulent Core Accretion model, we note that it has been chosen as a fiducial case to represent {a typical example of} massive star formation. Given expected star formation efficiencies of about 50\%, to form massive stars, i.e., with $10\lesssim m_{*f}/M_\odot\lesssim 100$, requires $20\lesssim M_c/M_\odot\lesssim 200$. The $M_c=60\:M_\odot$ case sits near the geometric middle of this range, i.e., within about a factor of 3 of all such models. Similarly, the observed environments of massive star formation span a range of $0.1\lesssim \Sigma_{\rm cl} / {\rm g\:cm}^{-2} \lesssim 10$ with most regions in our Galaxy within a factor of 3 of $\Sigma_{\rm cl}=1\, {\rm g\, cm^{-2}}$ \citep{2014prpl.conf..149T}. Thus we consider that the results of \citet{2023ApJ...947...40S} and the post-processing calculations we present in this paper {may} have general utility for understanding the massive star formation process. 

In this paper, Paper III of the series, we post-process the MHD simulations of \citet{2023ApJ...947...40S} using radiative transfer techniques to create synthetic observational data for molecular lines, such as \co, \13co, and \c18o. The resulting synthetic data is then analyzed in similar ways as real systems to understand how well certain intrinsic properties can be measured. In adddition, the results are also compared with observations of actual massive protostars. A brief overview of the MHD simulations and the details of the methodology for generating synthetic observations are explained in \S\ref{Data and Method}.
In \S\ref{Results}, we analyze the molecular lines to estimate the outflow mass, momentum and energy and compare it with the actual properties in the simulation. We compare the synthetic spectra with the observed outflow spectra obtained from ALMA in \S\ref{Outflow Observations by ALMA}. Finally, \S\ref{Conclusions} summarizes the conclusions of our study.

\section{Data and Method}
\label{Data and Method}

\subsection{Magnetohydrodynamics Simulations}
\label{Magnetohydrodynamics Simulations}

We analyze the 3D ideal MHD simulations by \citet{2023ApJ...947...40S}, which utilized the ZEUS-MP \citep{2000RMxAC...9...66N} code to simulate the disk wind outflow from a massive protostar forming from a 60~\msun\ core embedded in a clump with mass surface density of $\Sigma_{\rm cl}=1, {\rm g, cm^{-2}}$ following the Turbulent Core Accretion (TCA) model \citep{2003ApJ...585..850M}. The simulations are designed to follow the protostellar evolutionary sequence self-consistently, with the central protostellar mass $m_{*}$ growing from 1~\msun\ to over 24 \msun over a period of about 100,000 years. The simulated outflow is limited to one hemisphere, with the domain extending from 100 au above the disk midplane, where the disk wind is injected, to 26,500 au along the outflow ($x_1$) axis, and out to $\pm$16,000 au perpendicular to the outflow axis (i.e., parallel to the accretion disk axes, $x_2$ and $x_3$). {The rationale for the 100 au injection scale is extensively discussed by \citet{2019ApJ...882..123S,2023ApJ...947...40S}. In simulations of this kind, where self-consistent outflow launching is impractical due to the need for resolution down to the stellar surface, an injection surface is necessary. Given that the 100 au scale is significantly smaller than the 26,500 au scale of the domain, the exclusion of this small volume has limited impact on the overall outflow properties. Additionally, the choice of domain size, as explained by \citet{2023ApJ...947...40S}, was determined by computational cost limitations, aiming to have cells and timesteps that cover approximately 10 au in size in the simulation.} In this analysis, we include the missing side of the outflow by mirroring the domain across the disk midplane (i.e.\ $x_1 = 0$), resulting in a bipolar outflow extending to $\pm$26,500 au along the outflow axis. Further details on the setup of MHD simulations are described by \citet{2023ApJ...947...40S}.

\subsection{Radiative Transfer Post-Processing}
\label{Synthetic Observations}

\subsubsection{Continuum Radiative Transfer and Dust Temperature}
\label{Dust Temperature}

We use \radmc\ \citep{2012ascl.soft02015D} to calculate the dust temperature structures of the simulations. {Although the full results of the the dust radiative transfer calculations, including dust emission images and spectral energy distributions (SEDs), will be presented by Ramsey et al. (in prep.), for completeness, we describe our approach here.} The models assume a blackbody for the protostellar input spectrum, with radius and total luminosity at a given mass prescribed by the evolutionary tracks of \citet{2014ApJ...788..166Z}{, and summarized in Table \ref{tab:Table_protostellar_properties} in Appendix \ref{Structure of Outflow Density and Velocity}}. 
A gas-to-dust mass ratio of 100 is adopted. The dust opacities employed in the dust temperature calculation are taken from \citet{2011ApJ...733...55Z} and \citet{2013ApJ...766...86Z}, albeit only for the components included in the simulations of \citet{2023ApJ...947...40S}, i.e., the outflow and envelope components. {For the envelope, we use opacities specifically used for the envelope in \cite{2003ApJ...598.1079W}.} The opacity used for the outflow component is similar to the envelope opacity, but with smaller grains and no ice mantles \citep{2013ApJ...766...86Z}. Scattering is assumed to be purely isotropic.

Within the simulation domain, cells are defined to be part of the outflow if they have a forward ($x_1$) velocity exceeding 100 km\, s$^{-1}$. This division based on velocity was confirmed to give a good match to outflow structures identified in maps of the density structure. We present the sliced density and velocity structures of outflows with varying protostar masses in Appendix~\ref{Structure of Outflow Density and Velocity}{, which indicates the various methods used to delineate the outflow.}

{To ensure a converged dust temperature, particularly in the inner regions near the base of the outflow, we employ a large number of photon packages, $10^8$, and the modified random walk algorithm \citep{2010A&A...520A..70R} when calculating the dust temperature for all snapshots. We directly use the grid defined by the MHD simulations (see \citealt{2023ApJ...947...40S}) for the dust temperature calculations.}

Two choices of dust distribution in the outflow cavity have been investigated by Ramsey et al. (in prep.), i.e., dusty and dust-free scenarios. In the dusty case, dust is assumed to have a standard gas-to-dust mass ratio of 100 everywhere in the outflow cavity. In the dust-free case it is assumed that there is effectively no dust in the outflow cavity (for practical purposes, the dust density in the outflow cavity is actually reduced by a factor of $10^6$ from the standard value). 

However, we consider that a model with standard dust in the outflow is the more realistic scenario. In the semi-analytic models of \cite{2018ApJ...864...76Z} a boundary between dusty and dust-free outflow was set to be the streamline in the disk wind that originated from the disk surface where the temperature was equal to the dust sublimation temperature. This led to a relatively narrow dust-free region along the axis of the outflow, but with most of the volume of the outflow cavity occupied by the dusty streamlines. Thus, the approximation of a fully dusty outflow cavity is closer to the models of \cite{2018ApJ...864...76Z}. Therefore, we have chosen to utilize the dust temperature obtained from the dusty scenario as the fiducial model of our paper. As described below, we will assume that the gas has the same temperature as the dust.

\begin{figure*}[hbt!]
\centering
\includegraphics[width=0.99\linewidth]{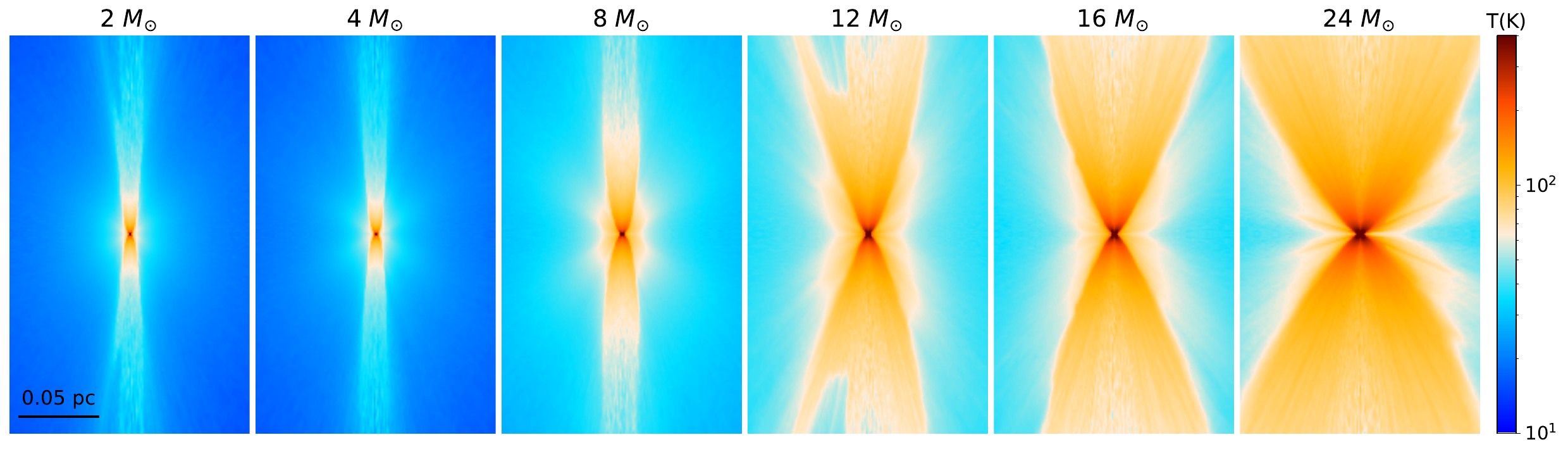}
\caption{Temperatures of dust in a slice along the outflow axis for different protostellar masses. 
}
\label{fig.temperature_outflow_compare_slice}
\end{figure*}

The distributions of dust (and gas) temperatures in example slices through the simulation domain for various evolutionary stages are shown in Figure~\ref{fig.temperature_outflow_compare_slice}. In general, the material in the more optically thin outflow cavity is much warmer ($\gtrsim 100\:$K) than that in the infall envelope ($\sim 10-50\:$K). In addition, as the protostellar mass increases, outflow and infall envelope components are heated to higher temperatures.

\subsubsection{\co, \13co\ and \c18o\ Emission}
\label{radmc}


In order to model the line emission of multiple \co, \13co, and \c18o rotational transitions from the simulated outflows, we employ the publicly available radiation transfer code \radmc\, \citep{2012ascl.soft02015D}. \radmc\ calculates level populations in accordance with the local density and temperature{, which operates under the assumption of non-LTE conditions.} The main simplifying assumptions we adopt in this modeling are that gas kinetic temperature is equal to the dust temperature, calculated above, and that the abundance of CO and its isotopologues is spatially constant. The assumption that the gas kinetic temperature is well-coupled to the dust temperature is generally expected to be the case in molecular clouds with densities $n_{\rm H} \gtrsim 10^5\:{\rm cm}^{-3}$, which applies to most of the regions in the simulation domain.

However, in photo-dissociation regions (PDRs) we expect there to be greater difference between dust and gas temperatures, as well as large variations in CO abundance. Full PDR modeling of the outflow structures has been carried out by Oblentseva et al. (in prep.) using a modified version of 3D-PDR (Bisbas et al. 2012). The importance of the PDR becomes greater at later evolutionary stages. The results presented in our paper should be considered the limiting cases when the PDR has only minor impact. A full comparison of our results with the PDR modeling results will be presented by Oblentseva et al. (in prep.).

We adopt a microturbulent line width of 1 \kms\ to correspond to the typical turbulent velocity of a 60~\msun\ pre-stellar core in the TCA model. 
Instead of performing a complete non-LTE radiative transfer calculation, \radmc\ employs an approximate Large Velocity Gradient (LVG) method \citep{1997NewA....2..365O} to solve the statistical equilibrium equation at each position. The density, temperature and velocity distributions used as inputs to \radmc\, are derived from the simulation data and the dust radiative transfer calculations (see above). We assume that \h2\ is the dominant collisional partner of CO, set the abundance ratios of \co /H nuclei to a fiducial value of $10^{-4}$, and the ratio of the isotopologues \13co\ and \c18o\ to \co\ to fiducial values of 62 and 500, respectively \citep{2006ApJ...646.1070A}.

\section{Results}
\label{Results}

\subsection{Synthetic \co, \13co\ and \c18o\ Maps}
\label{Synthetic CO Maps}


\begin{figure*}[hbt!]
\centering
\includegraphics[width=0.97\linewidth]{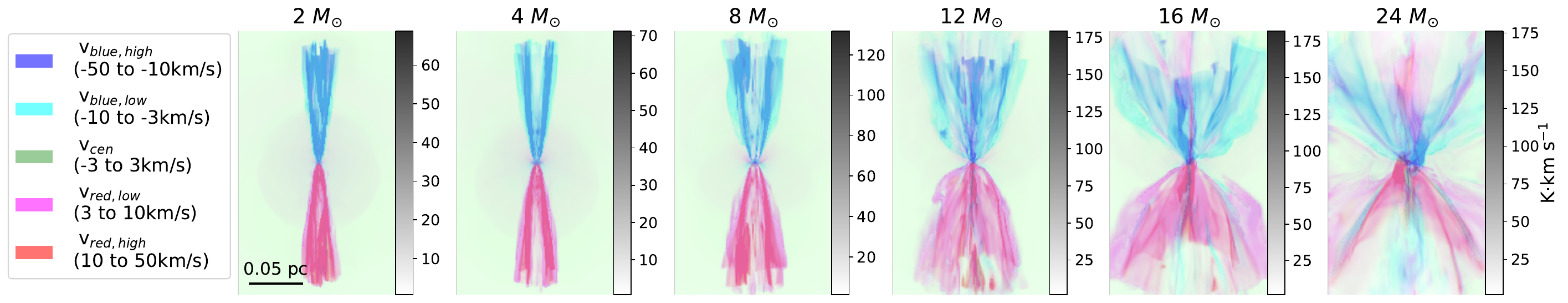}
\caption{Synthetic \co~(2-1) emission of outflows at different evolutionary stages, i.e., different protostellar mass, and viewed at an inclination angle of 58\deg to the outflow axis. Different colors indicate different velocity ranges of the integrated emission: blue-shifted high-velocity component (-50 to -10 \kms); blue-shifted low-velocity component (-10 to -3 \kms); central velocity component (-3 to 3 \kms); red-shifted low-velocity component (3 to 10 \kms); and red-shifted high-velocity component (10 to 50 \kms). The scalebar shows the grayscale intensity scale for each color map, and all colors in a given panel are assigned the same range of values.}
\label{fig.co21-cos0525-allmass}
\end{figure*} 


\begin{figure*}[hbt!]
\centering
\includegraphics[width=0.91\linewidth]{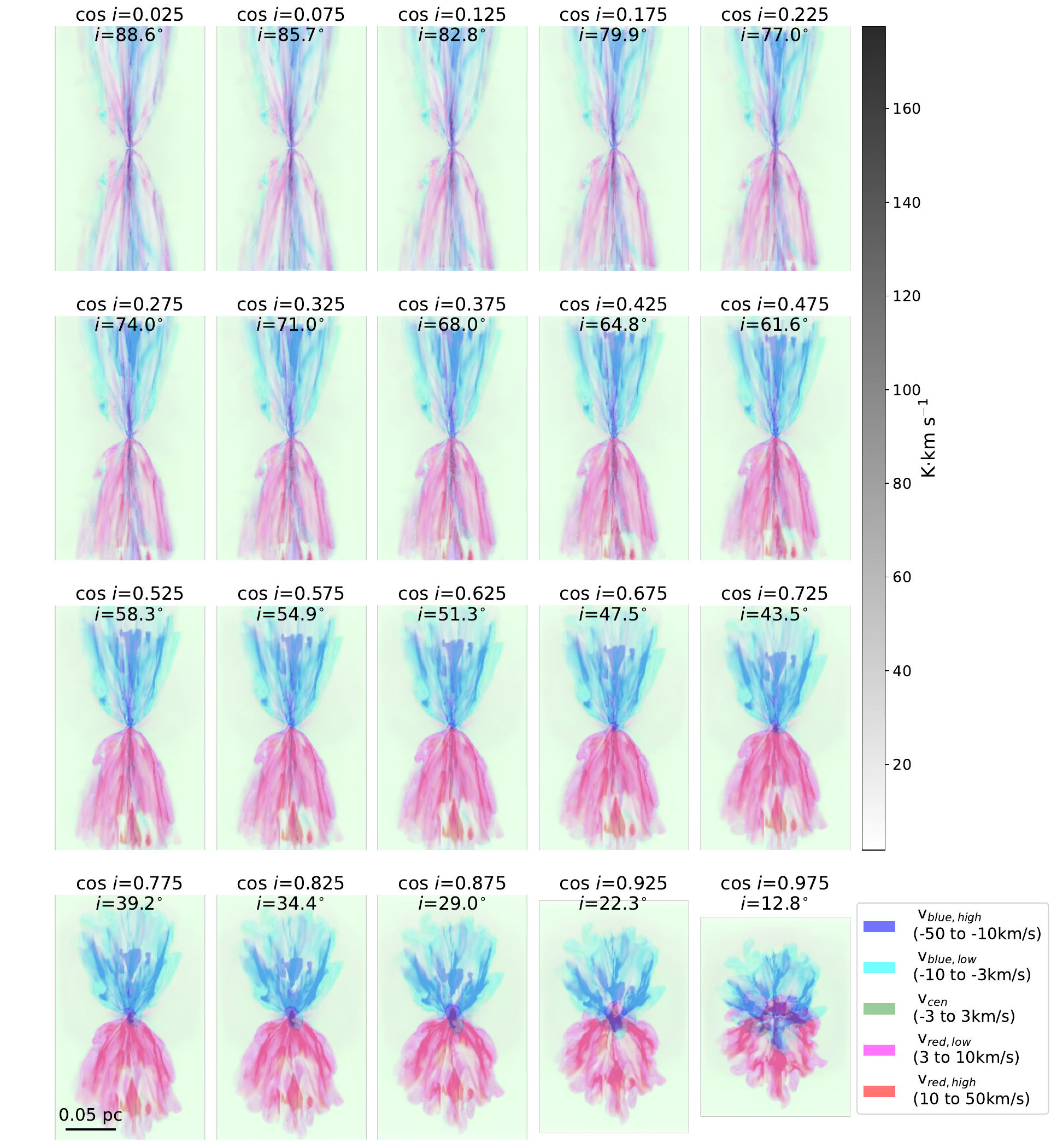}
\caption{Synthetic \co\ (2-1) emission of the outflow from a 12~\msun\ protostar at different inclination angles. The color description is the same as Figure~\ref{fig.co21-cos0525-allmass}. $\cos i = 1$ corresponds to looking along the outflow (``face-on''), while $\cos i = 0$ is looking across the outflow (``edge-on'').}
\label{fig.co21-mass12-allinclination}
\end{figure*} 


\begin{figure*}[hbt!]
\centering
\includegraphics[width=0.91\linewidth]{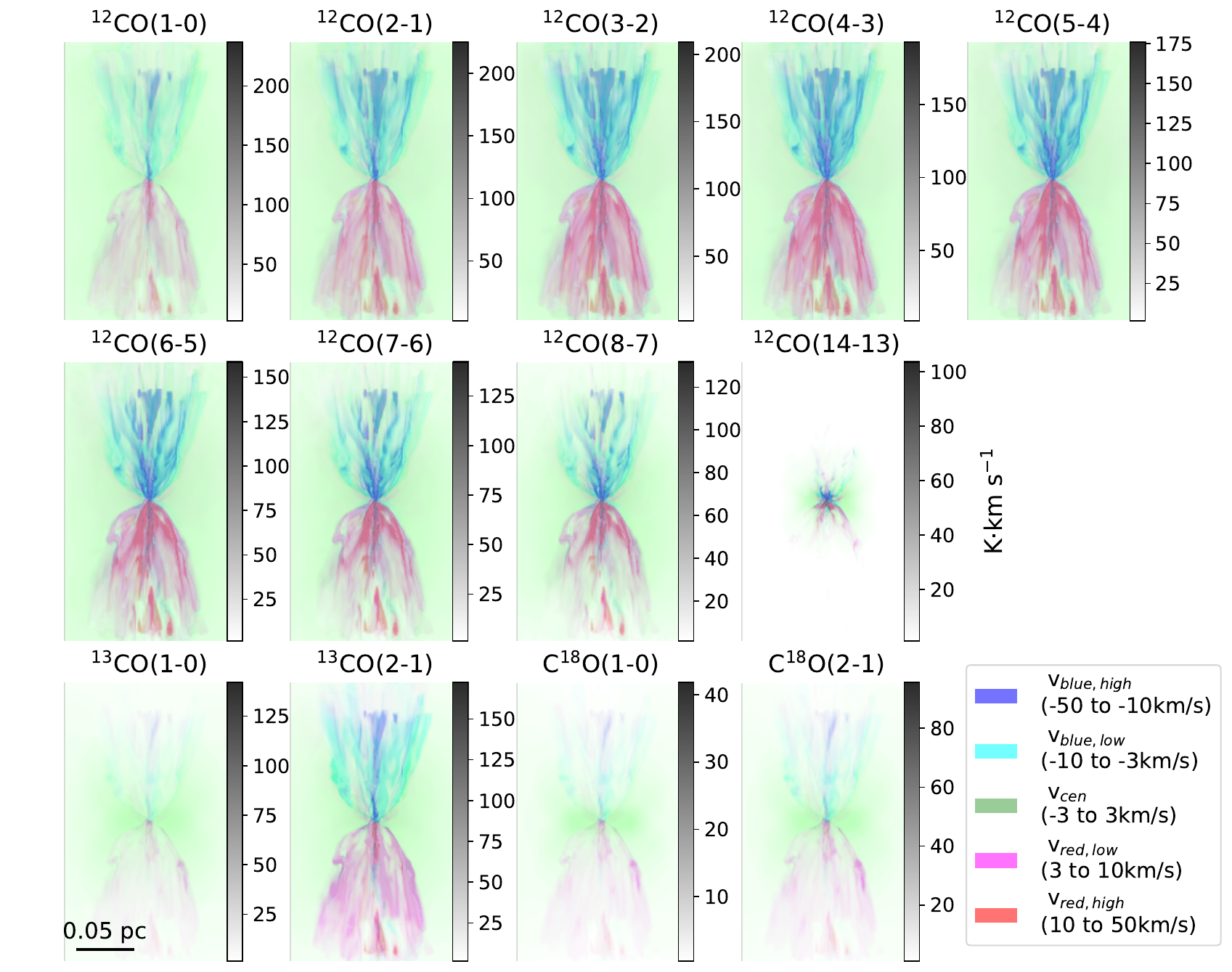}
\caption{Synthetic emission maps for multiple transitions of \co, \13co\, and \c18o\, from a 12~\msun\ star outflow at an inclination angel of 58\deg. The color description is the same as Figure~\ref{fig.co21-cos0525-allmass}.}
\label{fig.co21-mass12-cos0525-allco-4color}
\end{figure*} 






\begin{figure*}[hbt!]
\centering
\includegraphics[width=0.91\linewidth]{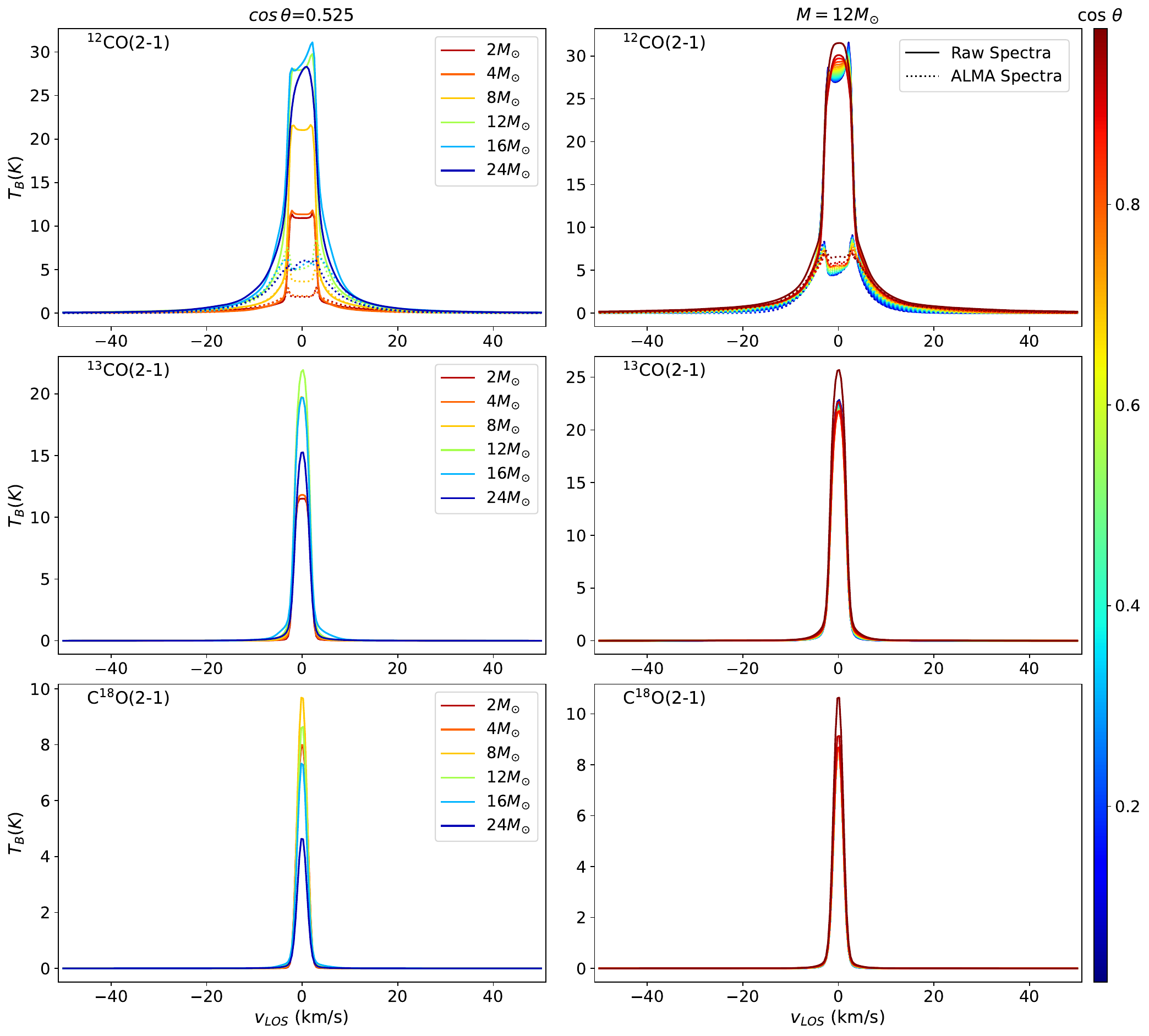}
\caption{{\it (a) Left column:} Synthetic \co~(2-1), \13co~(2-1), and \c18o~(2-1) spectra of outflows at different evolutionary stages viewed at an inclination angle of 58\deg. 
{\it (b) Right column:} Synthetic \co~(2-1), \13co~(2-1), and \c18o~(2-1) spectra of outflows {throughout the entire field of view} from a 12~\msun\ protostellar mass viewed at different inclination angles, as labeled by the color bar. The solid lines represent the raw synthetic spectra, whereas the dotted line represents the synthetic ALMA spectra post-processed with CASA. {Note that all these spectra are calculated within the same field of view, encompassing a rectangle of 15.6\arcsec$\times$25.3\arcsec, assuming a distance of 2 kpc.}}
\label{fig.co21-mass12-0525-spec-linear}
\end{figure*} 

\begin{figure*}[hbt!]
\centering
\includegraphics[width=0.91\linewidth]{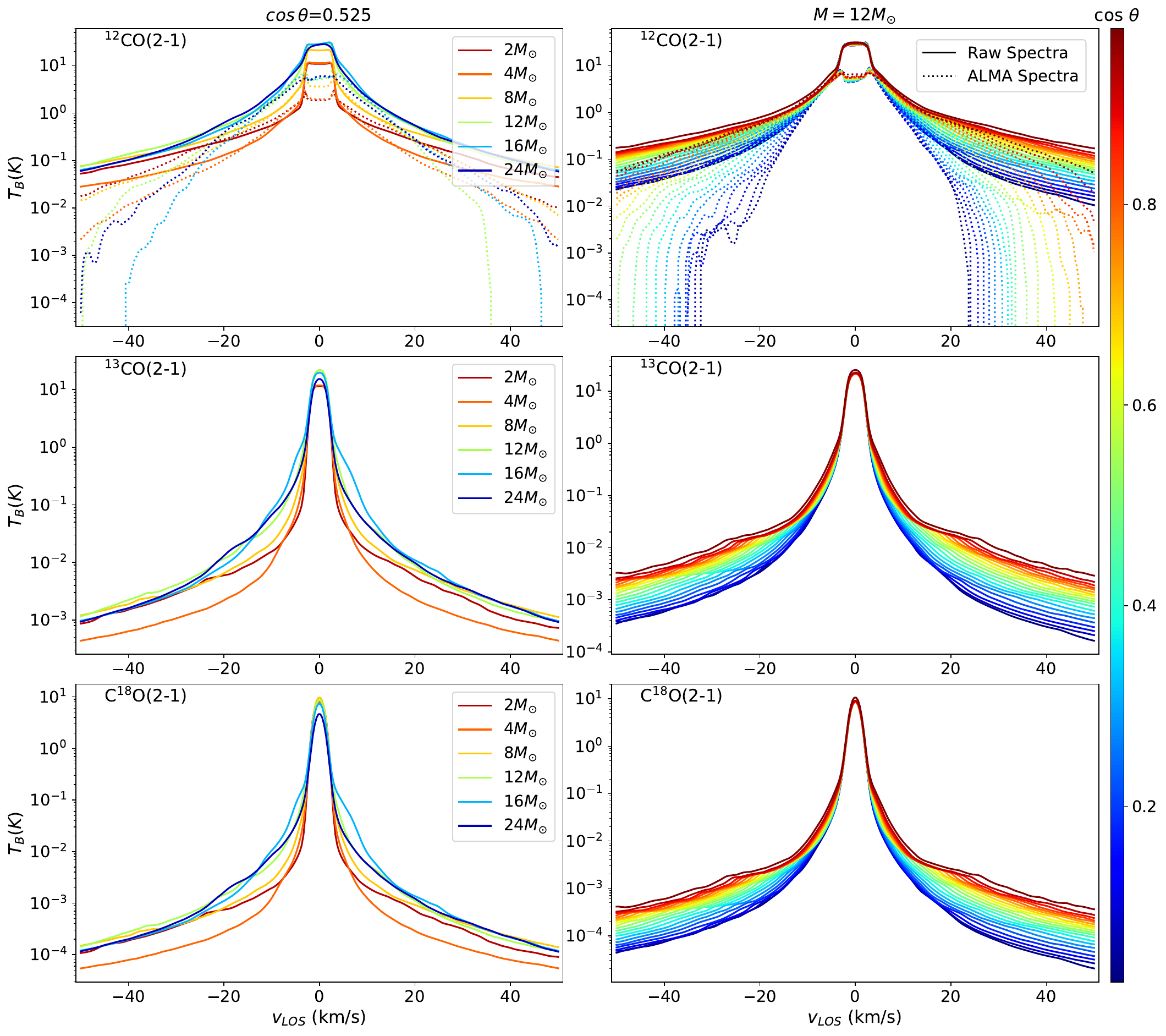}
\caption{Similar to Figure~\ref{fig.co21-mass12-0525-spec-linear}, but uses a logarithmic scale.}
\label{fig.co21-mass12-0525-spec-log}
\end{figure*}

\begin{figure*}[hbt!]
\centering
\includegraphics[width=0.91\linewidth]{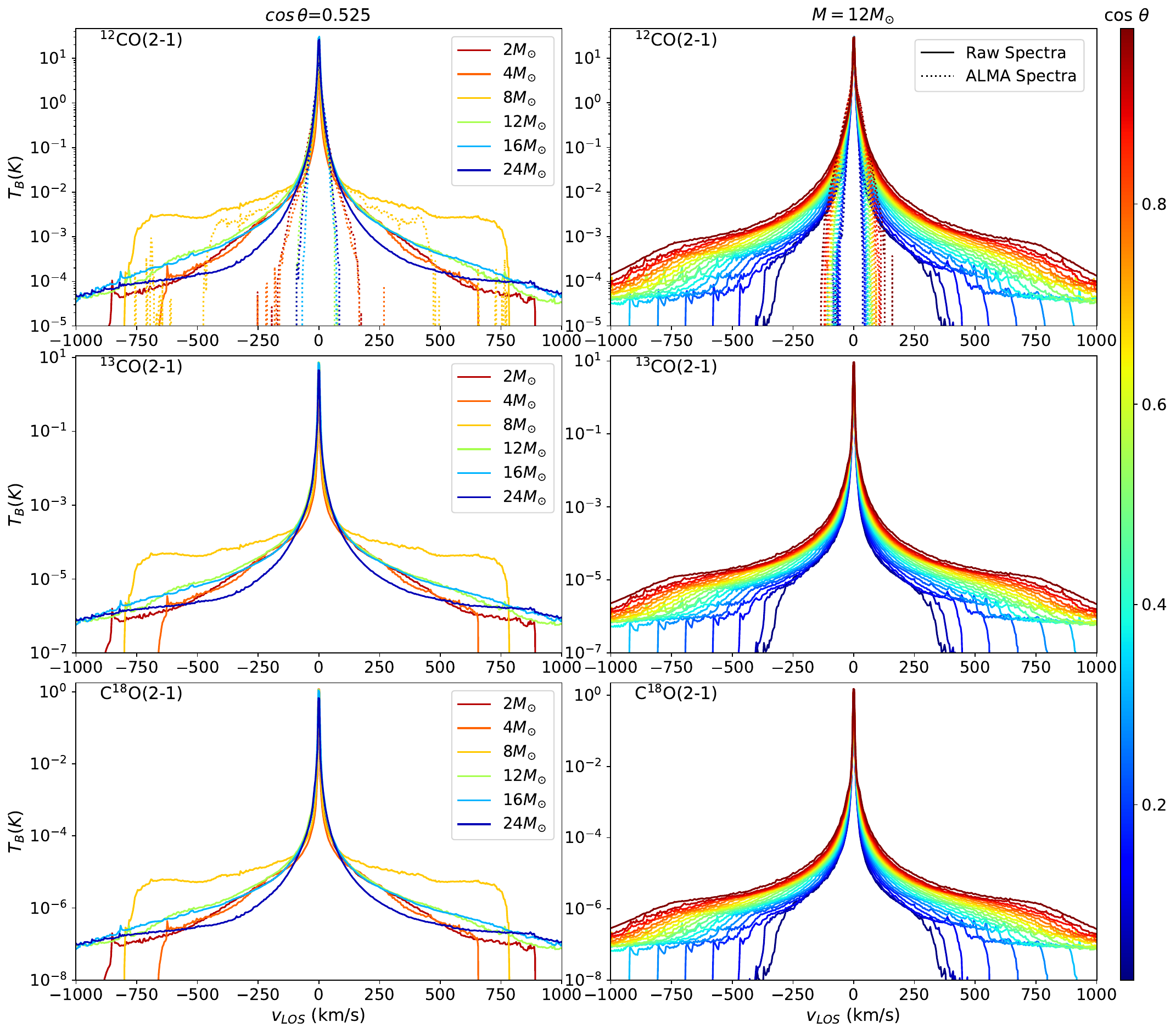}
\caption{Similar to Figure~\ref{fig.co21-mass12-0525-spec-linear}, but with a broader velocity range.}
\label{fig.12CO_13CO_12msun_0525_spec_v1000_log}
\end{figure*}

We present the results of the radiative transfer calculations for 20 different inclination angles, with the cosine values of inclination angle evenly spaced between 0.025 and 0.975. This choice is the same as that used in the continuum radiative transfer model grid of \citet{2018ApJ...853...18Z}. Figure~\ref{fig.co21-cos0525-allmass} shows the synthetic \co~(2-1) emission for the outflows at different evolutionary stages, i.e., different protostellar masses, with an inclination angle of 58\deg. To facilitate the visualization of different velocity components of the outflow emission, we divide the velocity channels into five parts: the blue-shifted high-velocity component (-50 to -10 \kms), the blue-shifted low-velocity component (-10 to -3 \kms), the central velocity component (-3 to 3 \kms), the red-shifted low-velocity component (3 to 10 \kms), and the red-shifted high-velocity component (10 to 50 \kms). As the protostellar mass increases, the opening angle of the outflow cavity increases, as do the overall velocities in the outflow  \citep[see][]{2023ApJ...947...40S}. Our synthetic \co\ emission maps reproduce these trends.

In Figure~\ref{fig.co21-mass12-allinclination}, we present synthetic \co~(2-1) emission of the outflow from the 12~\msun\ protostar at various inclination angles. Our results indicate that, as the observed inclination angle decreases, the visible high-velocity components of the emission become more prominent and exhibit a wider-angle, more overlapping spatial distribution. This is to be expected, as the majority of the momentum in the outflow is along the outflow axis and, at lower inclinations, more and more of this velocity is along the line of sight. 

In Figure~\ref{fig.co21-mass12-cos0525-allco-4color}, we present synthetic maps of multiple transitions of \co, \13co and \c18o, for the outflow from a 12~\msun\ protostar at a fixed inclination angle of 58\deg. Note that the emission from the ground transition $J=1-0$ of \co\ exhibits a greater contribution from the ambient gas compared to the other \co\ transitions. The morphology of the outflow appears consistent across various \co\ transitions, except for the extreme case of the \co~($14-13$) transition. In contrast to the other transitions, the \co~(14-13) transition shows a more centralized morphology, indicating that this highly excited gas is concentrated close to the central star. This behavior can be attributed to the high kinetic temperature needed to excite \co\ to such high energy levels.

Figures~\ref{fig.co21-mass12-0525-spec-linear}a, \ref{fig.co21-mass12-0525-spec-log}a and \ref{fig.12CO_13CO_12msun_0525_spec_v1000_log}a display the simulated \co~(2-1), \13co~(2-1), and \c18o~(2-1) spectra of the outflows {throughout the entire field of view} at various evolutionary stages, viewed at a fixed inclination angle of 58\deg to the outflow axis. In Figure~\ref{fig.co21-mass12-0525-spec-linear}a the spectra span a velocity range of -50 to 50 \kms, with a spectral resolution of 0.39 \kms, while Figure \ref{fig.co21-mass12-0525-spec-log}a shows the same information, but with a logarithmic intensity scale. As the mass of the star increases, there is a noticeable broadening of the high-velocity component. The emission at the ambient velocity of the core infall envelope also brightens as the protostellar mass increases, which is due to the warmer temperatures achieved at the later evolutionary stages. To examine the visibility of very high-velocity gas in the outflows, we present the outflow spectra with an extended velocity range of -1000 to 1000 \kms, with a velocity resolution of 3.9 \kms, in Figure~\ref{fig.12CO_13CO_12msun_0525_spec_v1000_log}a. Notably, when the protostellar mass exceeds 12 \msun, there is still emission present from very high-velocity gas at approximately $\pm$1000 \kms.

Figures \ref{fig.co21-mass12-0525-spec-linear}b, \ref{fig.co21-mass12-0525-spec-log}b and \ref{fig.12CO_13CO_12msun_0525_spec_v1000_log}b meanwhile show the simulated \co~(2-1), \13co~(2-1), and \c18o~(2-1) spectra of the outflow from a 12~\msun\ protostar at varying inclination angles. A decrease in the inclination angle also results in wider line wings, indicating an increase in the amount of high-velocity gas along the line of sight. 

In Appendix~\ref{Moment Maps of Synthetic Outflows}, a collection of moment maps for the synthetic outflows is displayed. {The effectiveness of lower and mid-$J$ \co\ transitions in delineating outflow lobes is evident, whereas extremely high-$J$ \co\ (14-13), \13co, and \c18o\ can primarily trace the relatively dense portions of the outflow envelope. This is attributed to the higher critical density for high-$J$ \co\ and the lower abundances of \13co\ and \c18o, which predominantly capture dense regions rather than the outflow lobes. Additionally, the velocity dispersion observed in the outflow lobes for lower and mid-J \co\ transitions is markedly higher than that derived from \co~(14-13), \13co, and \c18o\ lines.}

\subsection{Synthetic ALMA Observations}
\label{Synthetic ALMA Observations}

\begin{figure}[hbt!]
\centering
\includegraphics[width=0.97\linewidth]{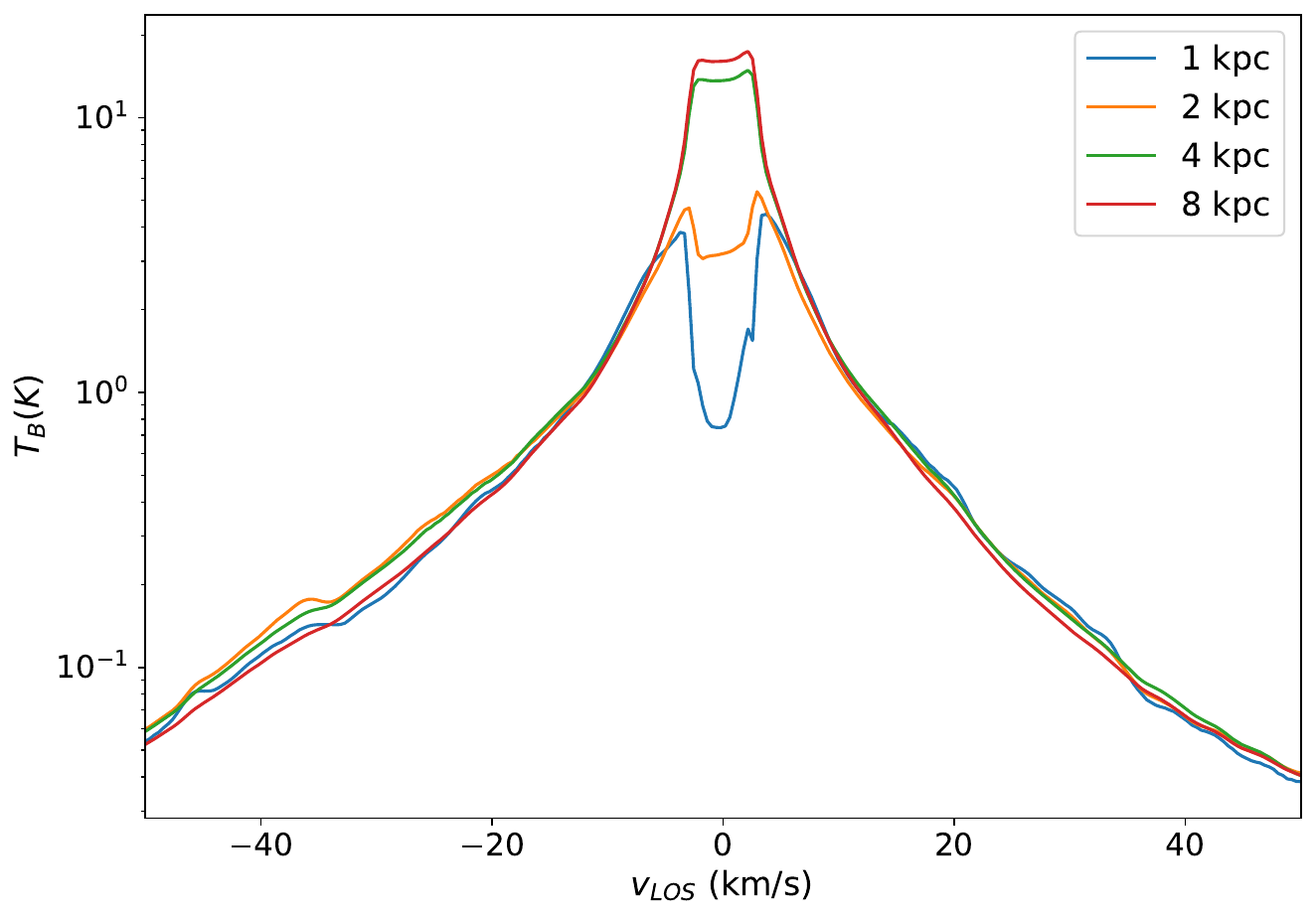}
\caption{Post-processed synthetic \co~(2-1) spectrum of the outflow generated by a 12 \msun\ protostar at an inclination angle of 58\deg\ using CASA, with different assumptions about the source's distance. {The brightness temperature is adjusted by the inverse square of the distance, assuming a reference distance of 2 kpc, to account for the filling factor.}}
\label{fig.12CO21_spec_alma_distance_all}
\end{figure} 

\begin{figure*}[hbt!]
\centering
\includegraphics[width=0.97\linewidth]{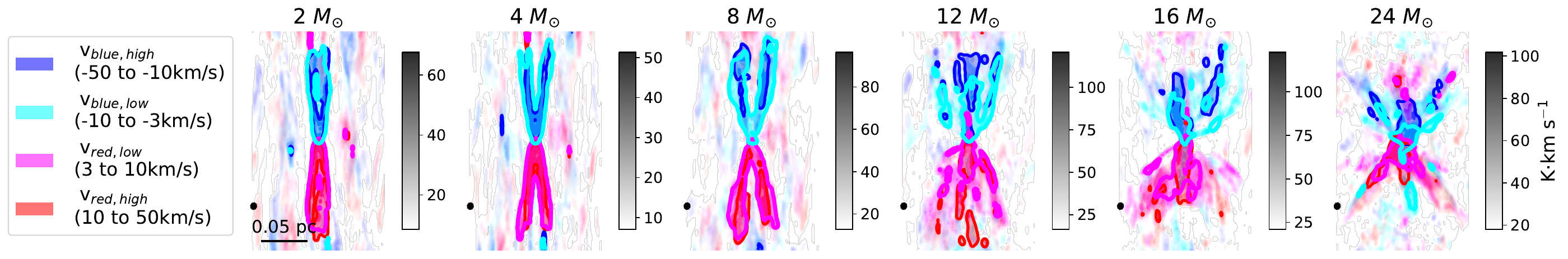}
\caption{Same as Figure~\ref{fig.co21-cos0525-allmass}, but post-processed with CASA and now not displaying the ambient gas. {We also present the 3$\sigma$ contour of each component for reference.}}
\label{fig.12CO21-cos0525-allmass-4color-casa-alma-dis2}
\end{figure*}

\begin{figure*}[hbt!]
\centering
\includegraphics[width=0.97\linewidth]{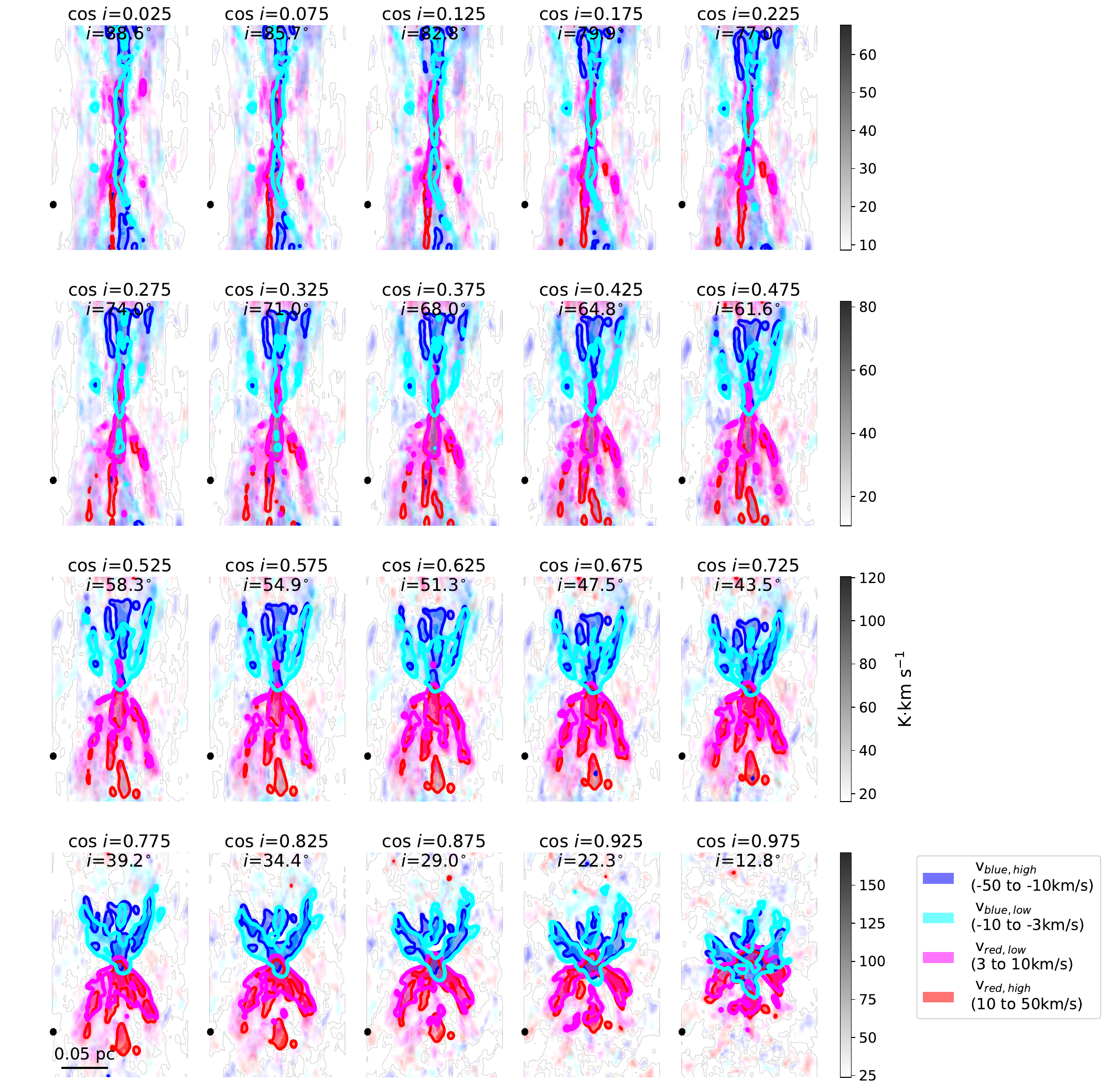}
\caption{Same as Figure~\ref{fig.co21-mass12-allinclination},  but post-processed with CASA. {We also present the 3$\sigma$ contour of each component for reference.}}
\label{fig.12CO21-12msun-allinclination-4color-casa-alma-dis2}
\end{figure*}

\begin{figure}[hbt!]
\centering
\includegraphics[width=0.97\linewidth]{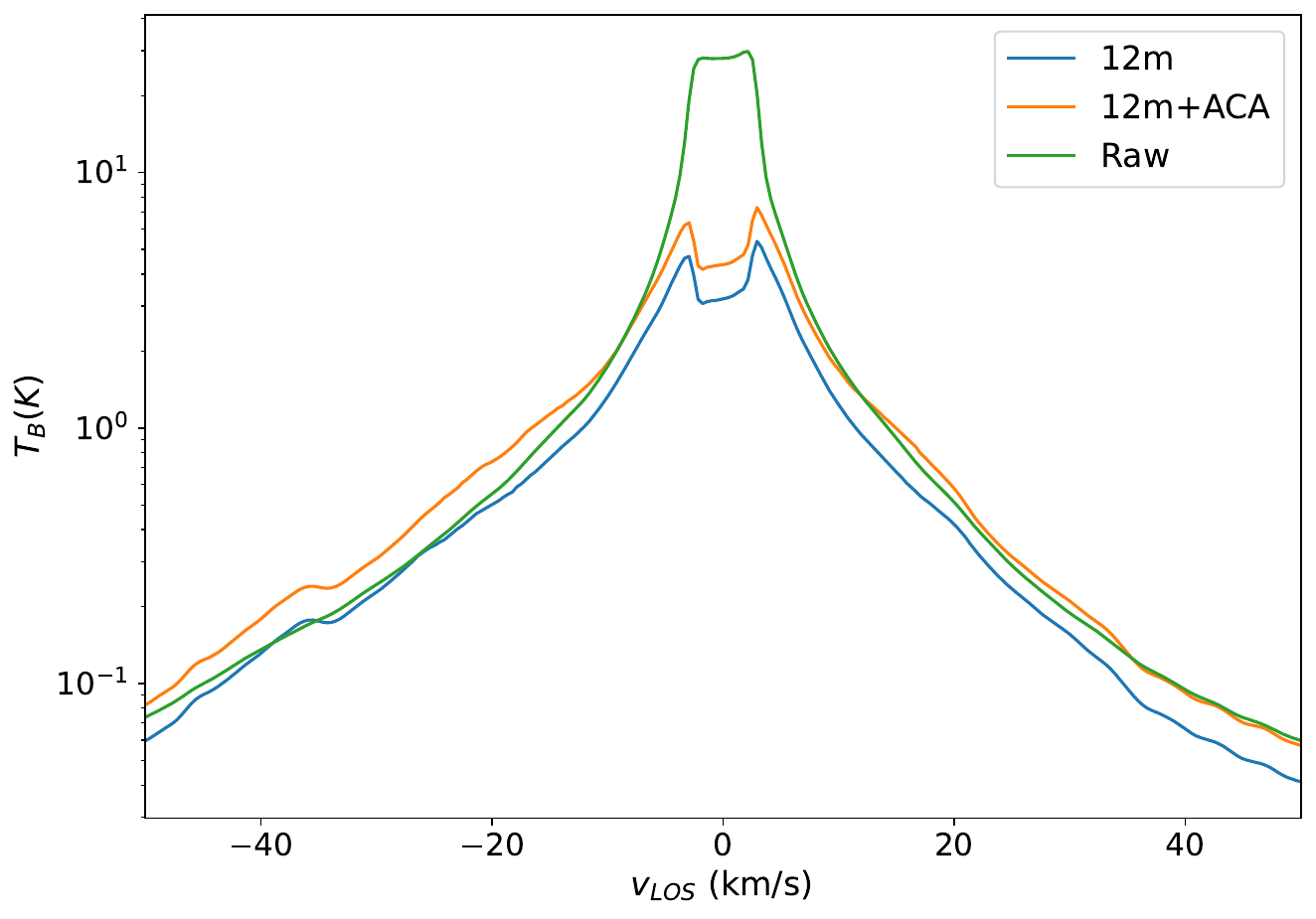}
\caption{Post-processed synthetic \co~(2-1) spectrum of the outflow produced by a 12 \msun\ protostar at an inclination angle of 58\deg\ using CASA, employing both the 12m array and the ACA, as well as using only the 12m array. The raw \co~(2-1) spectrum, without CASA processing, is also depicted. {Note that all these spectra are calculated within the same field of view, encompassing a rectangle of 15.6\arcsec$\times$25.3\arcsec, assuming a distance of 2 kpc.}}
\label{fig.12CO21_spec_alma_comp_ACA}
\end{figure}


To replicate the type of observations made by ALMA, we utilized CASA/\texttt{simalma} to post-process the synthetic molecular line data of \co~(2-1) generated in \S\ref{Synthetic CO Maps}. Our synthetic interferometry observations were conducted using the C36-3 12-m array configuration, which was also employed in the observations described in \S\ref{Outflow Observations by ALMA}. We ensured consistency by using the same integration time of 210 seconds. This integration time and array configuration match the observational data that will be utilized in \S\ref{Outflow Observations by ALMA}. We assumed that the source was located at a distance of 2~kpc, aligning with the outflow sources to be discussed in \S\ref{Outflow Observations by ALMA}. However, it is important to note that one of the observed outflow sources in \S\ref{Outflow Observations by ALMA} is situated at a distance of 8.4 kpc rather than around 2 kpc. To address concerns about inconsistency in distance assumptions during spectral fitting, we examined the impact of assumed distance on the spectral shape, particularly the line wings. As illustrated in Figure~\ref{fig.12CO21_spec_alma_distance_all}, the assumption of source distance affects the spectral shape, especially near the central velocity, but has minimal impact on the spectral wings. The variation in the spectral region near the rest frame velocity arises from the presence of ambient gas near the central velocity in the synthetic image. {When the source distance is larger, the synthetic image becomes smaller within the primary beam, and more of the larger-scale outflow emission close to zero velocities becomes recoverable by the interferometer.} 
{The default value of 0.5 mm is selected for the Precipitable Water Vapor, signifying excellent weather conditions for ALMA Band 6. This choice aims to mitigate the influence of significant noise in our determination of outflow properties and instead concentrate on the effects of interferometric observations. The synthetic beam dimensions are $0.63^{\prime\prime}\times 0.55^{\prime\prime}$. The original synthetic image has a cell resolution of $0.1^{\prime\prime}$ at a distance of 2~kpc, which is smaller than the synthetic beam size. We retain the resolution of $0.1^{\prime\prime}$ as the cell size for the tclean image.} Figure~\ref{fig.12CO21-cos0525-allmass-4color-casa-alma-dis2} illustrates the synthetic ALMA observations of \co\ (2-1) for outflows at different evolutionary stages, each with an inclination angle of 58\deg. The figure reveals the presence of artificial patterns resulting from the imperfect coverage in visibility space plus the CLEAN reconstruction process. It is important to note that interferometric observations typically filter out the emission from the ambient gas located at the central (or systemic) velocity, which we have omitted from the figure.

In addition, we present in Figure~\ref{fig.12CO21-12msun-allinclination-4color-casa-alma-dis2} the synthetic ALMA observations of \co\ (2-1) from the outflow from a 12~\msun\ star at different inclination angles. As also seen in Figure~\ref{fig.co21-mass12-allinclination}, we observe here that the high-velocity components of the emission become increasingly prominent as the inclination angle decreases.


Figures~\ref{fig.co21-mass12-0525-spec-linear}, \ref{fig.co21-mass12-0525-spec-log} and \ref{fig.12CO_13CO_12msun_0525_spec_v1000_log} display the \co~(2-1) spectra of outflows at various evolutionary stages post-processed using CASA/\texttt{simalma} in dotted lines for the case of an inclination angle of 58\deg. These figures also show the post-processed \co~(2-1) spectra of the outflow from a protostar of 12~\msun\ at different inclination angles in dotted lines. Similar to the raw synthetic spectra without post-processing using CASA/\texttt{simalma}, we observe a broader high-velocity component in the spectra as the protostellar mass increases, and wider line wings as the inclination angle decreases. Additionally, it is important to note that the drop in emission near the central velocity is a result of missing flux due to incomplete coverage of the $uv$-plane, i.e., missing short baselines. We also see a considerable intensity drop in the high-velocity channels when compared to the raw spectra. This drop is attributed to the weak emission near the noise level at these high-velocity channels, rendering them unobservable. Moreover, we conducted additional synthetic CASA observations using both ALMA 12m and Atacama Compact Array (ACA), excluding Total Power (TP) observations. The rationale for excluding TP observations is that, assuming a source distance of 2 kpc, the original domain size is {26.5$^{\prime\prime}$}, while the primary beam of the 12m telescope at the \co~(2-1) frequency is {25.3$^{\prime\prime}$}. {TP imaging with simobserve may not be accurate when the sky model is smaller than 2.5 times the primary beam\footnote{This factor of 2.5 is specified in the source code of CASA, particularly in the script task\_simobserve.py.}.} 
Therefore, we focused on the 12m+ACA combination as the complementary test. In Figure~\ref{fig.12CO21_spec_alma_comp_ACA}, we present the post-processed synthetic \co~(2-1) spectrum of the outflow generated by a 12 \msun\ protostar at an inclination angle of 58\deg\ using CASA, considering both the 12m array and the ACA, along with observations using only the 12m array. The spectral shapes are nearly identical between the 12m array observations and the combined 12m array with ACA observations, with a slight offset in the intensity of the gas, approximately 25\%. The distinct drop in emission near the central velocity remains evident, indicating that the ACA is insufficient to recover the emission. {It is essential to note that the 25\% increase in flux observed with the 12m+ACA configuration might be partially attributed to the residual artifacts in interferometry patterns, where the brightness temperature observed by 12m+ACA exceeds that of the raw spectrum on the blue-shifted side, possibly due to emission above 1 $\sigma$ in each channel.} We evaluate the estimates of mass, momentum, and energy for synthetic outflows obtained from synthetic ALMA 12m+ACA \co~(2-1) observations, along with the corresponding correction factor detailed in Appendix~\ref{Estimates of Mass, Momentum, and Energy of Synthetic Outflows using ALMA+ACA}. The analysis reveals a slight difference, with the mass, momentum, and energy consistently higher when utilizing 12m+ACA compared to employing only the 12m array, by approximately 25\%. 
{Therefore,} the increased 25\% flux might exert a minor influence on the calculation of outflow properties.


\subsection{Estimating the Mass, Momentum and Energy of Synthetic Outflows}
\label{Estimating the Mass, Momentum and Energy of Synthetic Outflows}


\begin{figure*}[hbt!]
\centering
\includegraphics[width=0.32\linewidth]{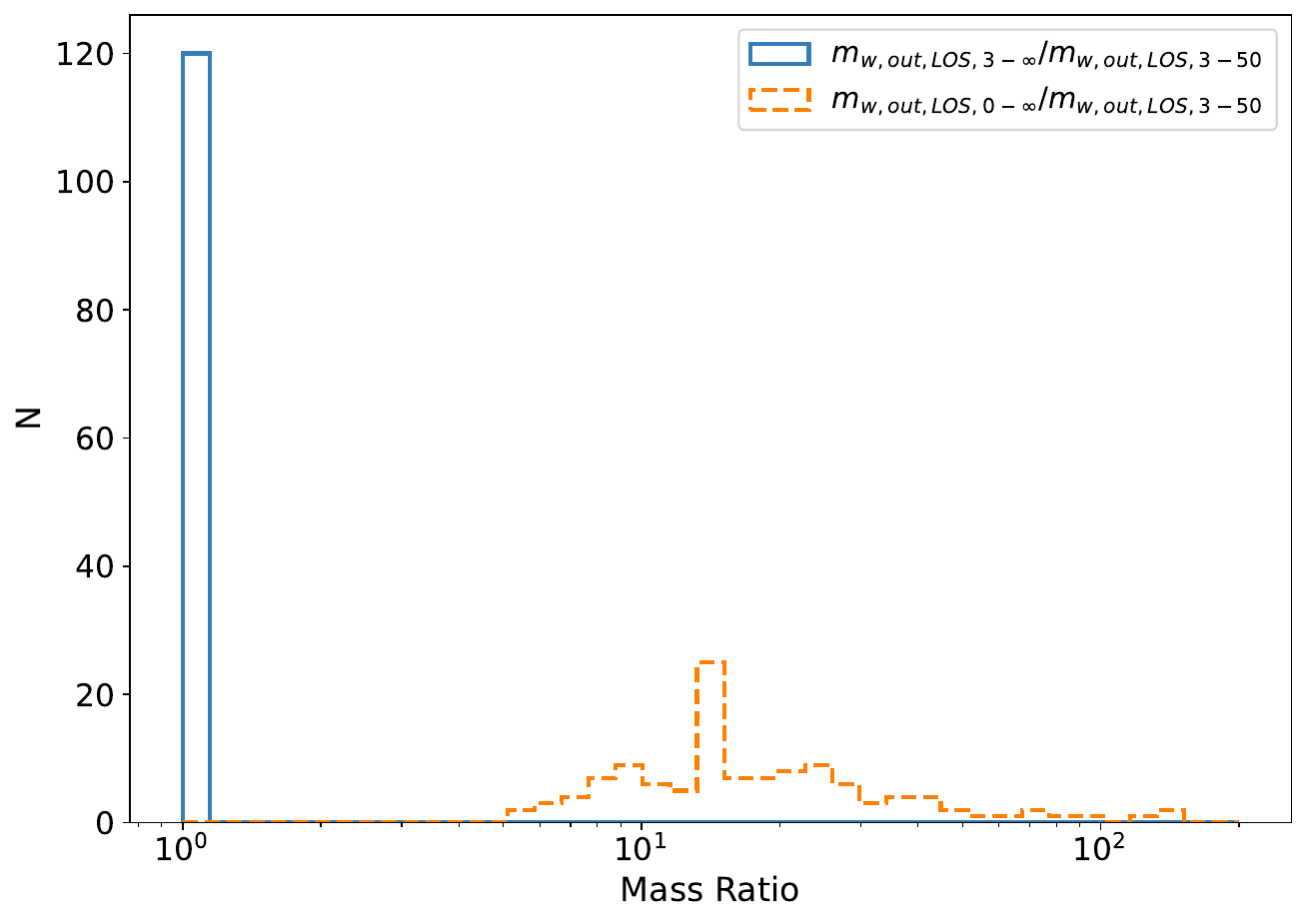}
\includegraphics[width=0.32\linewidth]{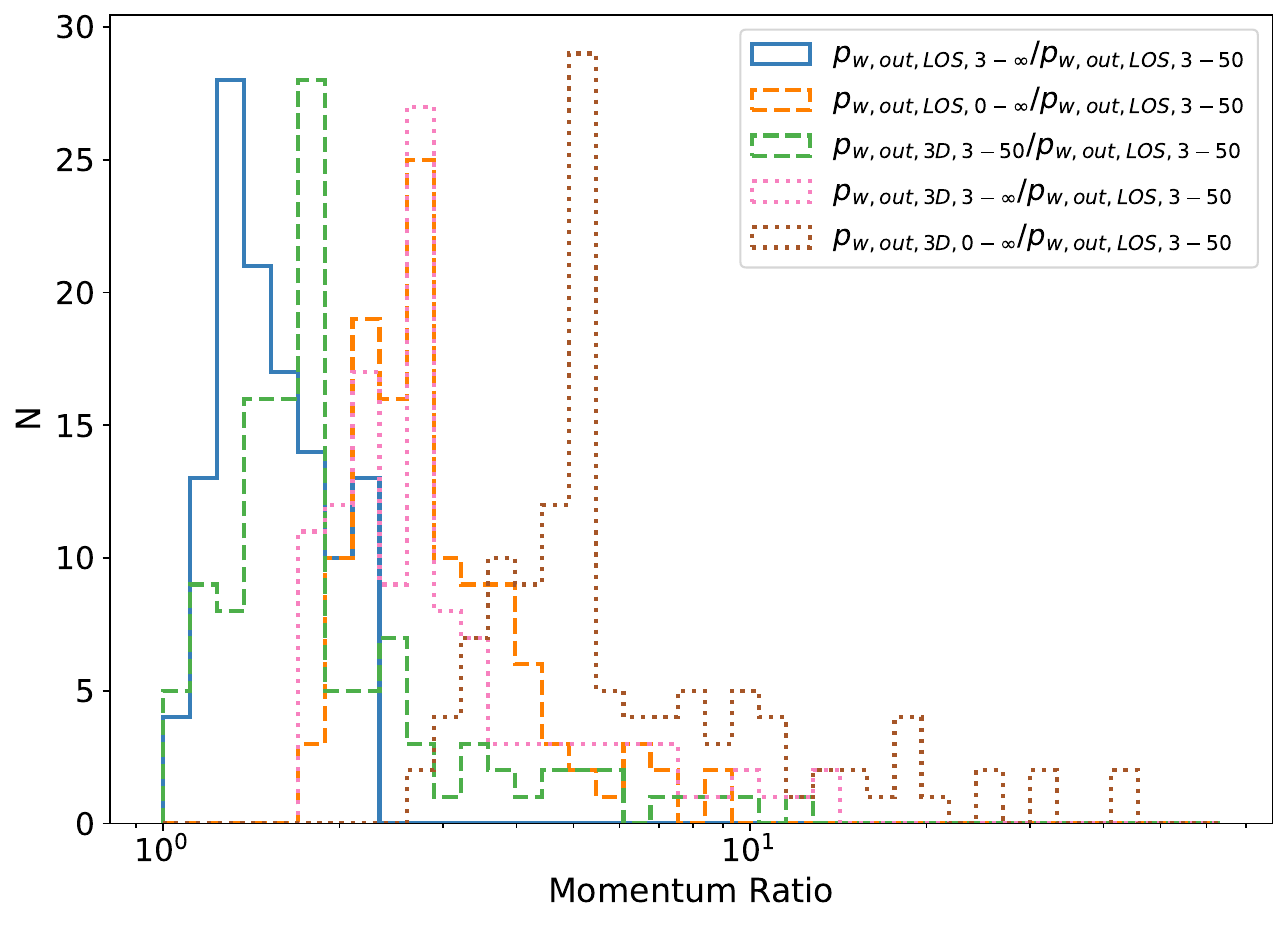}
\includegraphics[width=0.32\linewidth]{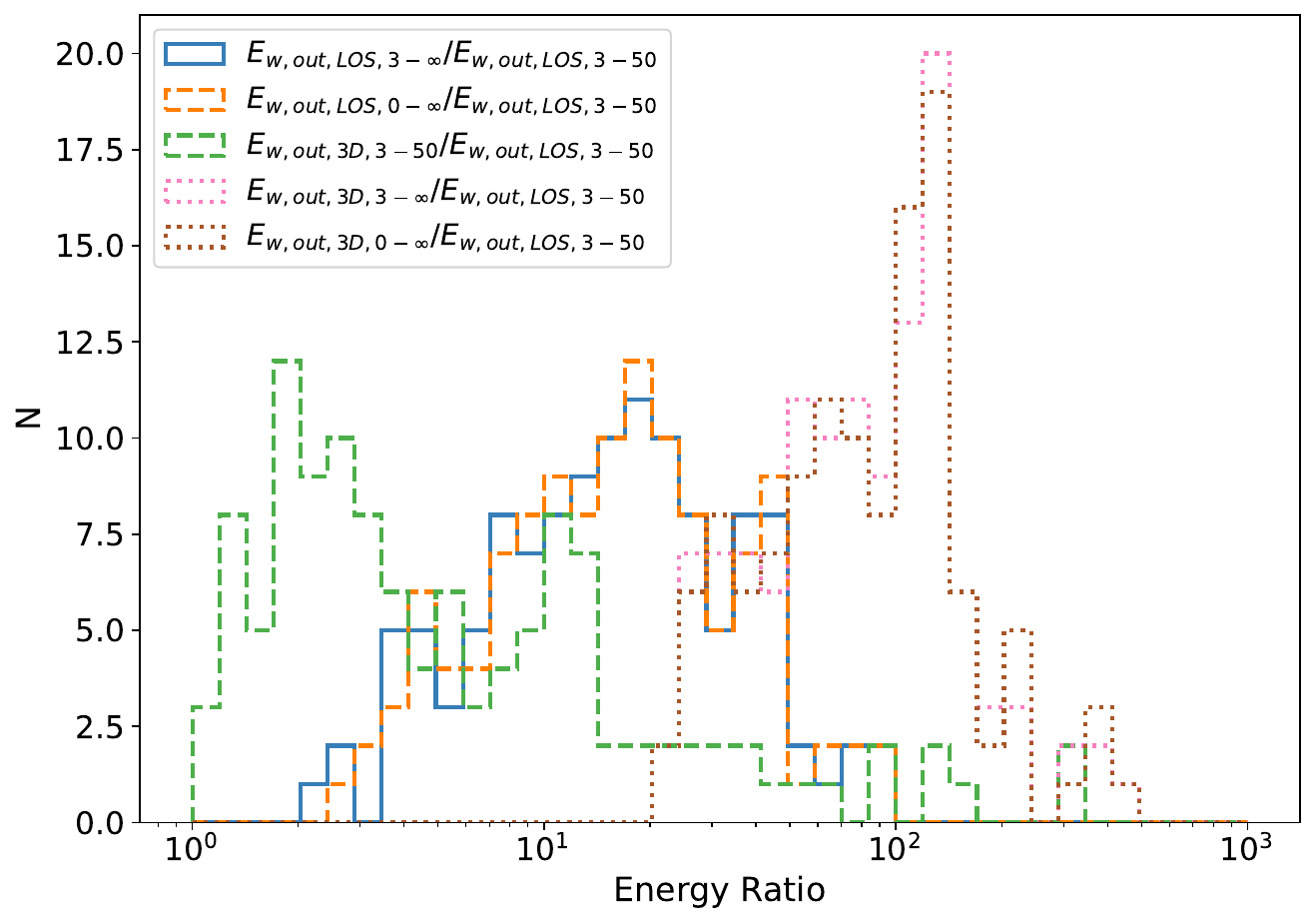}
\caption{{\it (a) Left:} Histogram showing the ratios of outflow mass for different LOS velocity cuts. {\it (b) Middle:} Histogram showing the ratios of outflow momentum for different LOS velocity cuts and 3D velocity cuts. {\it (c) Middle:} Histogram showing the ratios of outflow energy for different LOS velocity cuts and 3D velocity cuts.  }
\label{fig.hist_ratio_LOS_vrange_Mass_momentum_energy}
\end{figure*} 

\begin{figure*}[hbt!]
\centering
\includegraphics[width=0.32\linewidth]{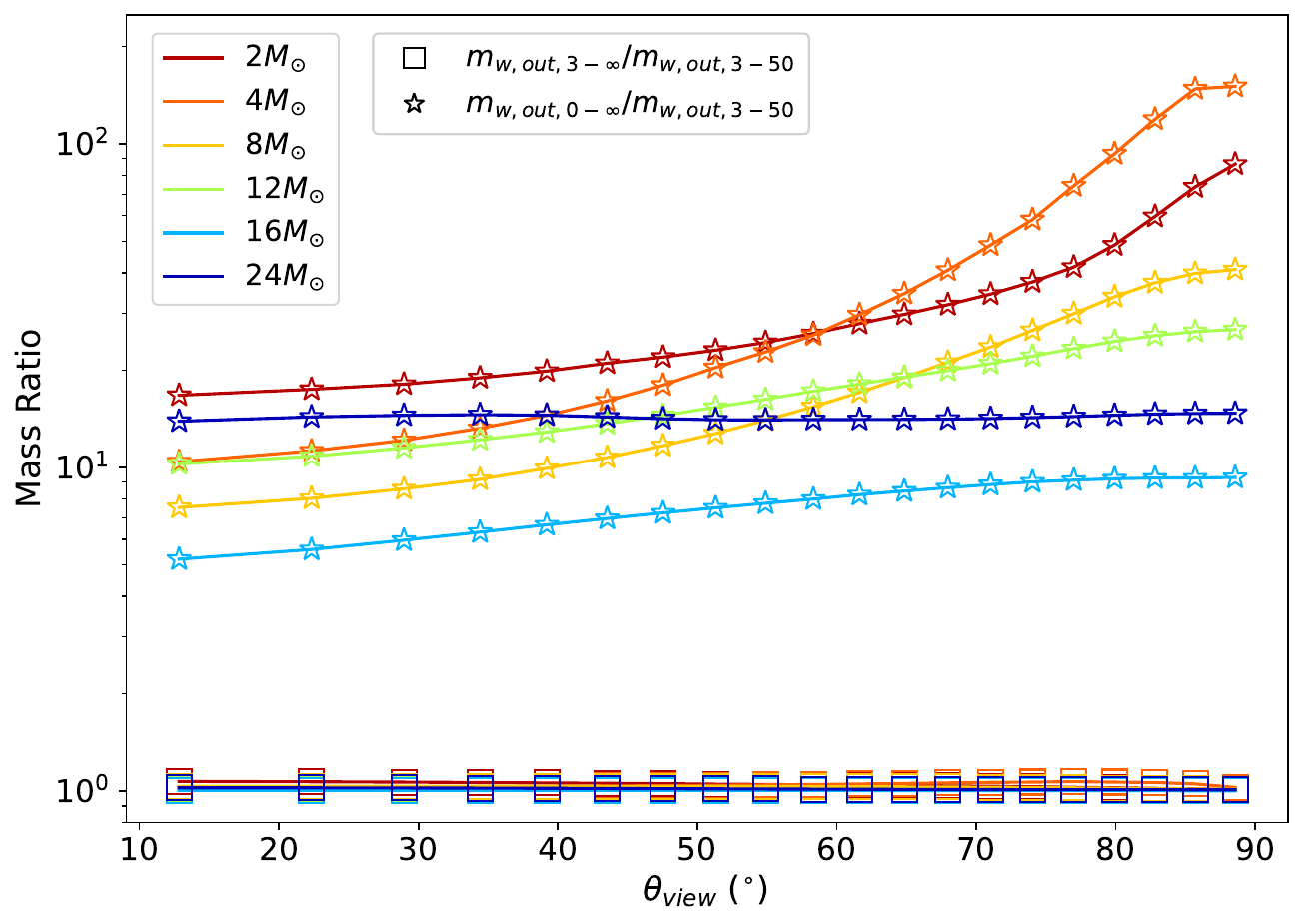}
\includegraphics[width=0.32\linewidth]{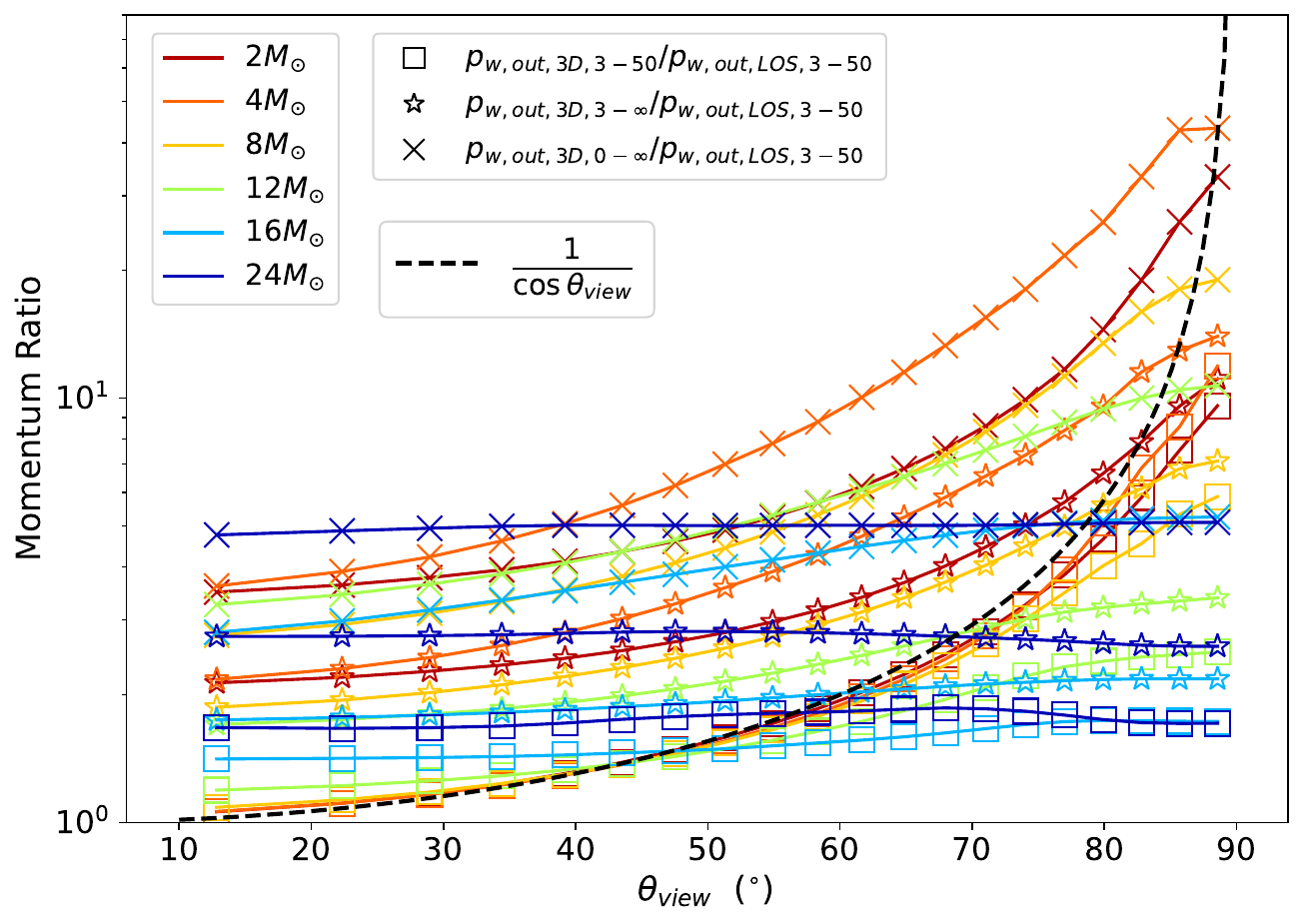}
\includegraphics[width=0.32\linewidth]{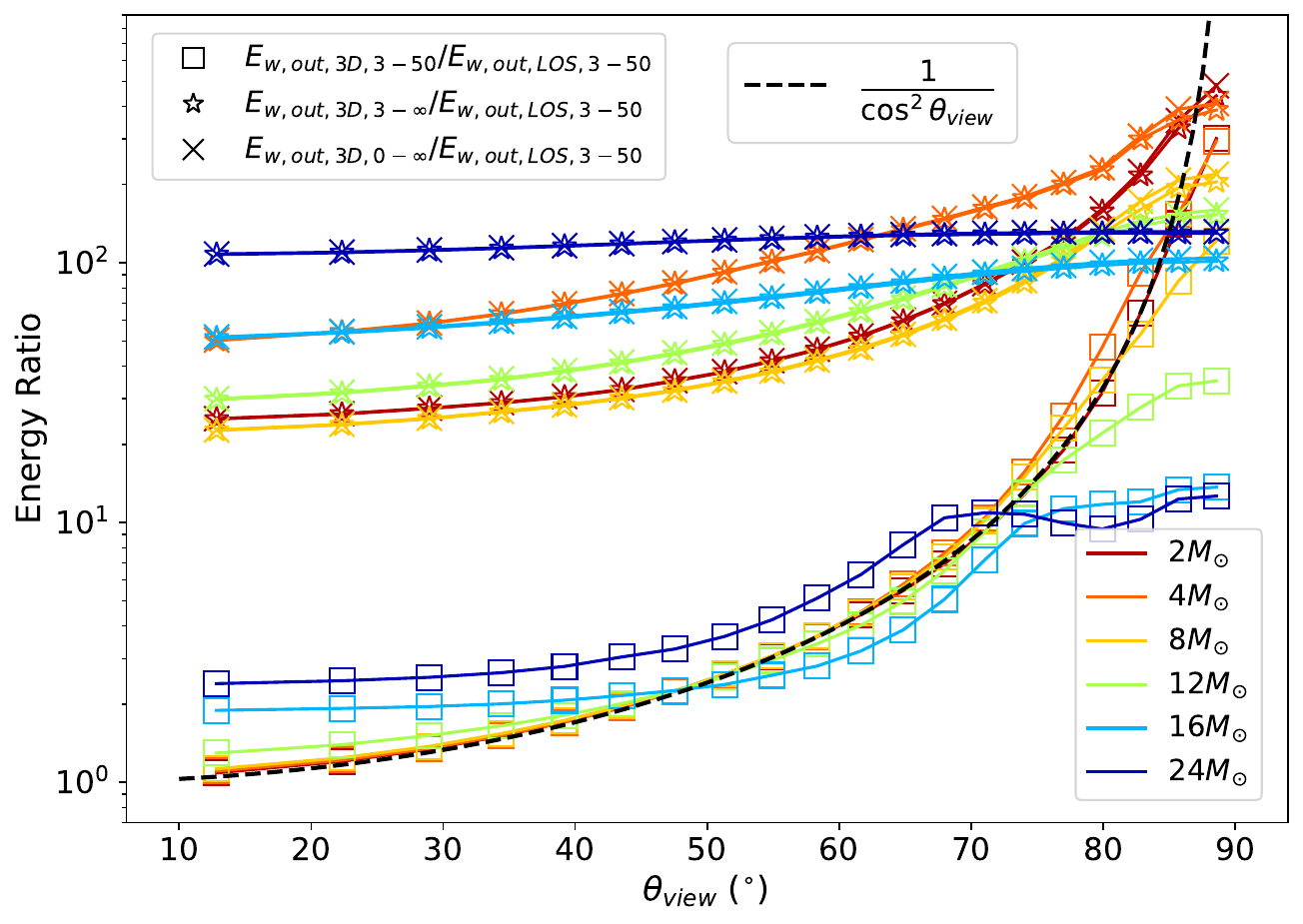}
\caption{{\it (a) Left:} Variation of the ratios of outflow mass for different LOS velocity cuts as a function of inclination of viewing angle $\theta_{\rm view}$. {\it (b) Middle:} Variation of the ratios of outflow momentum for different LOS velocity cuts as a function of inclination of viewing angle $\theta_{\rm view}$. {\it (c) Middle:} Variation of the ratios of outflow energy for different LOS velocity cuts as a function of inclination of viewing angle $\theta_{\rm view}$. }
\label{fig.scatter_outflow_mass_momentum_energy_frac_vs_incli}
\end{figure*} 



In this section, we evaluate the estimation of outflow mass, momentum, and energy derived from synthetic molecular line emission and compare it with the true values acquired from simulations. We delineate and quantify various factors {impacting the estimates of the physical} properties of outflows. These factors encompass different viewing angles and velocity cuts for outflows, denoted as ``Impact I," the {effects} stemming from radiative transfer when employing molecular line emission to compute outflow properties, referred to as ``Impact II," and the {effects introduced by} synthetic ALMA observations post-processed by CASA, labeled as ``Impact III." 

\subsubsection{Impact {\sc I}: Viewing Angle and Velocity Cut}

First, we investigate the impact of a line-of-sight (LOS) velocity cut on the estimation of outflow mass, momentum, and energy. We perform this analysis using only the simulation data without incorporating synthetic observations. In the simulations, the outflow gas can be defined as the gas moving outward from the central source above some given velocity threshold.
In real observations, 
it can be difficult to distinguish between the ambient gas and the outflow gas when the LOS velocity is close to the turbulent velocity of the cloud. 
In addition, it can be challenging to cover the entire velocity range of the outflow in the spectra with enough signal to noise to measure the highest velocity components.
To address these issues, we apply LOS velocity cuts as masks on the outflow gas and define the observed outflow mass as $m_{\rm w,out,LOS,3-50}$, representing the mass of outflowing gas with an absolute LOS velocity between 3 and 50 \kms. We also define $m_{\rm w,out,LOS,3-\infty}$ as the outflow mass with an absolute LOS velocity over 3 \kms, which helps quantify the missing mass at $v_{\rm LOS}>50$ \kms. Furthermore, we define the total outflow mass, denoted as $m_{\rm w,out,LOS,0-\infty}$, which encompasses the outflow gas covering the entire velocity range. The ratio $\rm m_{\rm w,out,LOS, 0-\infty}/m_{\rm w,out, LOS,3-50}$ provides insights into the missing mass near the rest frame velocity at $v_{\rm LOS}<3$ \kms\ and at $v_{\rm LOS}>50$ \kms. 

Figure~\ref{fig.hist_ratio_LOS_vrange_Mass_momentum_energy}a presents the distribution of $m_{\rm w,out, LOS,3-\infty}/m_{\rm w,out, LOS,3-50}$ and $m_{\rm w,out,LOS, 0-\infty}/m_{\rm w,out,LOS, 3-50}$. Table~\ref{tab:Table_ratio_M_P_E_simulation} displays the statistical values for these ratios. It is evident that the outflow gas mass with LOS velocities exceeding 50 \kms\ is negligible compared to the mass with LOS velocities ranging from 3 to 50 \kms. 
However, there is a substantial amount of mass, approximately a factor of 10 or more, that is being overlooked near the rest frame velocity ($v_{\rm LOS}<3$ \kms).

We then establish the definition for observed outflow LOS momentum, denoted as $p_{\rm w,out,LOS,3-50}$, which encompasses outflow gas with an absolute LOS velocity ranging from 3 to 50 \kms. Additionally, $p_{\rm w,out,LOS,3-\infty}$ is defined as the outflow LOS momentum derived from gas with an absolute LOS velocity over 3 \kms, while $p_{\rm w,out,LOS,0-\infty}$ represents the total outflow LOS momentum. Considering that outflowing gas moves in 3D space and not solely along the LOS direction, we also define the outflow 3D momentum, denoted as $p_{\rm w,out,3D,3-50}$, $p_{\rm w,out,3D,3-\infty}$, and $p_{\rm w,out,3D, 0-\infty}$, representing the outflow 3D momentum with different LOS velocity cutoffs. Analogously, we define the outflow LOS energy and outflow 3D energy. 

Figure~\ref{fig.hist_ratio_LOS_vrange_Mass_momentum_energy}b illustrates the distribution of outflow momentum ratios. The ratio $p_{\rm w,out,LOS,3-\infty}/p_{\rm w,out,LOS,3-50}$ denotes the missing outflow LOS momentum due to the high-velocity cutoff ($v_{\rm LOS}>50$ \kms). On the other hand, $p_{\rm w,out,LOS,0-\infty}/p_{\rm w,out,LOS,3-50}$ indicates the missing outflow LOS momentum from the gas near the rest frame velocity ($v_{\rm LOS}<3$ \kms) and at high velocities ($v_{\rm LOS}>50$ \kms). Additionally, $p_{\rm w,out,3D,3-50}/p_{\rm w,out,LOS,3-50}$ signifies the conversion factor between the outflow 3D momentum and the 1D LOS momentum for the observed outflow gas with an absolute velocities from 3 and 50 \kms. Finally, $p_{\rm w,out,3D,0-\infty}/p_{\rm w,out,LOS,3-50}$ represents the total conversion factor between the outflow total 3D momentum and the observed 1D LOS momentum. We observed that the high-velocity gas with $v_{\rm LOS}>50$ \kms, while having almost negligible outflow mass, carries a noticeable amount of momentum due to its high velocity. Conversely, the gas near the rest frame velocity ($v_{\rm LOS}<3$ \kms), which accounts for a factor of ten or more in mass compared to the observed mass, contributes to the momentum by only a factor of 2. 
Overall, when estimating the total 3D momentum from the LOS momentum with an absolute LOS velocity ranging from 3 to 50 \kms, there is a likely underestimation of the total 3D momentum by a factor of {7} and up to {43}. 

Figure~\ref{fig.hist_ratio_LOS_vrange_Mass_momentum_energy}c displays the distribution of outflow energy ratios. Notably, the high-velocity gas with $v_{\rm LOS}>50$ \kms\ carries a significant amount of energy, while the gas near the rest frame velocity ($v_{\rm LOS}<3$ \kms) contributes negligible outflow energy. In general, when estimating the total 3D energy from the LOS energy with an absolute LOS velocity ranging from 3 to 50 \kms, there is a likely underestimation of the total 3D energy by a factor of {100} and up to {480}. We present the statistical values for all the ratios discussed above in Table~\ref{tab:Table_ratio_M_P_E_simulation}.

\begin{table*}[htbp]
  \centering
  \caption{Table of mass, momentum and energy ratios for different LOS velocity cuts}
    \begin{tabular}{|c|c|c|c|c|c|}
    \hline
   Ratios & Mean  & Median & Min   & Max   & Std \\\hline\hline
    $m_{\rm w,out,LOS, 3-\infty}/m_{\rm w,out,LOS, 3-50}$     & 1.025 & 1.020 & 1.003 & 1.072 & 0.019 \\\hline
    $m_{\rm w,out,LOS, 0-\infty}/m_{\rm w,out,LOS, 3-50}$    & 23.68 & 14.71 & 5.207 & 150.5 & 24.33 \\\hline
    $p_{\rm w,out,LOS, 3-\infty}/p_{\rm w,out,LOS, 3-50}$     & 1.563 & 1.473 & 1.085 & 2.308 & 0.340 \\\hline
    $p_{\rm w,out,LOS, 0-\infty}/p_{\rm w,out,LOS, 3-50}$     & 3.151 & 2.790 & 1.781 & 8.766 & 1.323 \\\hline
    $p_{\rm w,out,3D, 3-50}/p_{\rm w,out,LOS, 3-50}$    & 2.270 & 1.725 & 1.060 & 11.90 & 1.712 \\\hline
    $p_{\rm w,out,3D, 3-\infty}/p_{\rm w,out,LOS, 3-50}$     & 3.506 & 2.744 & 1.703 & 13.99 & 2.320 \\\hline
    $p_{\rm w,out,3D, 0-\infty}/p_{\rm w,out,LOS, 3-50}$     & 7.892 & 5.055 & 2.763 & 43.29 & 7.265 \\\hline
    $E_{\rm w,out,LOS, 3-\infty}/E_{\rm w,out,LOS, 3-50}$     & 21.23 & 16.50 & 2.375 & 92.08 & 17.44 \\\hline
    $E_{\rm w,out,LOS, 0-\infty}/E_{\rm w,out,LOS, 3-50}$     & 21.38 & 16.66 & 2.748 & 92.15 & 17.40 \\\hline
    $E_{\rm w,out,3D, 3-50}/E_{\rm w,out,LOS, 3-50}$    & 17.63 & 3.945 & 1.088 & 301.0 & 44.22 \\\hline
    $E_{\rm w,out,3D, 3-\infty}/E_{\rm w,out,LOS, 3-50}$     & 98.82 & 83.54 & 22.64 & 439.3 & 72.24 \\\hline
    $E_{\rm w,out,3D, 0-\infty}/E_{\rm w,out,LOS, 3-50}$     & 101.5 & 85.24 & 22.76 & 480.3 & 77.30 \\\hline
    \end{tabular}%
  \label{tab:Table_ratio_M_P_E_simulation}%
\end{table*}%

We then investigate how the viewing angle $\theta_{\rm view}$ affects the previously studied ratios. Figure~\ref{fig.scatter_outflow_mass_momentum_energy_frac_vs_incli} illustrates the variation of the ratios of outflow mass, momentum, and energy for different LOS velocity cuts as a function of the viewing angle $\theta_{\rm view}$. As $\theta_{\rm view}$ increases, there is an increase in the amount of missed mass, momentum, and energy, primarily caused by most of the outflowing gas moving perpendicular to the LOS, making them unobservable in spectra with a LOS velocity between 3 and 50 \kms. It is important to highlight that the cases involving 16~\msun\ and 24~\msun\ stars exhibit a more flattened trend in the ratios with respect to $\theta_{\rm view}$. This behavior can be attributed to the larger opening angle in these cases compared to the lower mass ones. As a result, the velocity distribution becomes more isotropic, with multiple launching directions rather than a single dominant one. By analyzing Figure~\ref{fig.scatter_outflow_mass_momentum_energy_frac_vs_incli}, we can derive correction factors to adjust the true 3D values based on the observed 1D LOS values for different outflows with various inclination angles and evolutionary stages.

\subsubsection{Impact {\sc I}$+$Impact {\sc II}: Radiative Transfer}

\begin{figure*}[hbt!]
\centering
\includegraphics[width=0.49\linewidth]{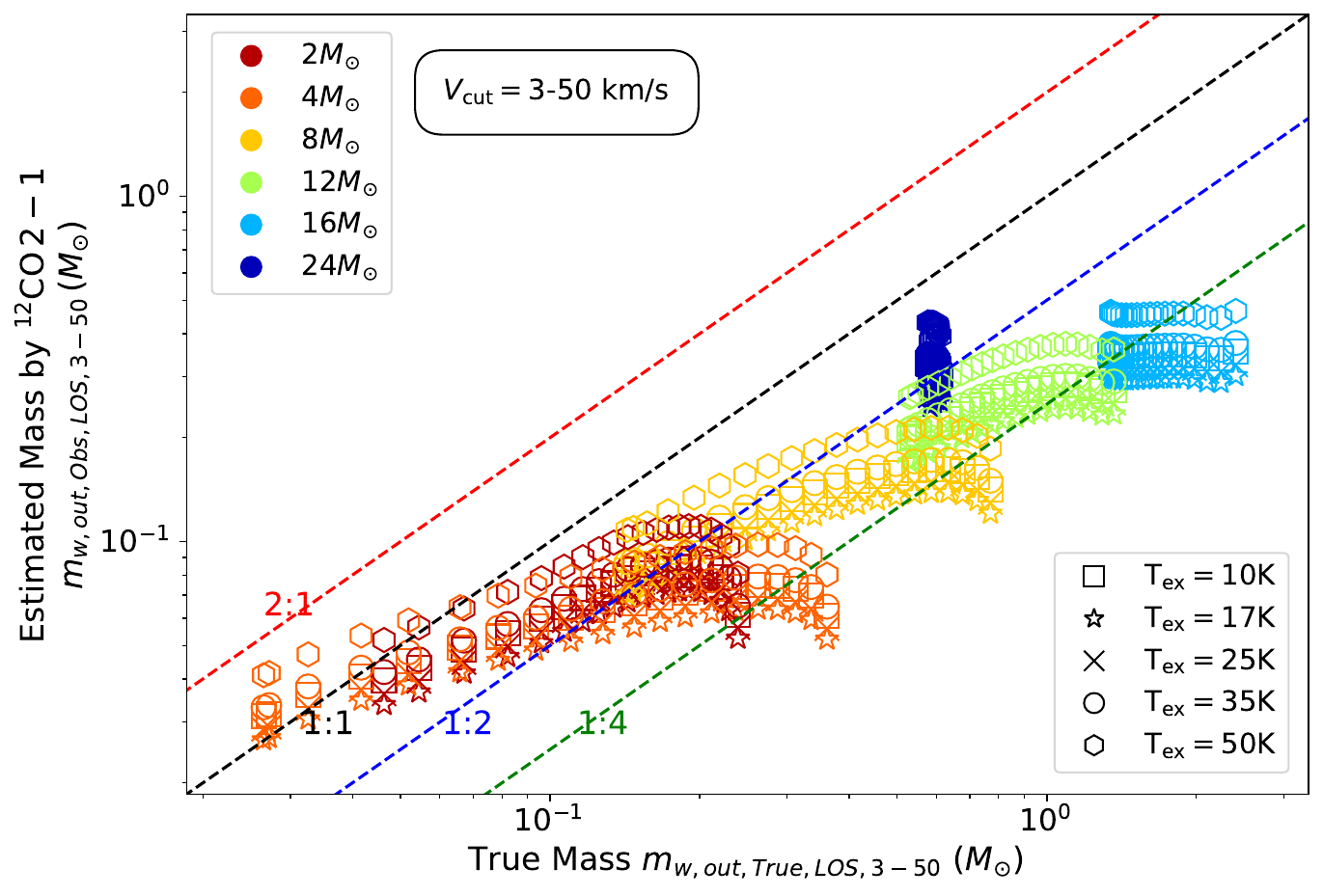}
\includegraphics[width=0.47\linewidth]{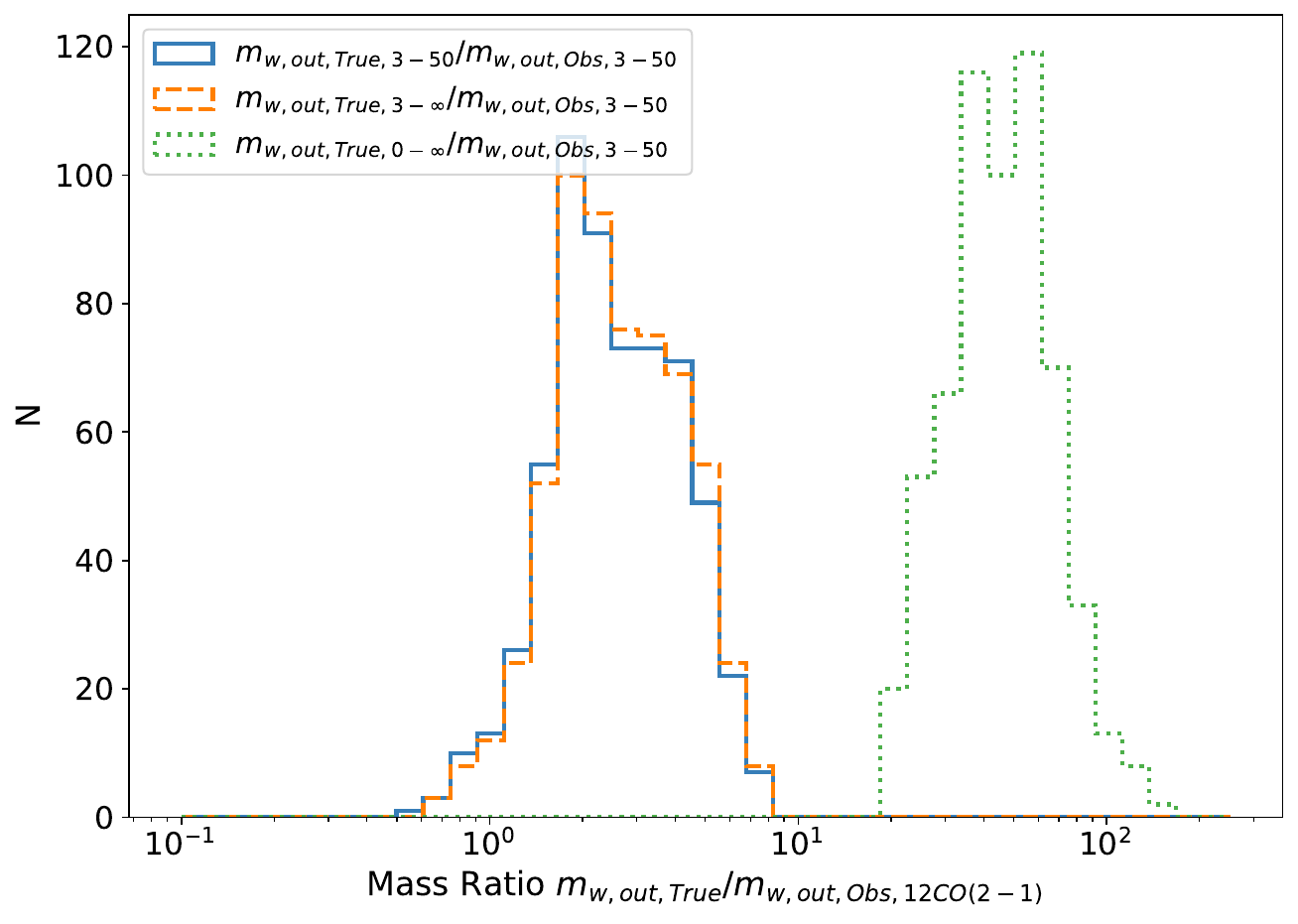}
\includegraphics[width=0.49\linewidth]{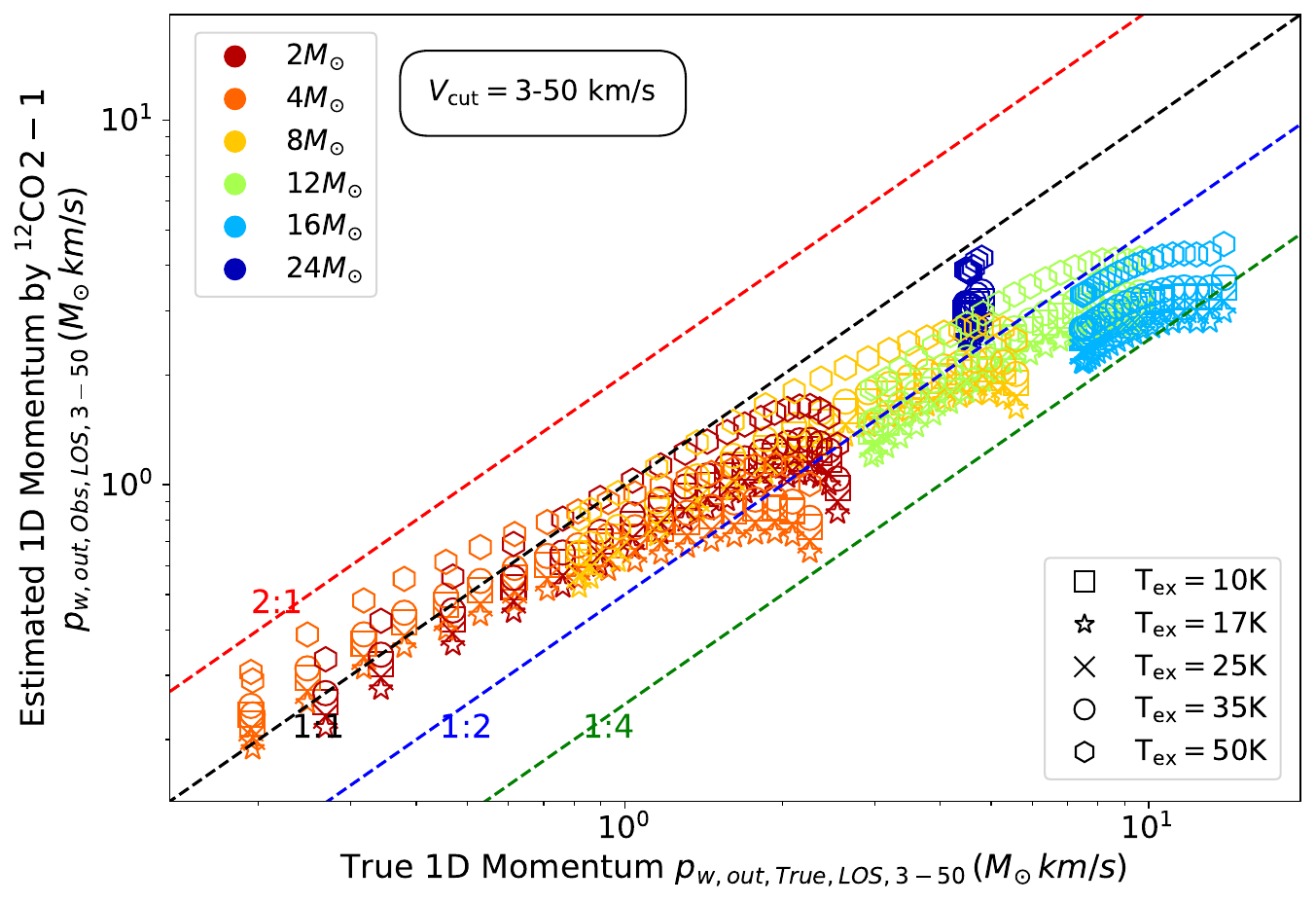}
\includegraphics[width=0.47\linewidth]{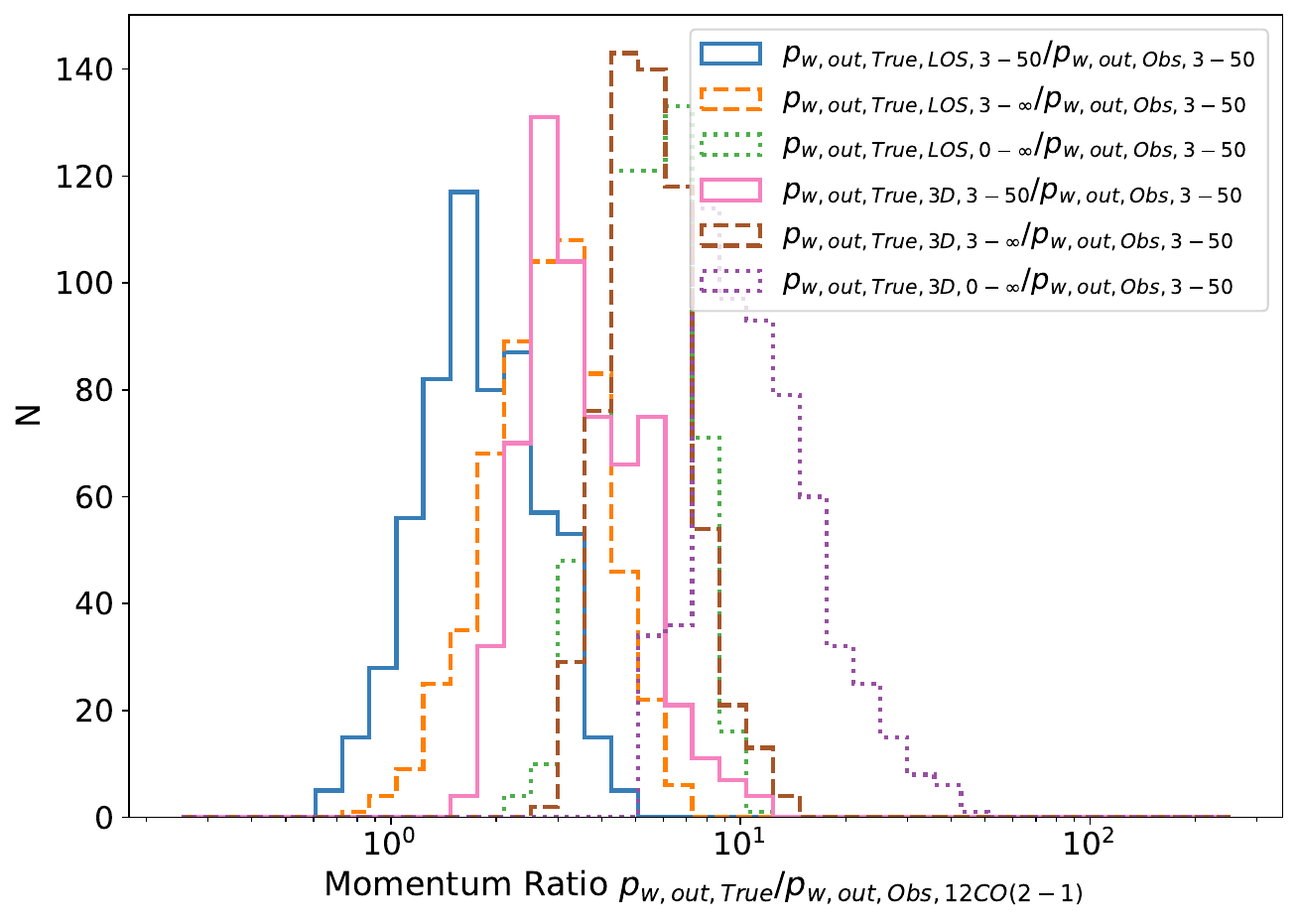}
\includegraphics[width=0.49\linewidth]{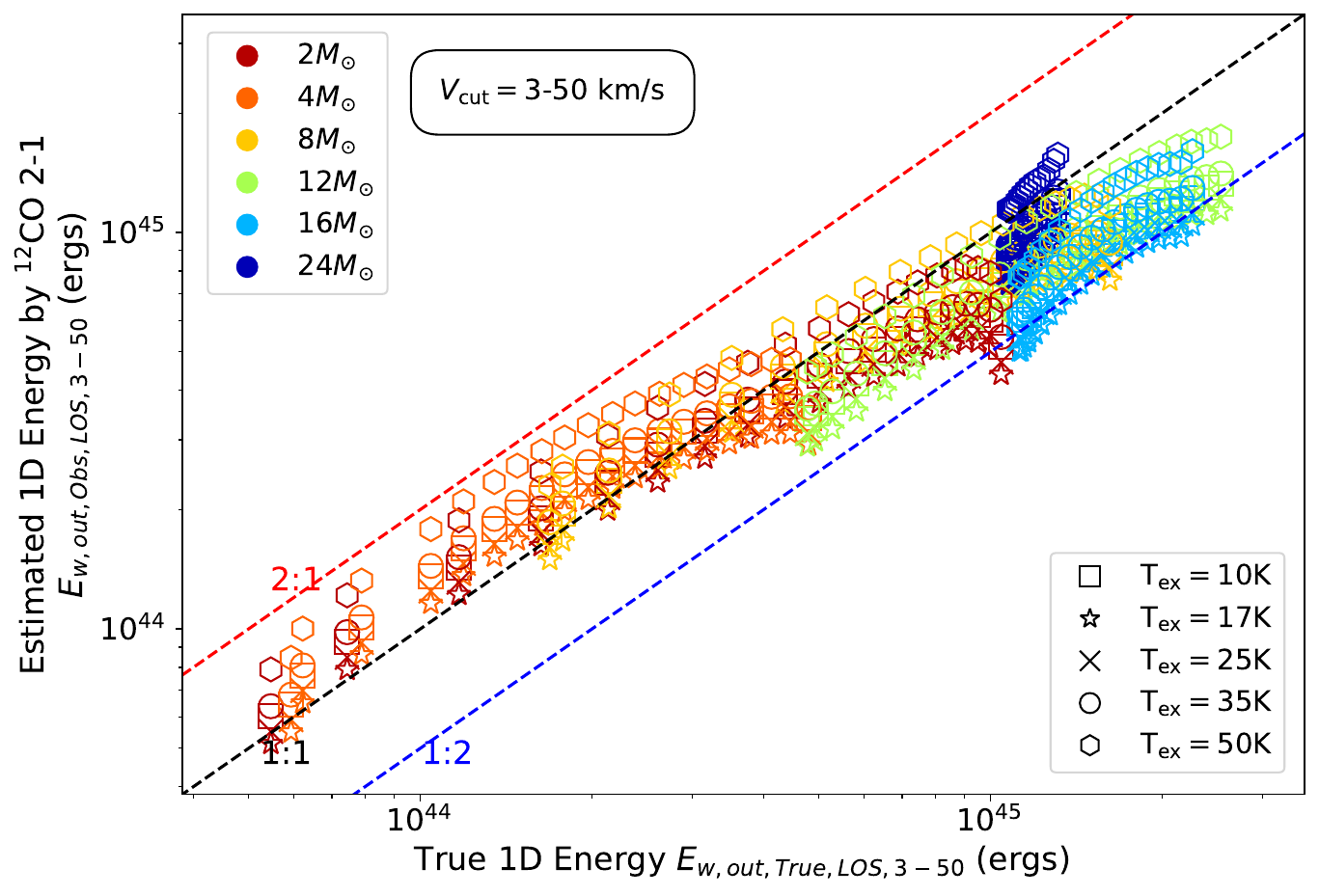}
\includegraphics[width=0.47\linewidth]{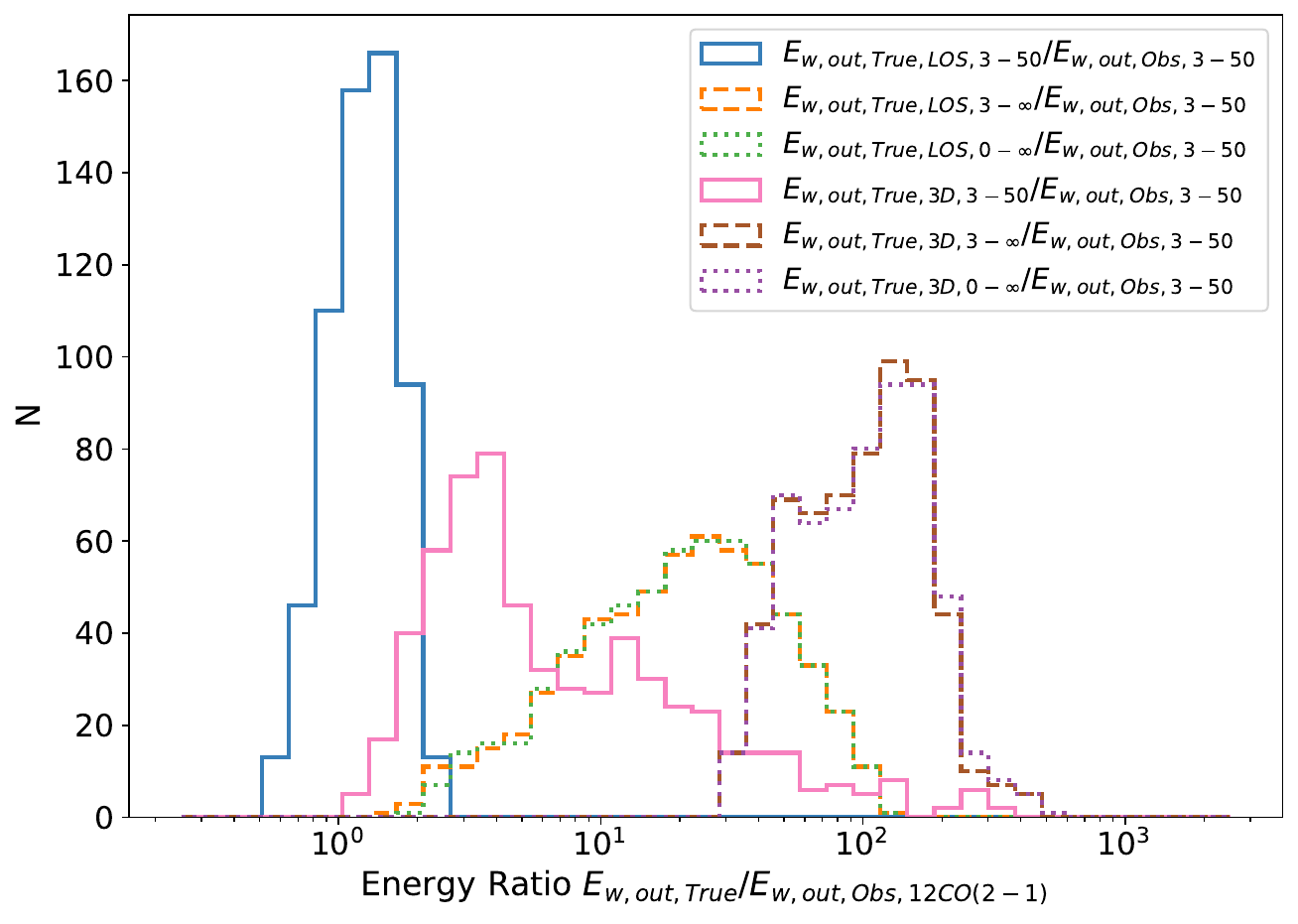}
\caption{{\it Left column:} Comparison of the mass ($1^{st}$ row), momentum ($2^{nd}$ row) and energy ($3^{rd}$ row) estimates obtained for different inclination angles using \co~(2-1) with a LOS velocity cutoff between 3 to 50 \kms, and the corresponding true outflow values with the same velocity cutoff. {\it Right column:}  the ratios between \co~(2-1) calculated mass ($1^{st}$ row), momentum ($2^{nd}$ row) and energy ($3^{rd}$ row) and the true simulation-derived values for different LOS velocity cuts.}
\label{fig.scatter_outflow_mass_P_E_comp_true_12co21_vcut}
\end{figure*} 

\begin{figure*}[hbt!]
\centering
\includegraphics[width=0.47\linewidth]{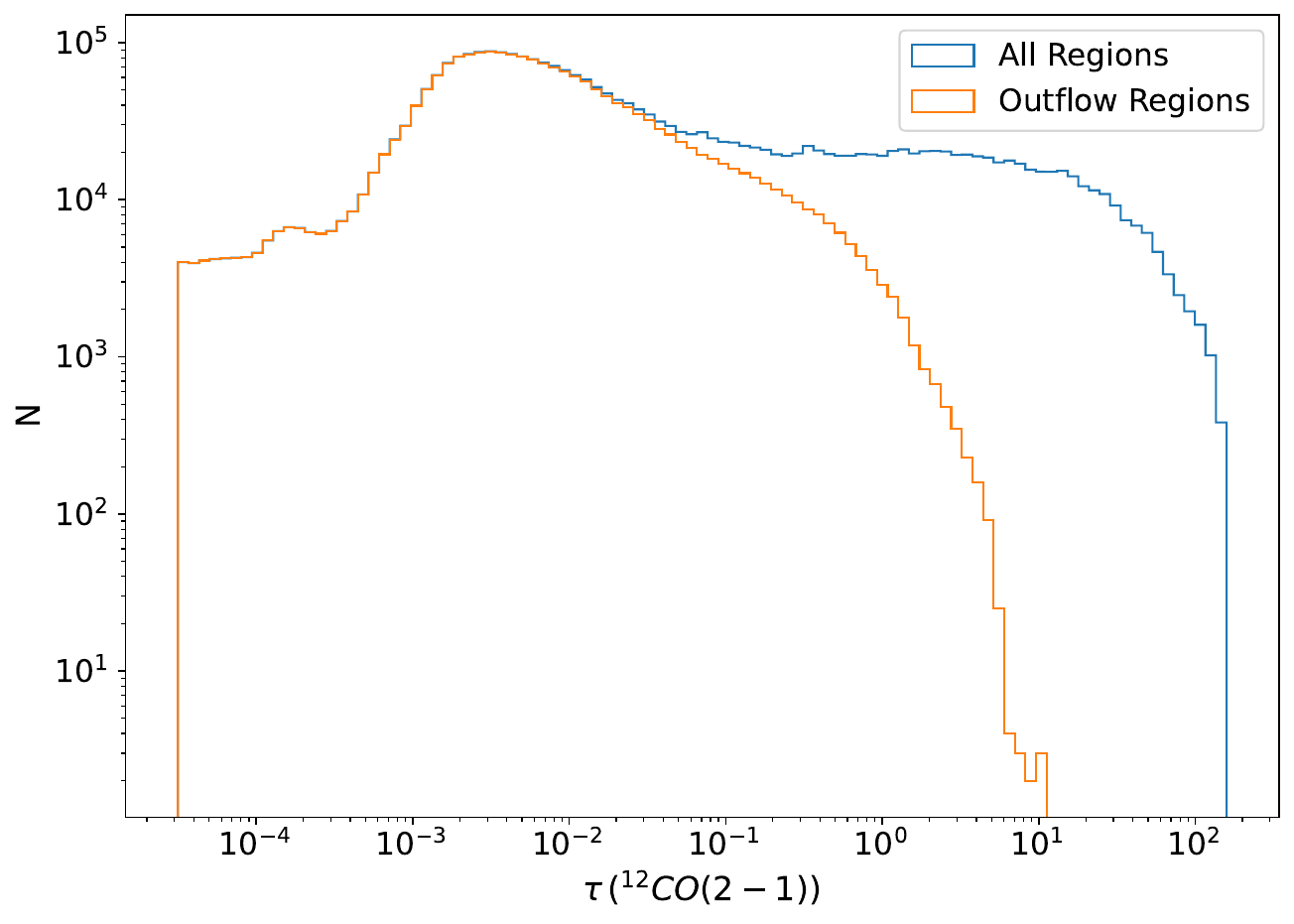}
\includegraphics[width=0.47\linewidth]{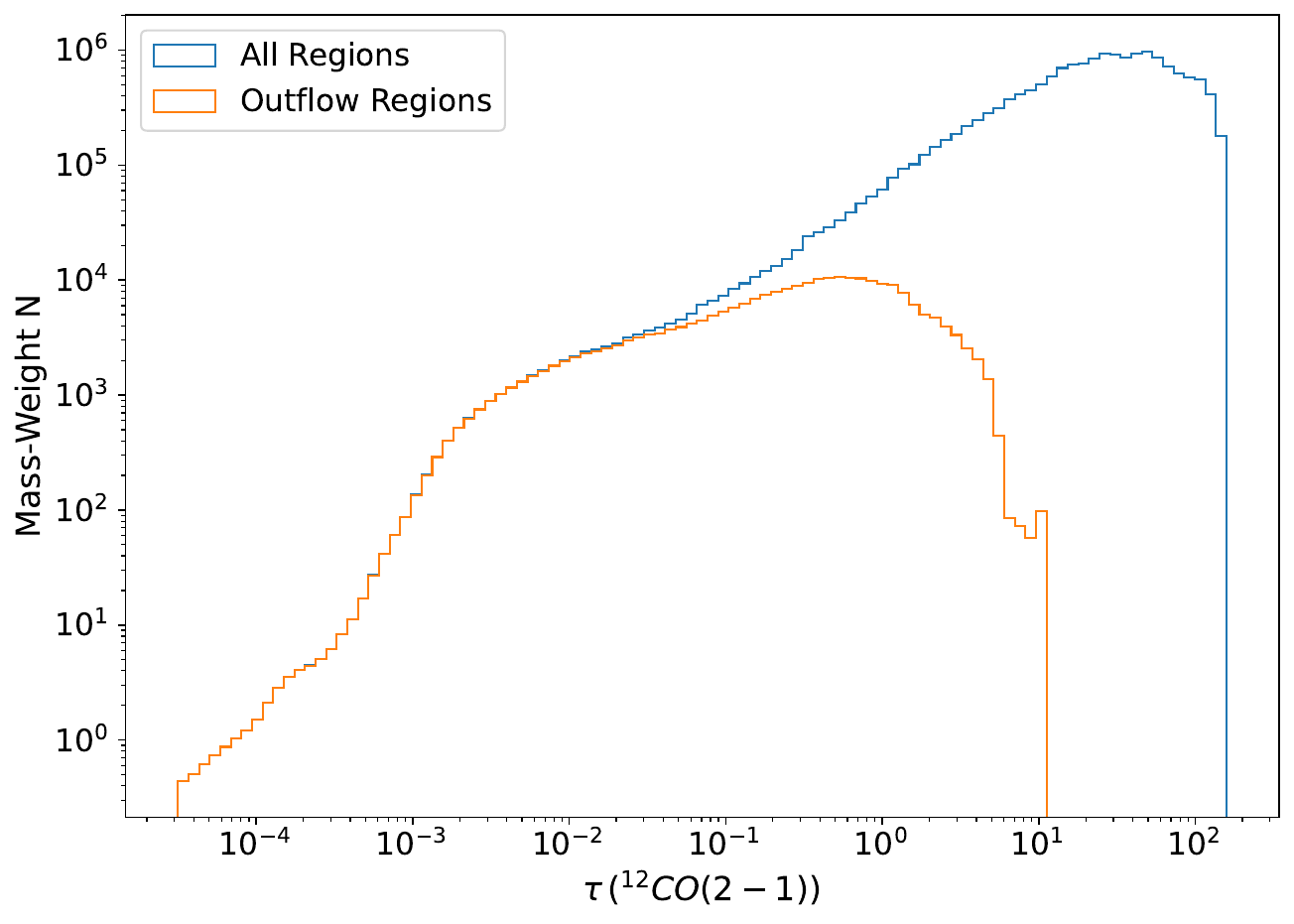}
\caption{Distribution of the optical depth for \co~(2-1) in the outflow generated by a 12 \msun\ protostar at an inclination angle of 58\deg, presented in terms of both volume-weighted (left) and mass-weighted (right).}
\label{fig.12CO21_optical_depth_hist}
\end{figure*}

We next determine the outflow mass, momentum and energy from the raw synthetic molecular spectra and compared it to the true outflow values obtained from simulations. To ensure a fair comparison, only the mass, momentum and energy with a LOS velocity above a certain threshold were considered in the true outflow values. This means that to obtain the total outflow mass momentum and energy, we must account for the missing fraction as previously mentioned. Our primary goal in this study is to assess the uncertainty in estimating outflow mass momentum and energy from molecular spectra.

The column density of CO molecules that are in the upper level of a transition from level $u \rightarrow l$ can be expressed as
\begin{equation}
N_{u,\rm CO}=\frac{8\pi k\nu^2}{hc^3 A_{ul}}\int T_{b} (V) dV
\label{equ-Nu}
\end{equation}
where $k$ is Boltzmann's constant, $\nu$ is the frequency of the transition from level $u \rightarrow l$,  $h$ is Planck's constant, $c$ is the speed of light, $A_{ul}$ is the spontaneous decay rate from upper level $u$ to lower level $l$, $T_b$ is the brightness temperature{, and $dV$ represents integration over all velocity channels. It is worth noting that Equation~\ref{equ-Nu} assumes the molecular line is optically thin to derive the column density.} The total CO column density $N_{\rm tot}$ is related to the upper level column density $N_{u}$ through 
\begin{equation}
N_{\rm tot, CO}=f_{u} f_b N_{u,\rm CO}.
\label{eq-Ntotal}
\end{equation}
In the equation above, the level correction factor $f_u$ can be determined analytically under the assumption of local thermodynamic equilibrium (LTE) as
\begin{equation}
f_{u}=\frac{Q(T_{\rm ex})}{g_u \exp\left(-\frac{h\nu}{kT_{\rm ex}}\right)}
\label{eq-fu}
\end{equation}
where $g_u$ is the statistical weight of the upper-level. $T_{\rm ex}$ is the excitation temperature and $Q(T_{\rm ex})=kT_{\rm ex}/hB_0$ is the LTE partition function, with $B_0$ being the rotational constant. The correction for the background is given by
\begin{equation}
f_b=\left[1-\frac{e^{\frac{h\nu}{kT_{\rm
ex}}}-1}{e^{\frac{h\nu}{kT_{\rm bg}}}-1}\right]^{-1}
\end{equation}
where $T_{\rm bg}$ is the background temperature, assumed to be 2.7K. The final mass is calculated as follows:
\begin{equation}
m_{\rm tot} = \sum_i  \mu m_{\rm H} d^{2} \frac{N_{\rm tot, CO}}{X_{\rm CO-H}},
\end{equation}
where $\mu = 1.4$ takes into account the mass of helium per H-nucleus, $m_{\rm H}$ is the mass of an H-nucleus, $d$ represents the physical scale of each pixel in the map, $X_{\rm CO-H}$ is the abundance ratio between CO and H-nuclei, and $\sum_i$ indicates the summation over all pixels.

The 1D LOS momentum and energy of outflows are calculated using the following equations:
\begin{equation}
\label{momentum-eq1}
p =\sum_{i,j,k}{X_{\rm CO-H}}m_{{\rm CO},i,j,k}|v_{i,j,k}-\overline{v_{\rm c}}|,
\end{equation}
\begin{equation}
\label{energy-eq1}
E =\sum_{i,j,k}\frac{1}{2}{X_{\rm CO-H}}m_{{\rm CO},i,j,k}(v_{i,j,k}-\overline{v_{\rm c}})^{2} ,
\end{equation}
where $m_{{\rm CO},i,j,k}$ represents the mass of CO in the $(i,j,k)$ grid cell in the position-position-velocity (ppv) cube, $X_{\rm CO-H}$ is the abundance ratio between CO and H-nuclei, $v_{i,j,k}$ represents the LOS velocity of the CO in that grid cell, and $\overline{v_{\rm c}}$ is the global central LOS velocity of the outflow.

\begin{figure*}[hbt!]
\centering
\includegraphics[width=0.49\linewidth]{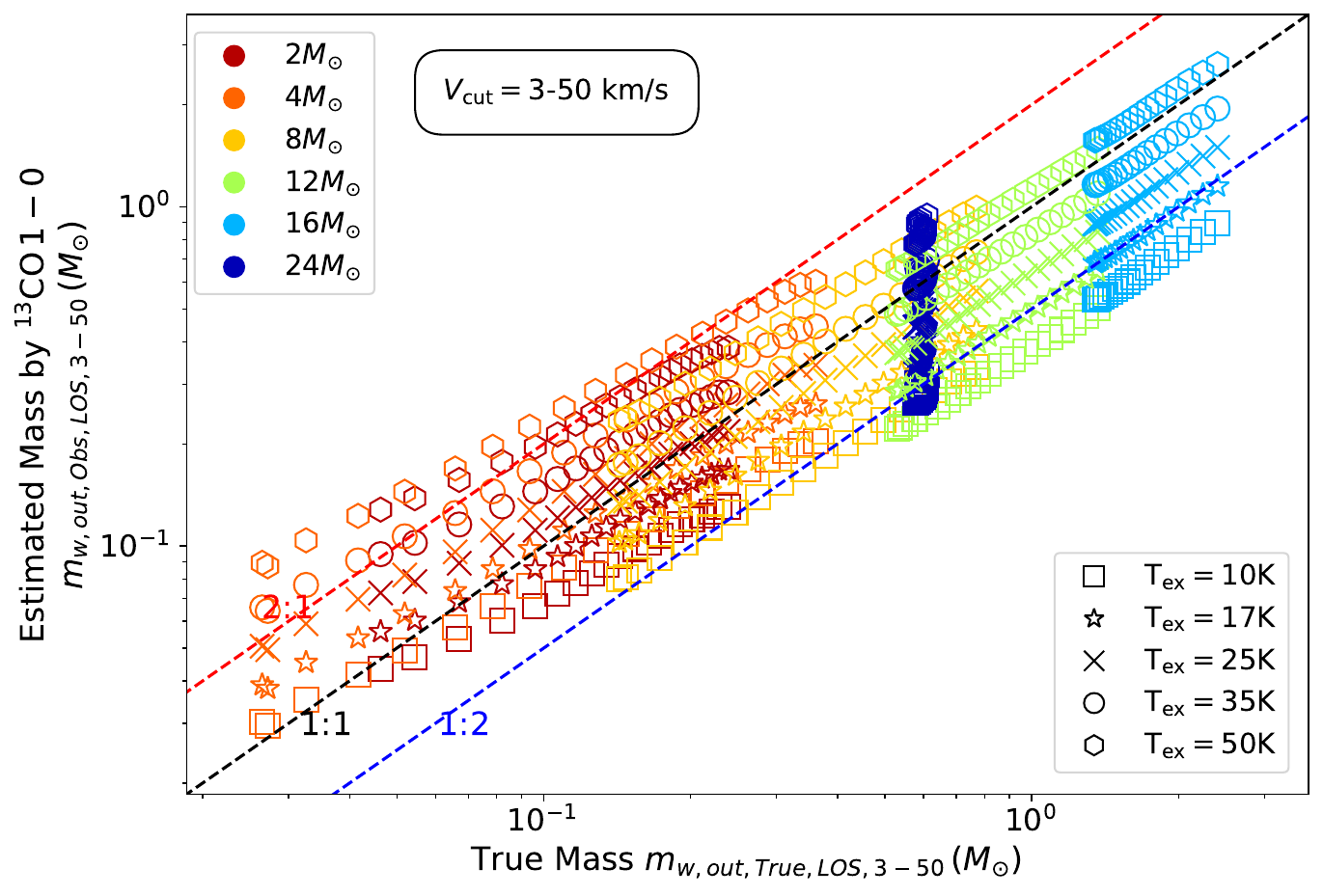}
\includegraphics[width=0.47\linewidth]{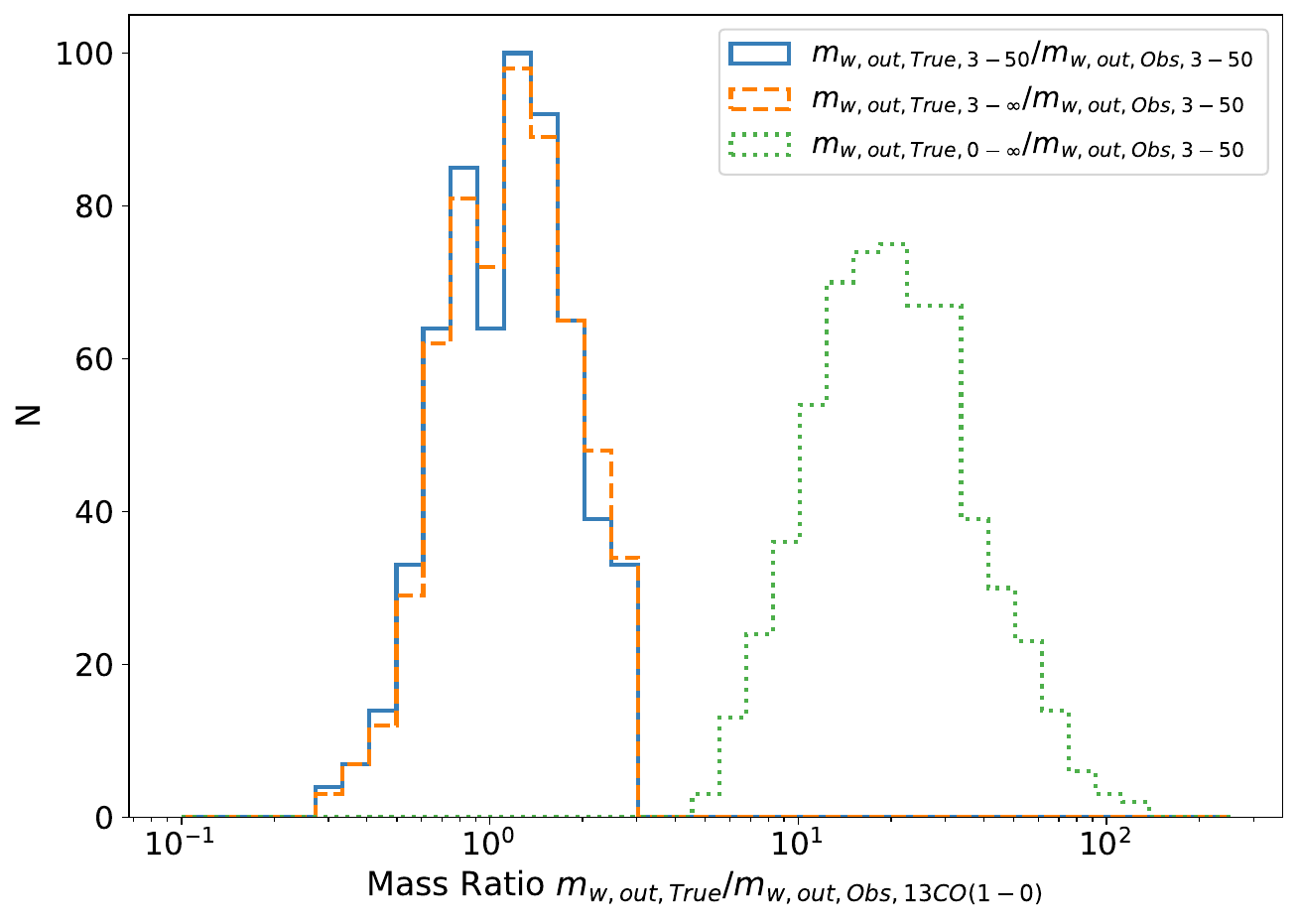}
\includegraphics[width=0.49\linewidth]{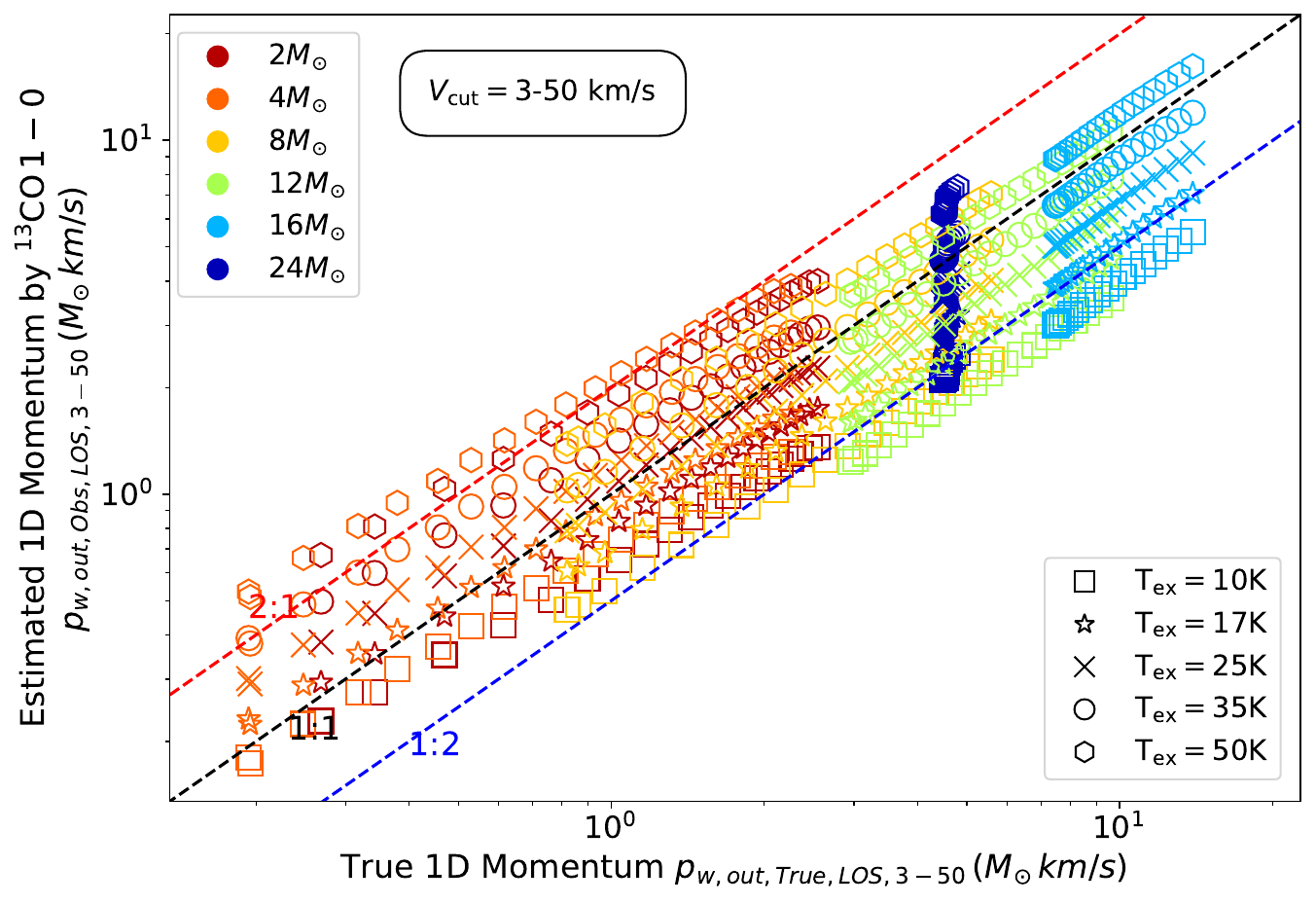}
\includegraphics[width=0.47\linewidth]{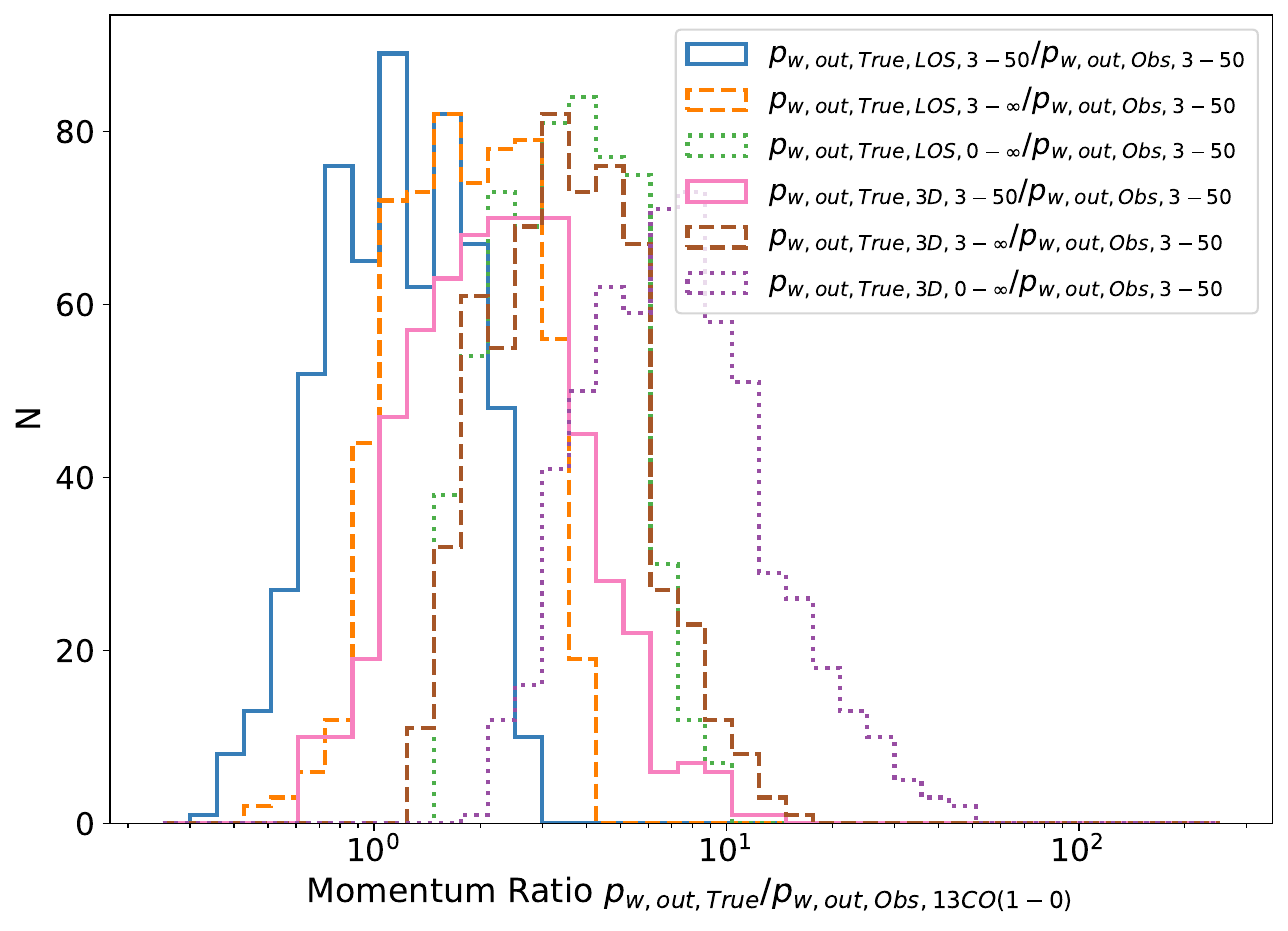}
\includegraphics[width=0.49\linewidth]{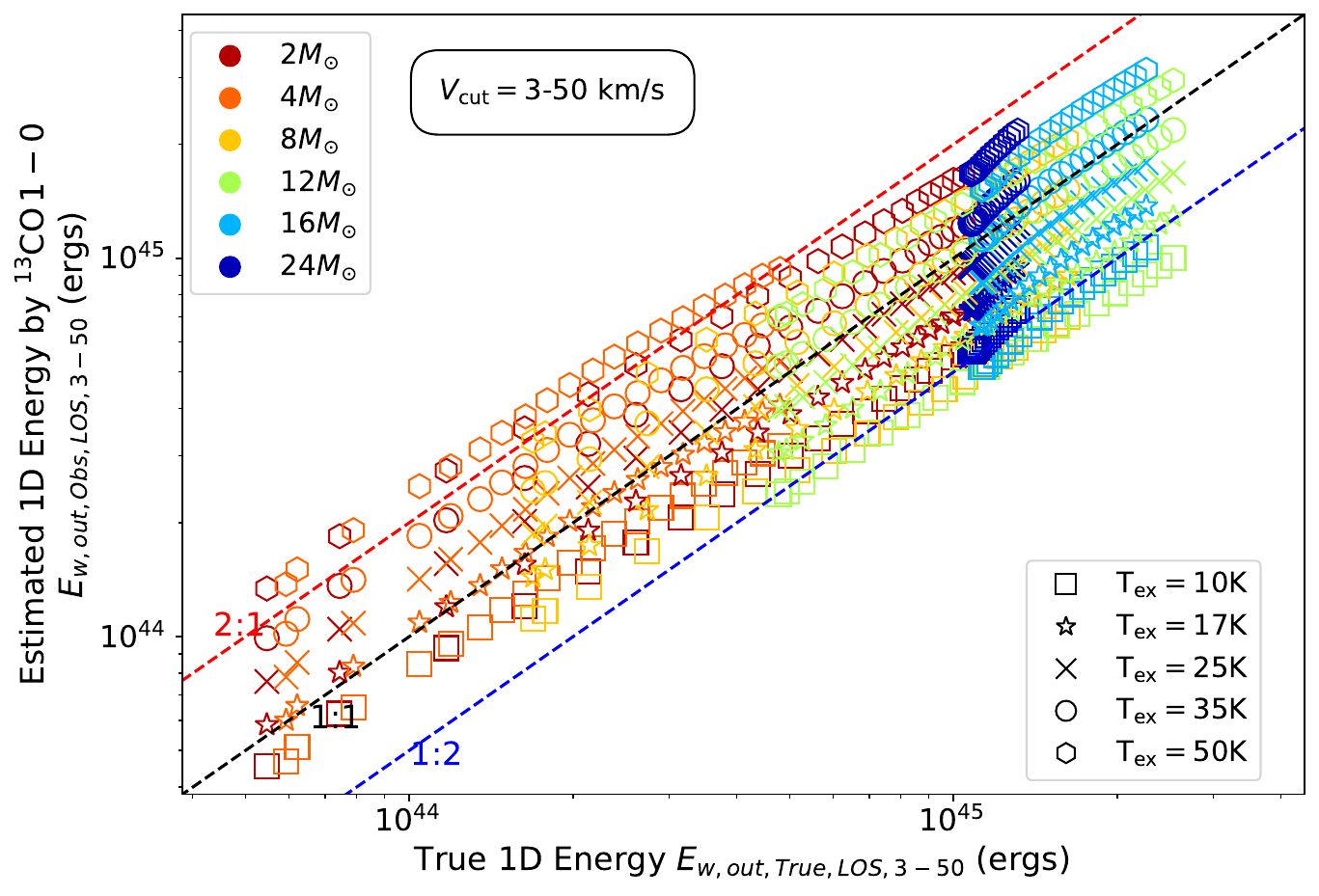}
\includegraphics[width=0.47\linewidth]{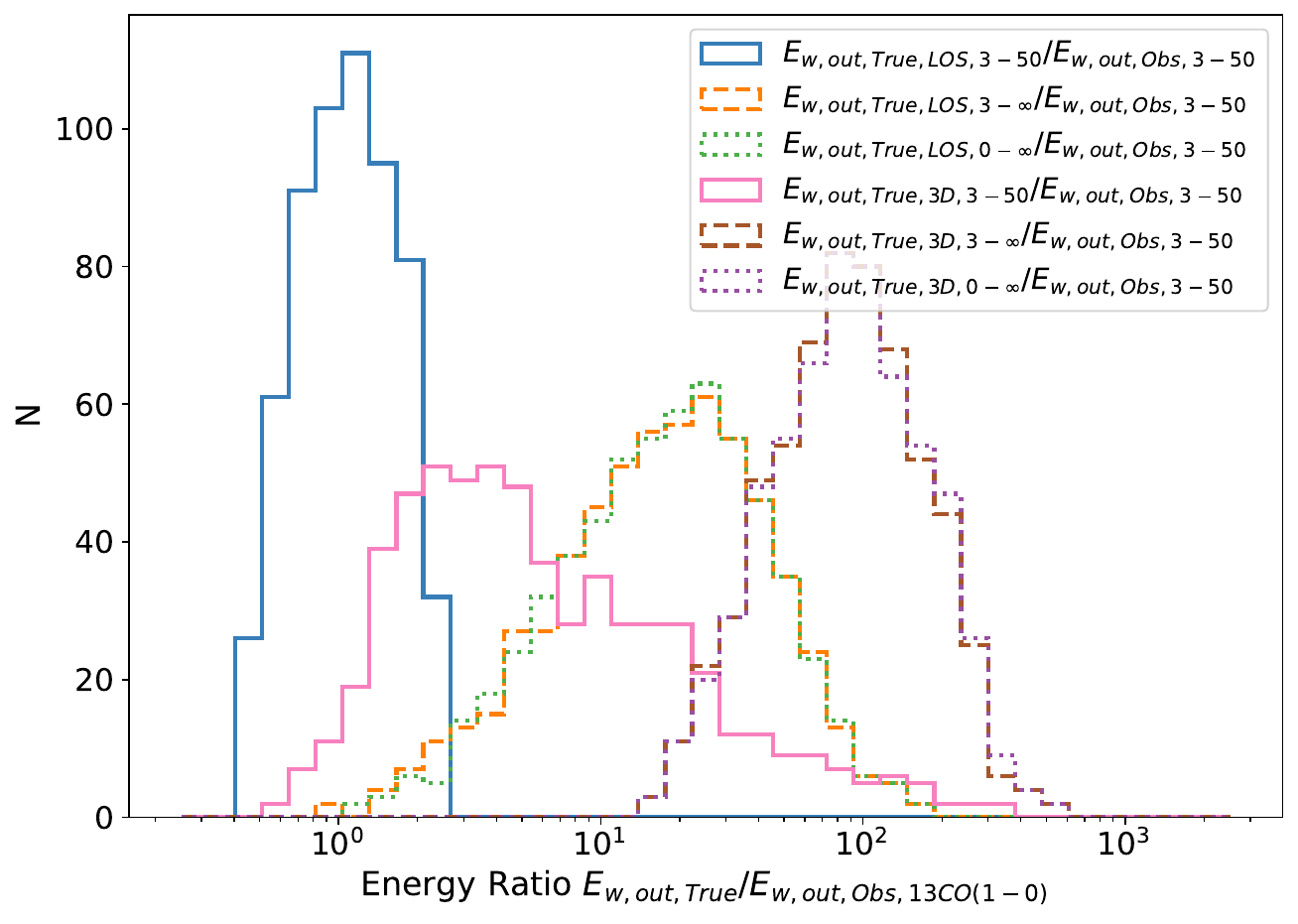}
\caption{{\it Left column:} Comparison of the mass ($1^{st}$ row), momentum ($2^{nd}$ row) and energy ($3^{rd}$ row) estimates obtained for different inclination angles using \13co~(1-0) with a LOS velocity cutoff between 3 to 50 \kms, and the corresponding true outflow values with the same velocity cutoff. {\it Right column:}  the ratios between \13co~(1-0) calculated mass ($1^{st}$ row), momentum ($2^{nd}$ row) and energy ($3^{rd}$ row) and the true simulation-derived values for different LOS velocity cuts.}
\label{fig.scatter_outflow_mass_P_E_comp_true_13co_vcut}
\end{figure*} 

\begin{figure*}[hbt!]
\centering
\includegraphics[width=0.49\linewidth]{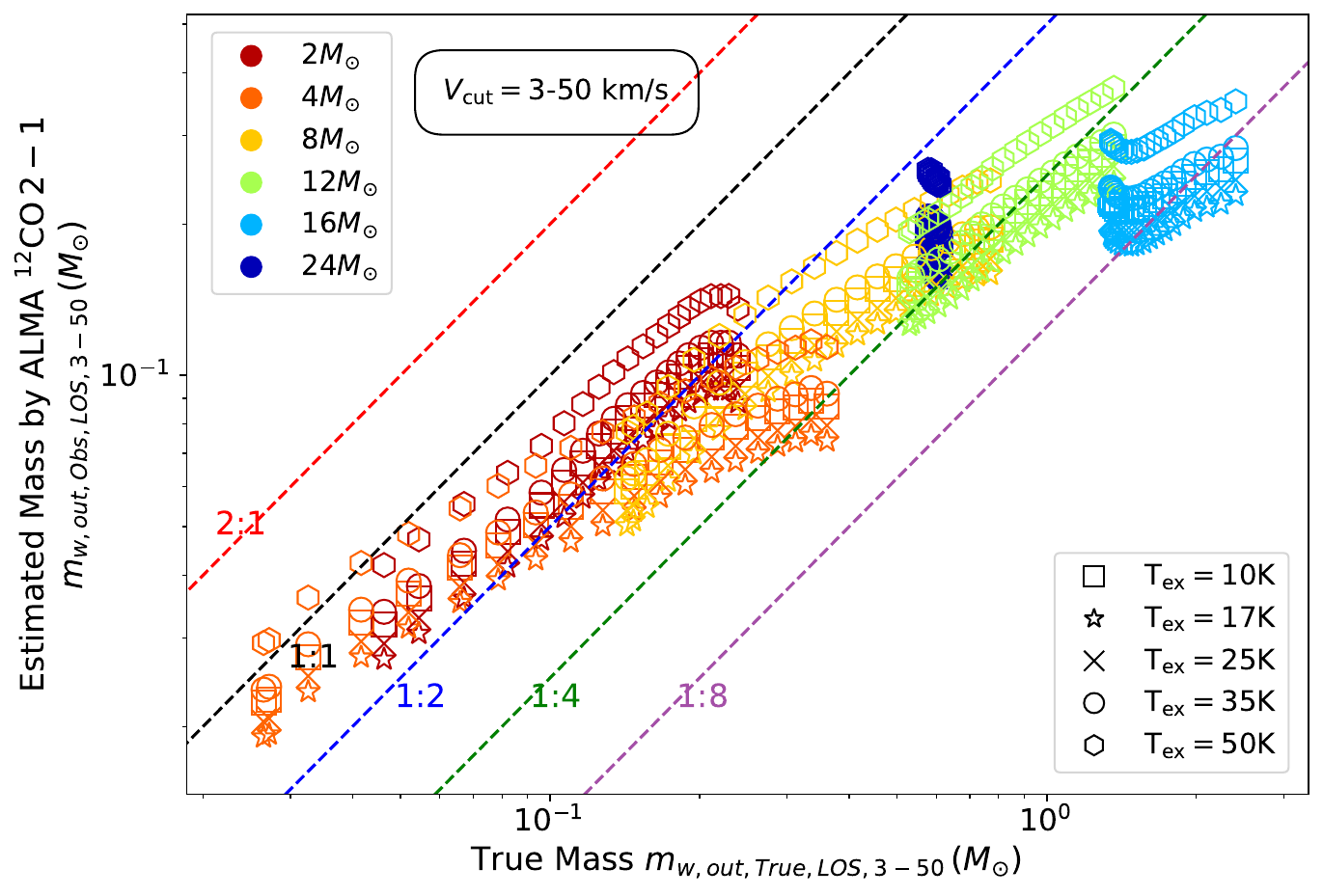}
\includegraphics[width=0.47\linewidth]{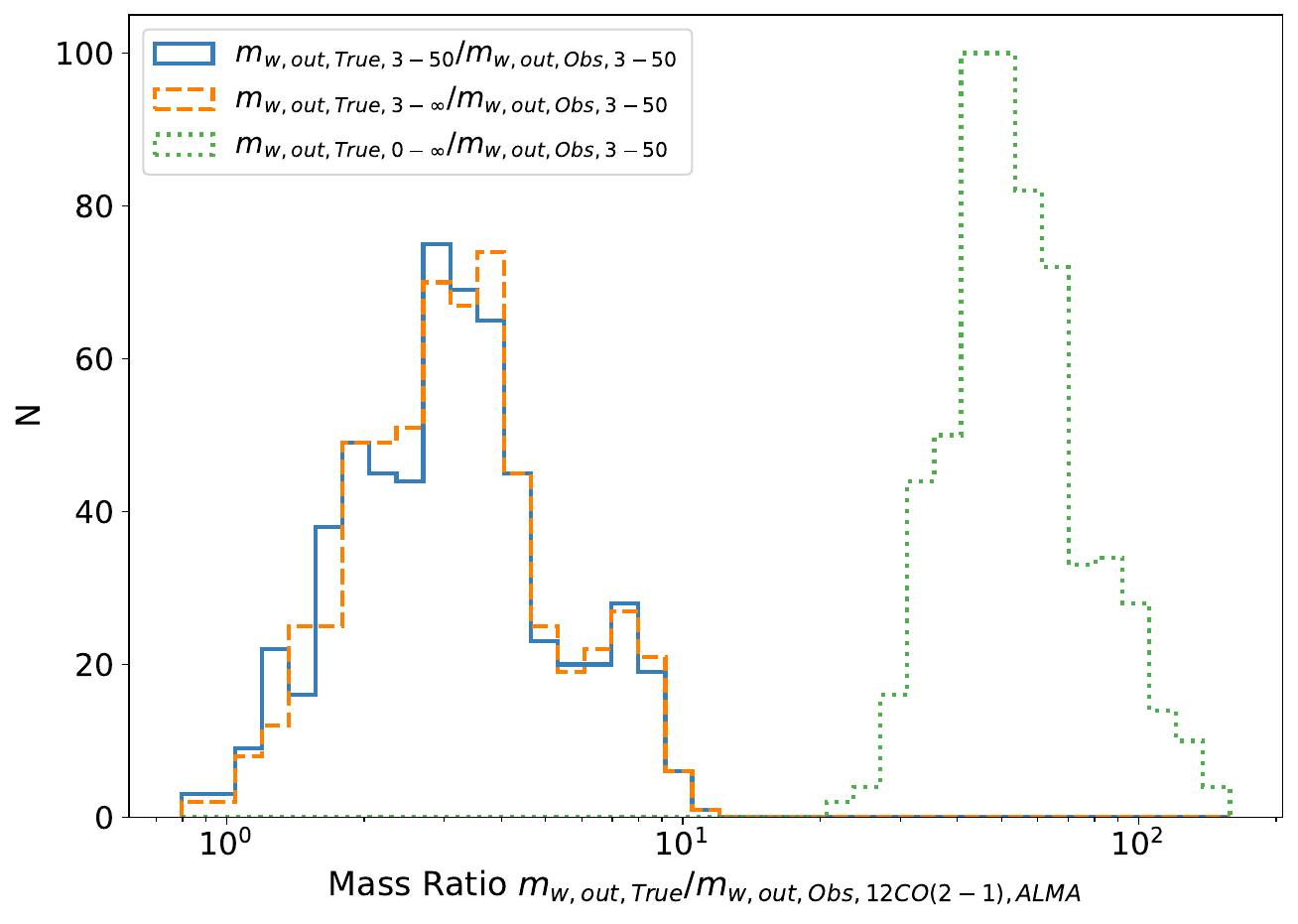}
\includegraphics[width=0.49\linewidth]{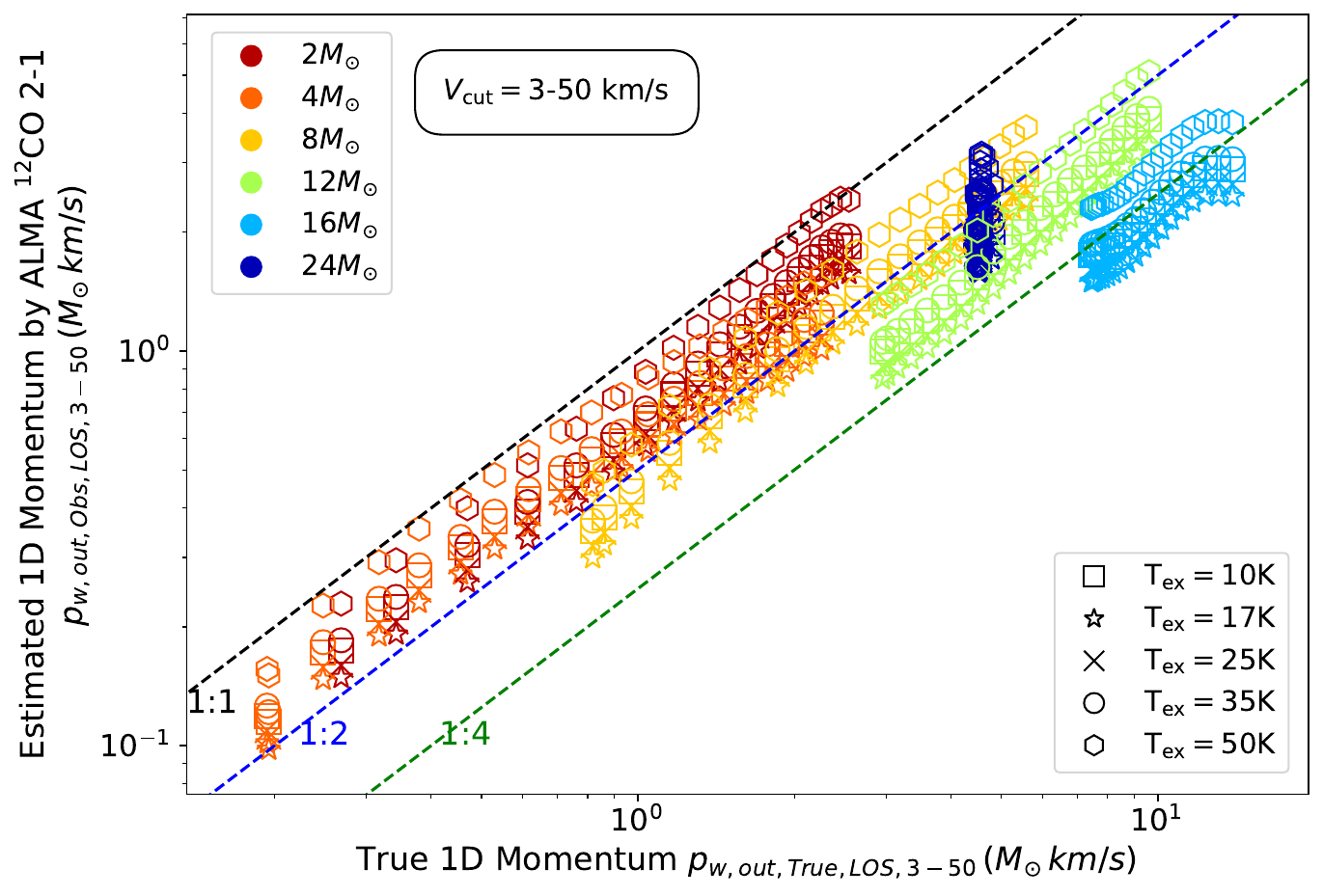}
\includegraphics[width=0.47\linewidth]{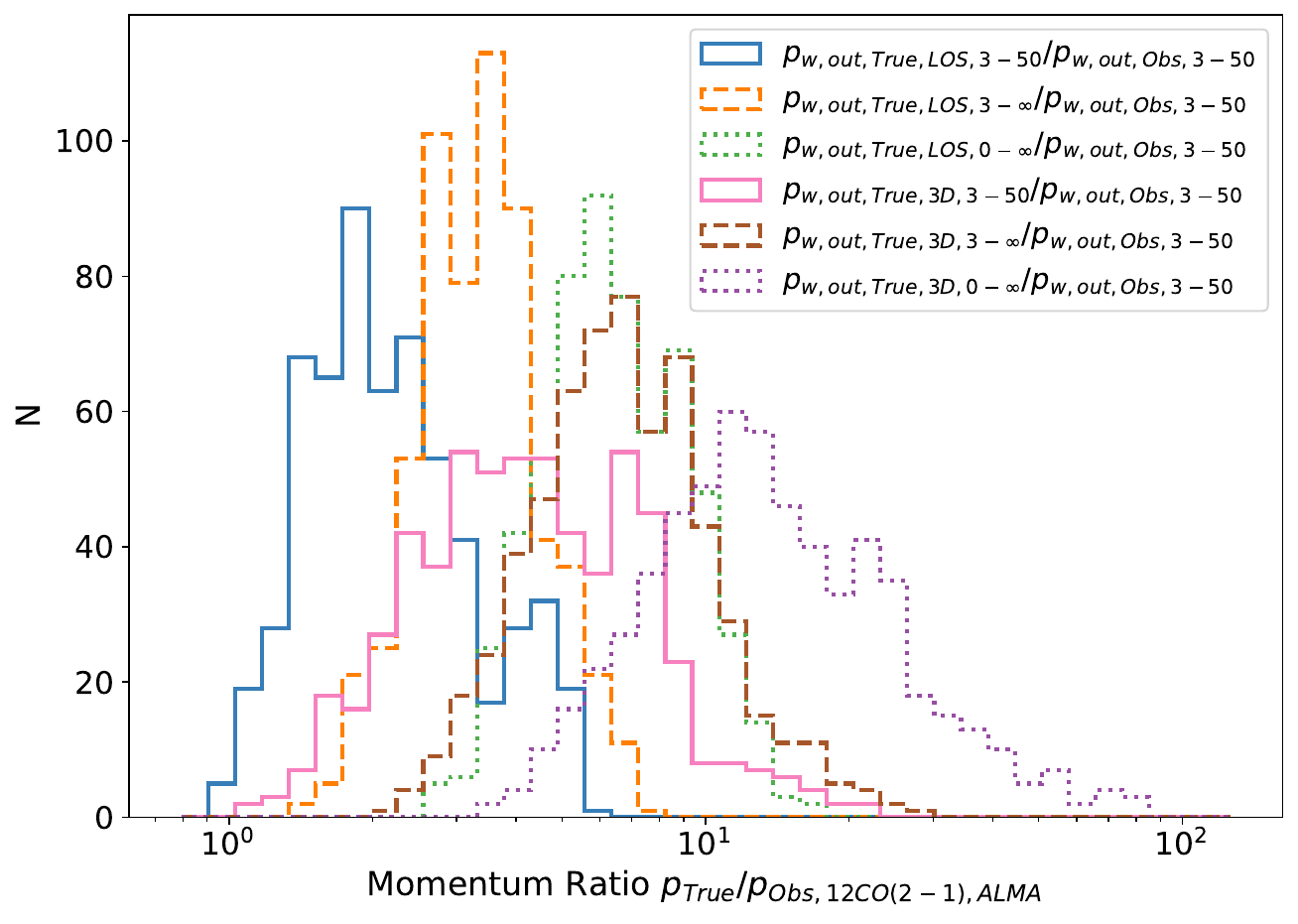}
\includegraphics[width=0.49\linewidth]{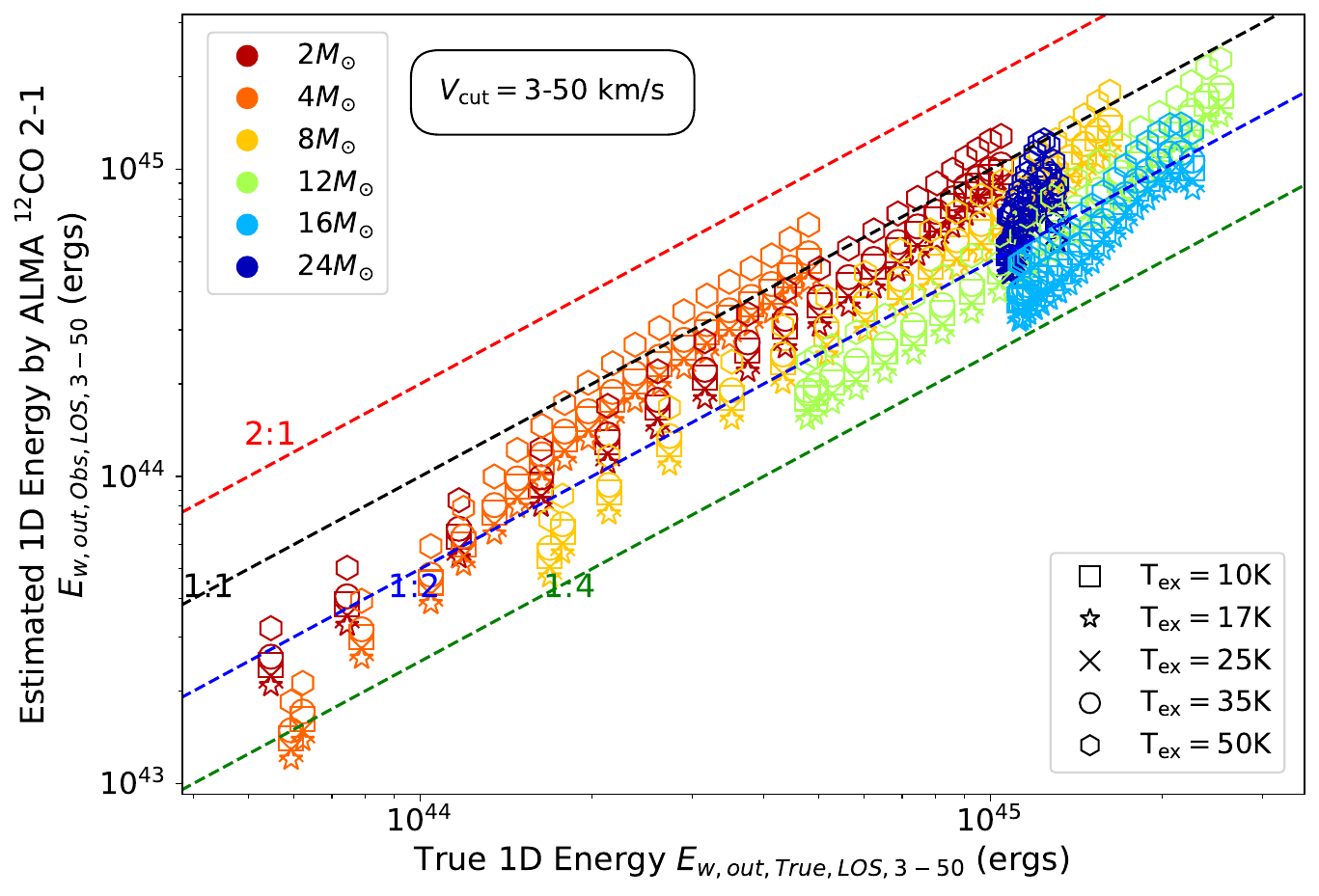}
\includegraphics[width=0.47\linewidth]{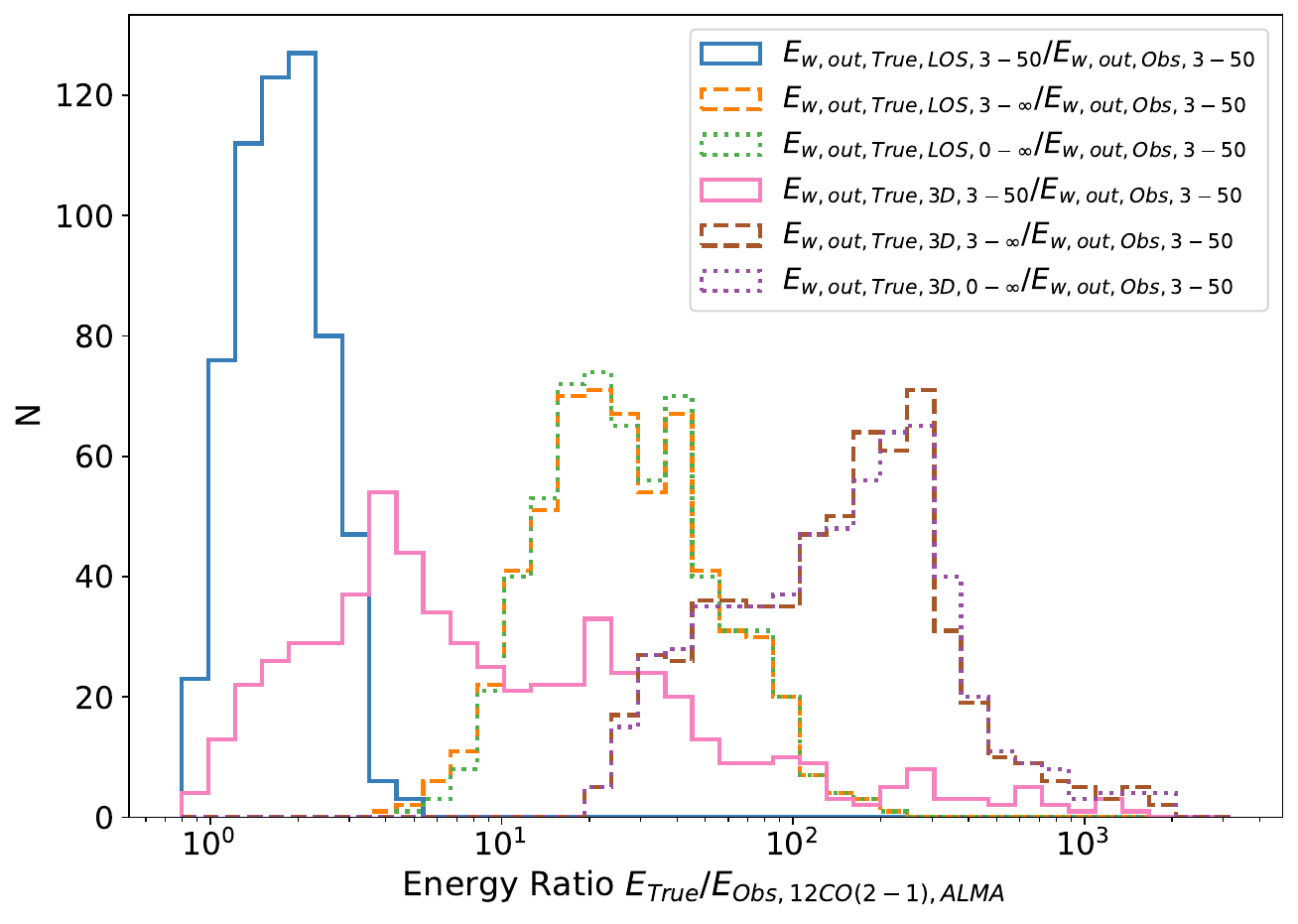}
\caption{{\it Left column:} Comparison of the mass ($1^{st}$ row), momentum ($2^{nd}$ row) and energy ($3^{rd}$ row) estimates obtained for different inclination angles using synthetic ALMA \co~(2-1) post-processed with CASA/\texttt{simalma} with a LOS velocity cutoff between 3 to 50 \kms, and the corresponding true outflow values with the same velocity cutoff. {\it Right column:} the ratios between synthetic ALMA \co~(2-1) calculated mass ($1^{st}$ row), momentum ($2^{nd}$ row) and energy ($3^{rd}$ row) and the true simulation-derived values for different LOS velocity cuts.}
\label{fig.scatter_outflow_mass_P_E_comp_true_12co21_casa_vcut}
\end{figure*} 

We determine the mass of the outflow using \co~(2-1) and \13co~(1-0) lines separately. Since it is not possible to determine the excitation temperature based on a single molecular line observation, certain assumptions must be made regarding the excitation temperature. We consider multiple excitation temperatures for the outflow gas, ranging from 10 K to 50 K. {According to Equations \ref{eq-Ntotal} and \ref{eq-fu}, when selecting \13co~(1-0) lines, the column density exhibits nearly linear growth with the assumed excitation temperature. This indicates that assuming an excitation temperature of 10 K results in the lowest estimate of column density, while assuming 50 K leads to the highest estimate. However, when opting for \co~(2-1) lines, the column density follows a U-shaped curve concerning the assumed excitation temperature. Specifically, selecting an excitation temperature of 17 K results in the lowest estimated column density. This result aligns with the trends observed in Figure~\ref{fig.scatter_outflow_mass_P_E_comp_true_12co21_vcut} and \ref{fig.scatter_outflow_mass_P_E_comp_true_13co_vcut}.} 

Figure~\ref{fig.scatter_outflow_mass_P_E_comp_true_12co21_vcut} presents a comparison between the true outflow mass, momentum, and energy and the corresponding estimated values obtained from \co~(2-1) lines. The estimation is carried out using a LOS velocity cutoff between 3 and 50 \kms. The choice of 3 \kms\ as the lower cutoff mainly excludes the ambient gas, which exhibits a turbulent velocity of approximately 1 \kms. On the other hand, the upper cutoff of 50 \kms\ is selected since most outflow observations cover the velocity range up to 50 \kms, and beyond this limit, the outflow gas emission becomes too faint to be detected, as evident from the spectra of the three sources in \S\ref{Outflow Observations by ALMA}. The estimated outflow mass and momentum using \co~(2-1) spectra show a scatter within a factor of 4 compared to the true values. However, for higher outflow mass and momentum, there is a noticeable underestimation. This suggests that \co~(2-1) may be optically thick for certain outflow gas, leading to the trapping of emission and resulting in lower measured mass values. {In Figure~\ref{fig.12CO21_optical_depth_hist}, we present the distribution of optical depth for \co~(2-1) in the synthetic observation of the outflow produced by a 12 \msun\ protostar at an inclination angle of 58\deg. The optical depth remains primarily below 1, indicating the optically thin nature of \co~(2-1) in most outflow regions. However, a subset of regions exhibits an optical depth between 1 and 10, signifying an optically thick regime. While these regions constitute a small volume fraction, their influence on mass estimation is non-negligible.} In contrast, the estimated outflow energy using \co~(2-1) spectra demonstrates a scatter within a factor of 2 compared to the true values. This indicates that the relatively low-velocity gas is likely to be optically thick, while the high-velocity gas is optically thin. The optically thin high-velocity gas significantly contributes to the outflow energy. Figure~\ref{fig.scatter_outflow_mass_P_E_comp_true_12co21_vcut} also presents the distribution of outflow mass, momentum, and energy ratios between the true values and the observed values from \co~(2-1). Notably, the outflow gas at high-velocity channels ($v_{\rm LOS}>50$ \kms) contributes negligible mass but significant momentum and energy. In summary, the true outflow mass is underestimated by approximately a factor of 50 when compared to the mass calculated from \co~(2-1) spectra with a LOS velocity between 3 and 50 \kms. The true 3D outflow momentum is approximately 10 times larger than the LOS momentum calculated from \co~(2-1) spectra with the same velocity cutoff. The true 3D outflow energy is approximately 150 times larger than the LOS energy calculated from \co~(2-1) spectra with a LOS velocity between 3 and 50 \kms. 

Figure~\ref{fig.scatter_outflow_mass_P_E_comp_true_13co_vcut} presents a comparison between the true outflow mass, momentum, and energy and the corresponding estimated values obtained from \13co~(1-0) lines. The estimated outflow mass, momentum and energy using \13co~(1-0) spectra show a scatter within a factor of 2 compared to the true values. And there is no significant underestimation when the outflow mass is high, which implies that \13co~(1-0) is more optically thin compared to \co~(2-1). In addition we find that an assumption between 17 to 35 K is likely sufficient for accurately deriving the outflow mass, momentum and energy from \13co~(1-0) spectra. Figure~\ref{fig.scatter_outflow_mass_P_E_comp_true_13co_vcut} also presents the distribution of outflow mass, momentum, and energy ratios between the true values and the observed values from \13co~(1-0). In summary, the true outflow mass is underestimated by approximately a factor of 20 when compared to the mass calculated from \13co~(1-0) spectra with a LOS velocity between 3 and 50 \kms. The true 3D outflow momentum is approximately 7 times larger than the LOS momentum calculated from \13co~(1-0) spectra with the same velocity cutoff. The true 3D outflow energy is approximately 100 times larger than the LOS energy calculated from \13co~(1-0) spectra with a LOS velocity between 3 and 50 \kms. 

Furthermore, we provide mass, momentum and energy estimates for synthetic outflows utilizing the \co~(1-0) line emission and assess their precision compared to the true values attained from simulations in Appendix~\ref{Estimates of Mass, Momentum, and Energy of Synthetic Outflows using co 1-0}.

\subsubsection{Impact {\sc I}$+$Impact {\sc II}$+$Impact {\sc III}: Synthetic ALMA Observations}

In Figure~\ref{fig.scatter_outflow_mass_P_E_comp_true_12co21_casa_vcut}, we compare the true outflow mass, momentum, and energy with the corresponding estimated values obtained from synthetic ALMA \co~(2-1) observations generated in \S\ref{Synthetic ALMA Observations}. It is important to note that we use a 1 $\sigma$ threshold in the calculation to prevent significant missing flux of the outflow. The estimated outflow mass using synthetic ALMA \co~(2-1) spectra shows a scatter within a factor of 8 compared to the true values. Likewise, the estimated outflow momentum and energy using synthetic ALMA \co~(2-1) spectra show a scatter within a factor of 4 compared to the true values. However, it is worth noting that there is a systematic underestimation of mass, momentum, and energy when using synthetic ALMA \co~(2-1) spectra compared to the raw \co~(2-1) spectra in Figure~\ref{fig.scatter_outflow_mass_P_E_comp_true_12co21_vcut}. This discrepancy is likely due to the interferometry missing flux of the gas with a LOS velocity between 3 and 50 \kms, especially near 3 \kms. Figure~\ref{fig.scatter_outflow_mass_P_E_comp_true_12co21_casa_vcut} also presents the distribution of outflow mass, momentum, and energy ratios between the true values and the observed values from synthetic ALMA \co~(2-1) spectra. In summary, the true outflow mass is underestimated by approximately a factor of 50 when compared to the mass calculated from synthetic ALMA \co~(2-1) spectra with a LOS velocity between 3 and 50 \kms. The true 3D outflow momentum is approximately 15 times larger than the LOS momentum calculated from synthetic ALMA \co~(2-1) spectra with the same velocity cutoff. The true 3D outflow energy is approximately 250 times larger than the LOS energy calculated from synthetic ALMA \co~(2-1) spectra with a LOS velocity between 3 and 50 \kms. 

We present estimates for the mass, momentum, and energy of synthetic outflows based on the \co~(1-0) and \13co~(2-1) line emissions in Appendix~\ref{Estimates of Mass, Momentum, and Energy of Synthetic Outflows using co 1-0}. {It is important to highlight that in the appendix, we examined impacts I and II in these tracers without incorporating synthetic ALMA observations.} We observe that \co~(1-0) shows some scatter and a slight underestimation of outflow mass, likely attributed to optical depth effects. In contrast, \13co~(2-1) displays a robust correlation between the estimated properties and the true values, affirming its reliability as a tracer for outflow gas. {The rationale behind \13co\ serving as a reliable outflow tracer lies in its optically thin nature, as assumed in Equation~\ref{equ-Nu}.}

\subsection{Estimating the Mass Outflow Rate and Momentum Outflow Rate}
\label{Estimating the Outflow Mass Injection Rate and Momentum Injection Rate}

In this section, we assess the mass outflow rate and momentum outflow rate using synthetic ALMA \co~(2-1) data. Initially, we quantify the true values of these rates by calculating the mass and momentum flux across a plane at specific heights from the central stars, as obtained from simulations. The mass and momentum flux are measured at a height of 25,000~au, i.e., at the top of the simulation domain. 
Table~\ref{tab.M_P_rate_summary} presents a summary of the mass outflow rates and momentum outflow rates, measured at 25,000 AU, along with the injection rates of the primary outflow reported in \citet{2023ApJ...947...40S}. Note that the mass outflow rate includes entrained gas, i.e., primary outflow gas launched directly by the disk wind plus swept-up secondary outflow material.
As emphasized in \citet{2023ApJ...947...40S}, the entrained gas can be up to nine times larger than the directly injected mass from the central source. 
On the other hand, due to the conservation of momentum, the momentum outflow rate is a more direct tracer of the input disk wind properties, i.e., momentum injection rate of the primary outflow.

\begin{table*}[htbp]
  \centering
  \caption{Summary of mass outflow rate and momentum outflow rate$^a$ }
    \begin{tabular}{cccccccc}
    \hline
    \hline
    $m_{*}$  & $h_{\rm max}$   & $\dot{m}_{\rm w,out}$    &  $\dot{m}_{\rm w,inj}$     &  $\dot{p}_{\rm w,out}$       &    $\dot{p}_{\rm w,inj}$   &   $\dot{m}_{\rm w,out}/\dot{m}_{\rm w,inj}$   & $\dot{p}_{\rm w,out}/\dot{p}_{\rm w,inj}$ \\
     ($M_{\odot}$)  & ($10^3\:$AU)   & ($10^{-5}\, M_{\odot} {\rm yr}^{-1}$)     &  ($10^{-5}\, M_{\odot} {\rm yr}^{-1}$)    &  ($10^{-3}\, M_{\odot} {\rm km\, s^{-1}\, yr}^{-1}$)     &     ($10^{-3}\, M_{\odot} {\rm km\, s^{-1}\, yr}^{-1}$)   &      &  \\\hline
    2  &  25.0 & 6.9   & 1.4   & 10.5   & 9.9   & 4.98   & 1.06 \\
    4  &  25.0 & 2.9   & 2     & 8.8   & 9.4   & 1.5   & 0.94 \\
    8  &  25.0  & 3.2  & 2.7   & 12.4  & 14.2  & 1.2   & 0.87 \\
    12 &  25.0  & 11.0  &   -    & 27.4  &  -     &   -    &  - \\
    16 & 25.0  & 13.0  & 3.2   & 45.0  & 41.2  & 4.1   & 1.09 \\
    24 & 25.0  & 19.6  & 3.3   & 55.3  & 49.5  & 5.9   & 1.12 \\\hline
    \multicolumn{8}{p{1.0\linewidth}}{Notes:}\\
\multicolumn{8}{p{1.0\linewidth}}{$^a$ Outflow mass injection rate ($\dot{m}_{\rm w,inj}$) and outflow momentum injection rate ($\dot{p}_{\rm w,inj}$) are reported in \citet{2023ApJ...947...40S}. The mass outflow rate ($\dot{m}_{\rm w,out}$) and momentum outflow rate ($\dot{p}_{\rm w,out}$) are measured at the maximum height in the simulation domain that captures the entire outflow moving across a surface in the $z$ direction (reported in the second column).
}
    \end{tabular}%
  \label{tab.M_P_rate_summary}%
\end{table*}%

\begin{figure}[hbt!]
\centering
\includegraphics[width=0.99\linewidth]{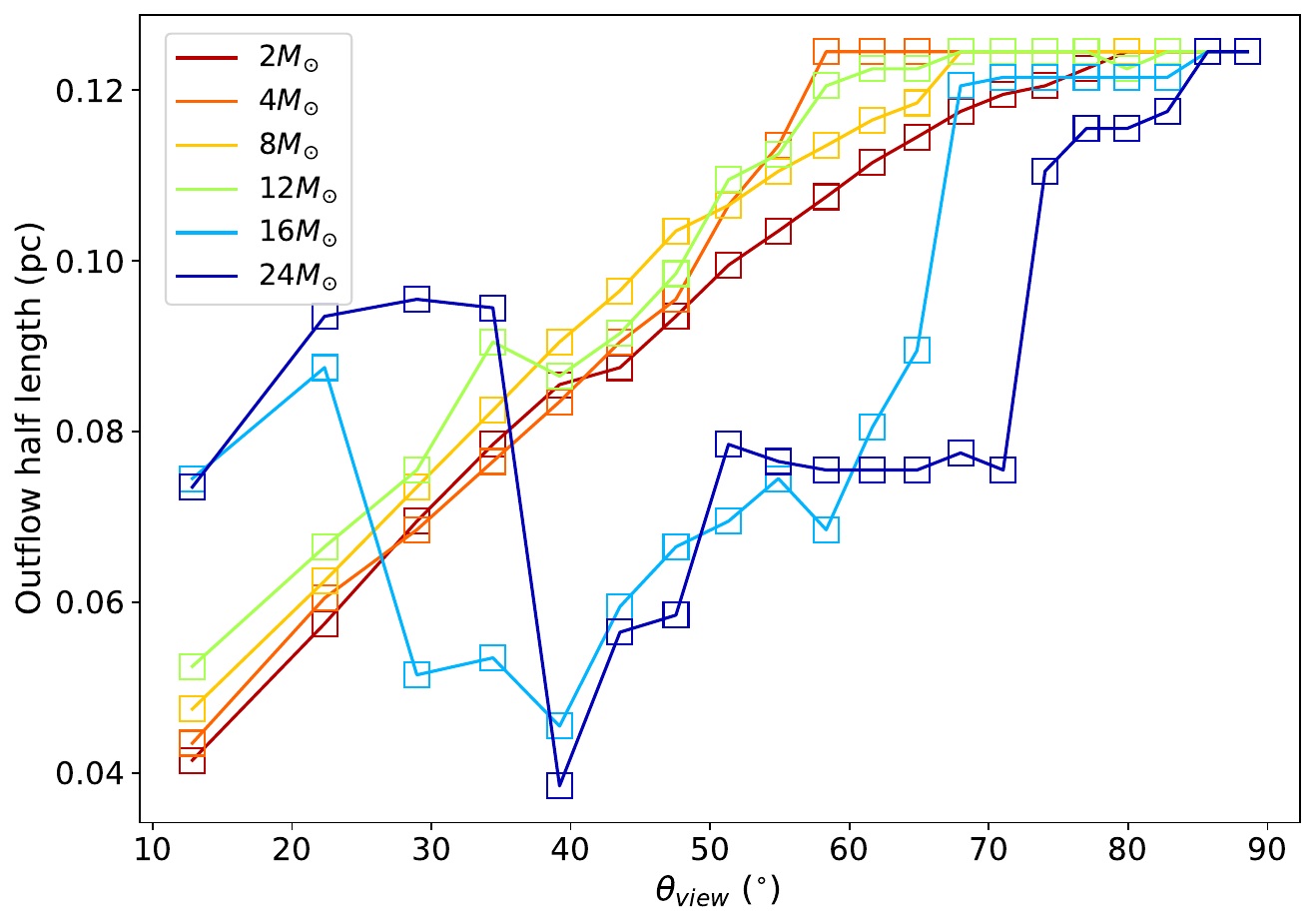}
\caption{Variation of the outflow length (half-length, i.e., one side) as a function of inclination of viewing angle $\theta_{\rm view}$. {It is important to highlight that the upper limit of the outflow length in this figure corresponds to the boundary of the simulation's half domain, set at 0.125 pc.} }
\label{fig.scatter_outflow_length_vs_incli}
\end{figure} 

\begin{figure*}[hbt!]
\centering
\includegraphics[width=0.99\linewidth]{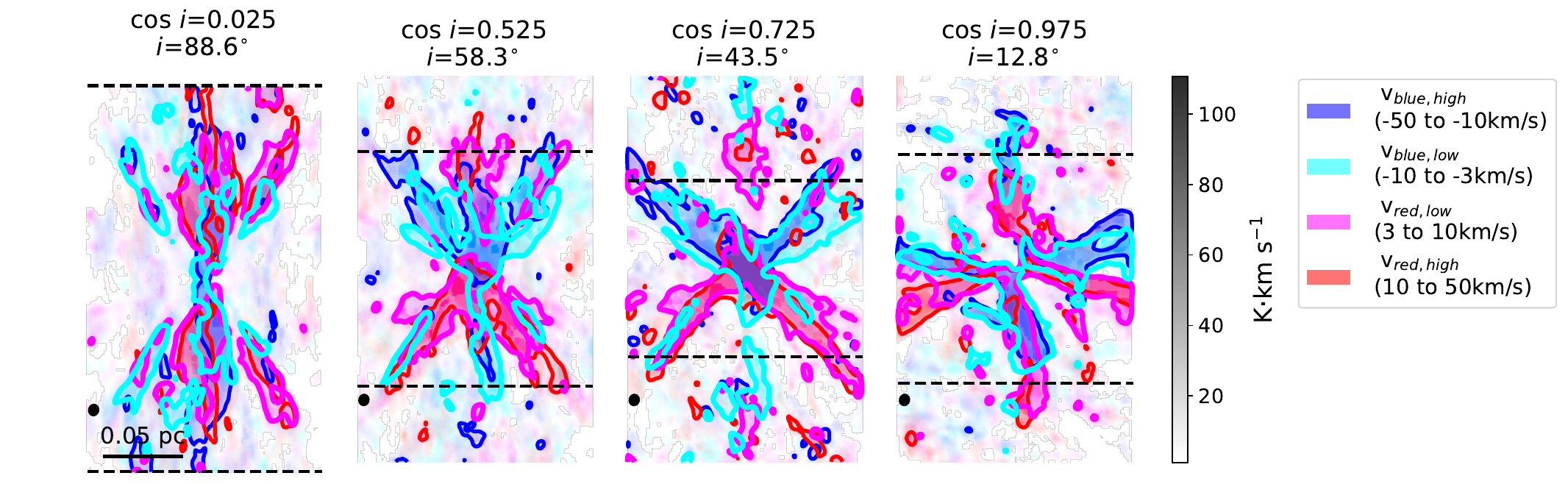}
\caption{Synthetic ALMA \co~(2-1) emission of the outflow from a 24~\msun\ protostar at different inclination angles. The black dashed lines represent the boundaries of the outflow length. The color description is the same as Figure~\ref{fig.co21-cos0525-allmass}. {We also present the 3$\sigma$ contour of each component for reference.}}
\label{fig.12CO21_24msun_allinclination_4color_casa_alma_dis2_4pannel}
\end{figure*}

\begin{figure*}[hbt!]
\centering
\includegraphics[width=0.89\linewidth]{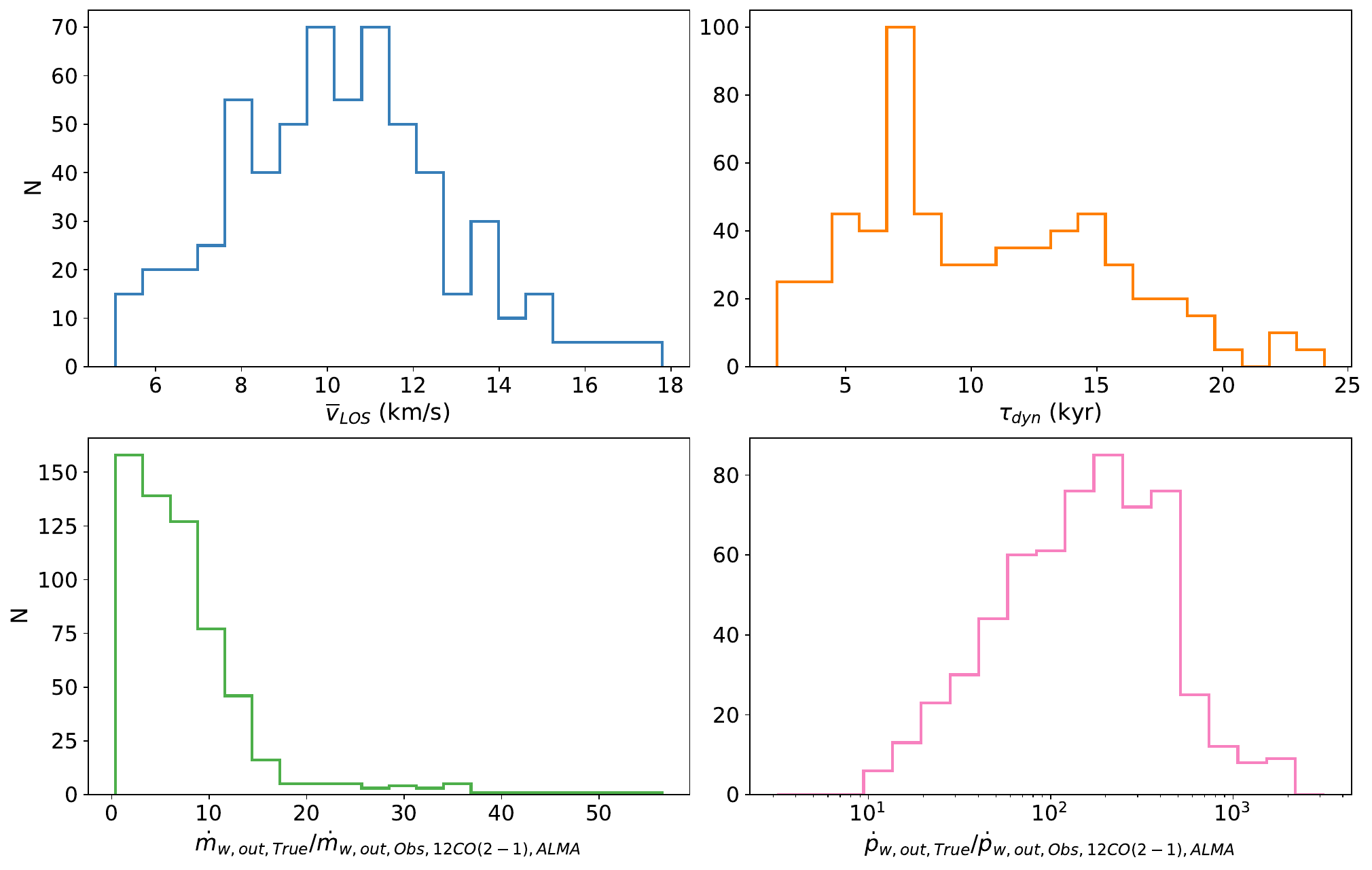}
\caption{{\it Upper left:} Histogram depicting the mass-weighted mean LOS velocity {of the post-processed synthetic outflows observed with ALMA}. {\it Upper right:} Histogram showing the outflow dynamic timescale. {\it Lower left:} Histogram displaying the ratio between the true mass outflow rate and the observed mass outflow rate. {\it Lower right:} Histogram presenting the ratio between the true momentum outflow rate and the observed momentum outflow rate. }
\label{fig.hist_outflow_vmean}
\end{figure*} 

\begin{figure*}[hbt!]
\centering
\includegraphics[width=0.98\linewidth]{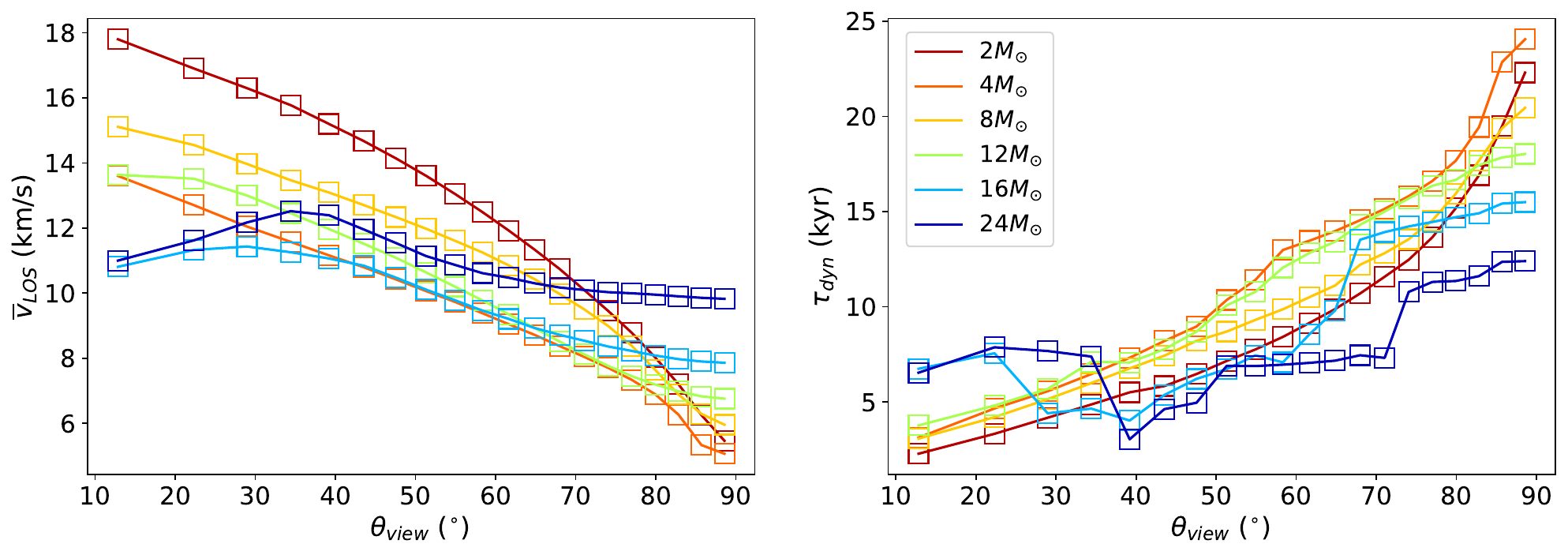}
\caption{{\it Left (a):} Variation of the mean LOS velocity as a function of inclination of viewing angle $\theta_{\rm view}$ of the post-processed synthetic outflows observed with ALMA. {\it Right (b):} Variation of the dynamic timescale as a function of inclination of viewing angle $\theta_{\rm view}$. }
\label{fig.scatter_outflow_vlos_tdyn_vs_incli}
\end{figure*} 

To estimate the mass outflow rate and momentum outflow rate from synthetic ALMA \co~(2-1) data, we adopt a similar observational approach as in \citet{2019ApJ...873...73Z}. We first determine the outflow length based on the integrated intensity of the synthetic ALMA \co~(2-1) data, considering only emission above 3$\sigma$ as valid. Using a box to enclose the outflow emission, we obtain the length of the box, representing the outflow length derived from synthetic ALMA \co~(2-1) data. Figure~\ref{fig.scatter_outflow_length_vs_incli} displays the variation of the outflow length (considering one sided lobe of the outflow, i.e., a half-length) as a function of the inclination angle $\theta_{\rm view}$. We observe that for protostellar masses between 2~\msun\ and 12 \msun, the outflow length increases with $\theta_{\rm view}$. However, we observe a non-uniform trend for the 16~\msun\ and 24~\msun\ cases, likely due to the large opening angle of the outflow, where the cavity wall is distinct, but the emission inside the cavity remains unobserved. An example of this behavior for the 24~\msun\ case at several different viewing angles is shown in Figure~\ref{fig.12CO21_24msun_allinclination_4color_casa_alma_dis2_4pannel}. 

Next, we compute the mass-weighted mean LOS velocity of the outflow using the formula:
\begin{equation}
\begin{split}
& \overline{v}_{\rm LOS,3-50}=\\
& p_{\rm w,out,obs,12CO(2-1),ALMA}/m_{\rm w,out,obs,12CO(2-1),ALMA},
\end{split}
\end{equation}
where $m_{\rm w,out,obs,12CO(2-1),ALMA}$  represents the observed outflow mass, and $p_{\rm w,out,obs,12CO(2-1),ALMA}$ is the observed 1D LOS momentum, {confined to the velocity range between 3 and 50 \kms.} Furthermore, we estimate the dynamical timescale $\tau_{\rm dyn}$ of the outflow by calculating the ratio between the half length of the outflow and the mass-weighted mean LOS velocity. Consequently, we can determine the observed mass outflow rate and momentum outflow rate by dividing $m_{\rm w,out,obs,12CO(2-1),ALMA}$ by $\tau_{\rm dyn}$ and $p_{\rm w,out,obs,12CO(2-1),ALMA}$ by $\tau_{\rm dyn}$, respectively.

Figure~\ref{fig.hist_outflow_vmean} illustrates the distribution of the mass-weighted mean LOS velocity and the dynamical timescale of outflows, estimated from synthetic ALMA \co~(2-1) data. {The typical dynamical timescale of the outflows is around 7 kyr in our synthetic outflows. It is important to consider that the dynamical timescale of the outflows can be sensitive to the chosen length of the outflow. In our study, our half domain is limited to 25,000 AU, which represents the lower limit of typical outflows.}

Figure~\ref{fig.hist_outflow_vmean} also displays the distribution of the ratio between the true mass outflow rate and the observed estimate of the mass outflow rate from \co~(2-1). Similarly, the figure also shows the ratio between the true momentum outflow rate and the observed estimate of the momentum outflow rate from\co~(2-1).
In general, the true mass outflow rate is about 7 times larger (but with wide variation in this factor, i.e., depending on viewing angle and $m_*$) than the observed mass outflow rate calculated from synthetic ALMA \co~(2-1) data. Similarly, the true momentum outflow rate is about 200 times larger (but with wide variation in this factor, i.e., depending on viewing angle and $m_*$) than the observed momentum outflow rate calculated from synthetic ALMA \co~(2-1) data. 

\begin{figure*}[hbt!]
\centering
\includegraphics[width=0.48\linewidth]{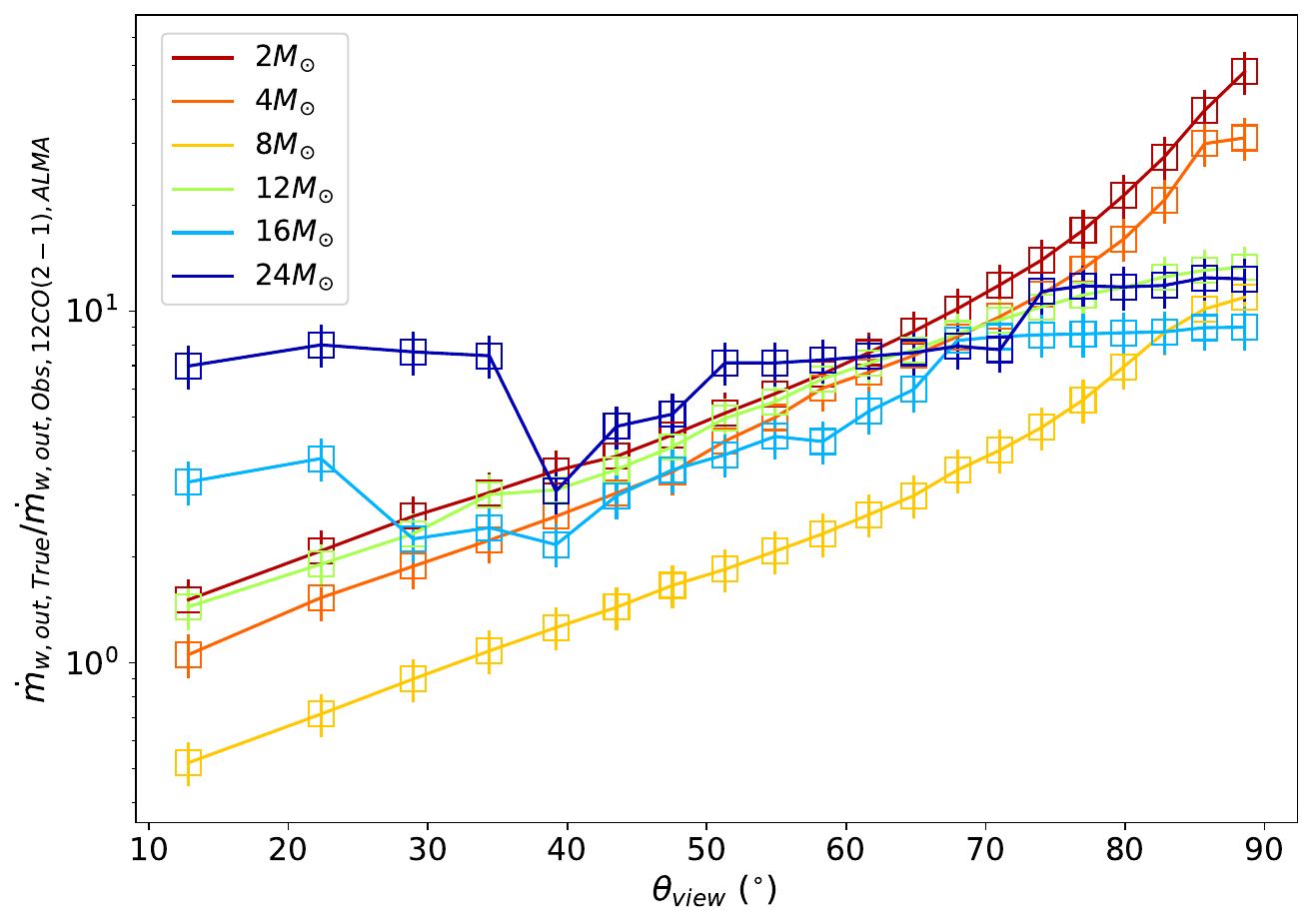}
\includegraphics[width=0.48\linewidth]{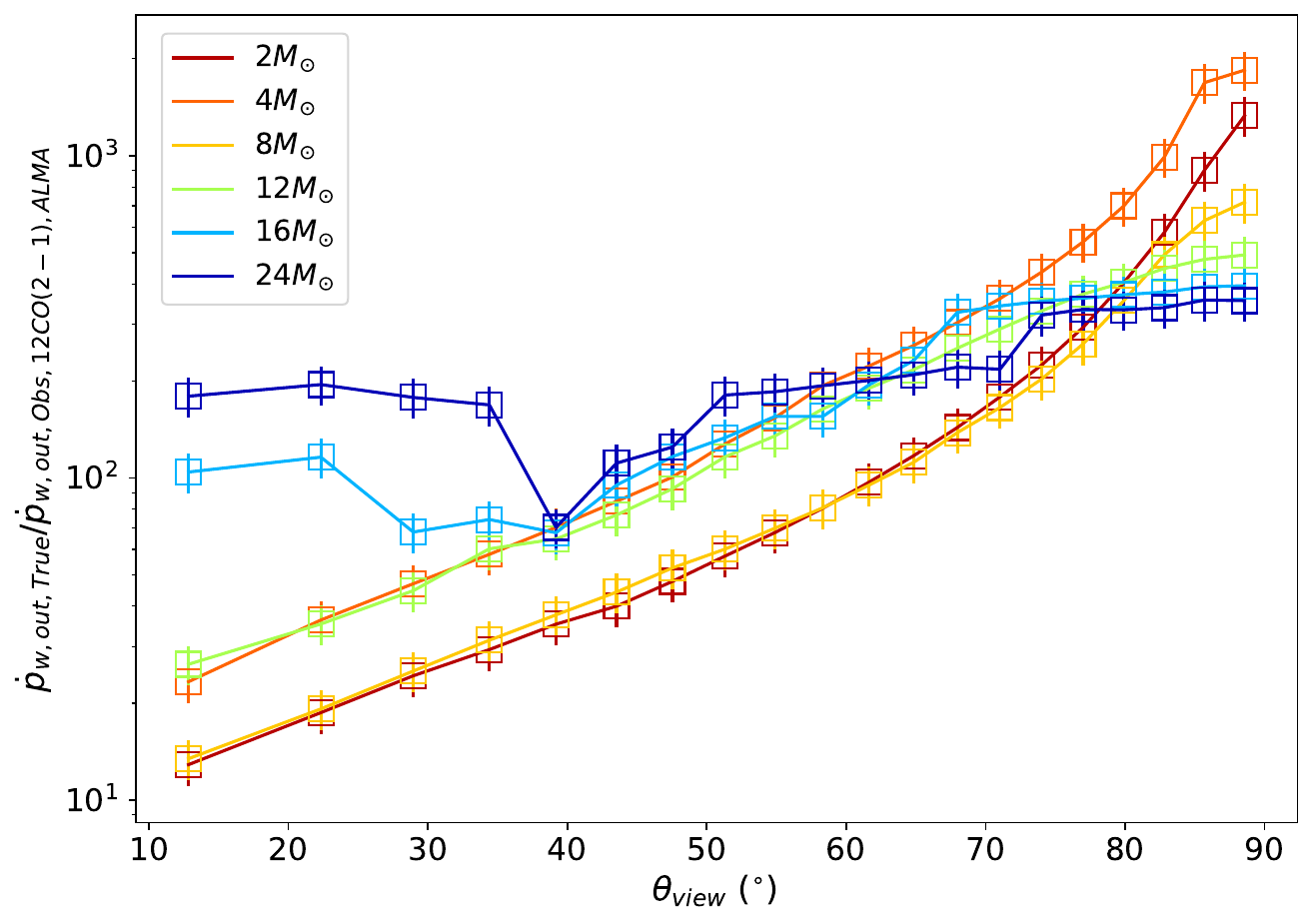}
\caption{{\it Left (a):} Variation of the ratio between the true mass outflow rate and the observed mass outflow rate as a function of inclination of viewing angle $\theta_{\rm view}$ {of the post-processed synthetic outflows observed with ALMA}. {\it Right (b):} Variation of the ratio between the true momentum outflow rate and the observed momentum outflow rate as a function of inclination of viewing angle $\theta_{\rm view}$. The error bars represent the standard deviation of the ratio obtained using different assumptions of excitation temperatures. }
\label{fig.scatter_outflow_ratio_mdot_pdot_vs_incli}
\end{figure*} 

{Figure~\ref{fig.scatter_outflow_vlos_tdyn_vs_incli}a illustrates how the mean LOS velocity is influenced by the viewing angle and protostellar mass. The mean LOS velocity decreases with an increasing viewing angle, reflecting a higher proportion of gas moving perpendicular to the LOS. Additionally, the outflow associated with a 2 \msun\ protostellar mass appears to exhibit a generally larger mean LOS velocity, possibly due to the smaller opening angle of the outflow. In Figure~\ref{fig.scatter_outflow_vlos_tdyn_vs_incli}b, the impact of the viewing angle and protostellar mass on the dynamical timescale is presented. With an increase in the viewing angle, the estimated dynamical timescale also rises. This is primarily attributed to a greater reduction in the mean LOS velocity rather than changes in the typical outflow scale.
}

Figure~\ref{fig.scatter_outflow_ratio_mdot_pdot_vs_incli}a demonstrates the impact of the viewing angle and protostellar mass on the ratio between the true mass outflow rate and the observed mass outflow rate. As the viewing angle increases, there is a significant rise in this ratio. This ratio remains relatively stable in the case of a 24~\msun\ protostellar mass, which possesses a wider opening angle, making it less influenced by the viewing angle. 
For relatively early evolutionary stages, i.e., $m_* \lesssim 12\:M_\odot$, there is a relatively tight relation between the mass flux correction factor and viewing angle, with a typical intermediate value of $\sim 10$.
Figure~\ref{fig.scatter_outflow_ratio_mdot_pdot_vs_incli}b demonstrates a similar trend for the ratio between the true momentum outflow rate and the observed momentum outflow rate. Like the mass outflow rate, this ratio increases with the viewing angle, with relatively simple monotonic behavior for $m_* \lesssim 12\:M_\odot$. For typical viewing angles, i.e., $\sim 60^\circ$, the momentum flux correction factor is $\sim 100$.


\section{Comparison to ALMA Observations of Outflows}
\label{Outflow Observations by ALMA}

\subsection{Overview of CO Morphologies of G35.30, G45.47 and G338.88}
\label{Outflow_morphologies}

\begin{figure*}[hbt!]
\centering
\includegraphics[width=0.97\linewidth]{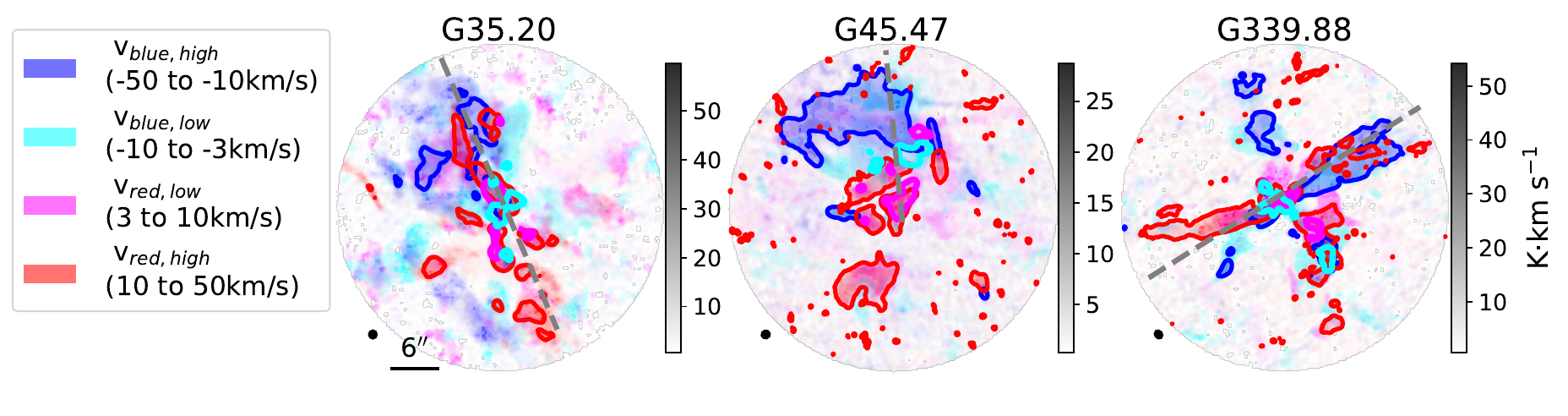}
\caption{\co~(2-1) emission of three observed outflows G35.20 \citep{2022ApJ...936...68Z}, G45.47 \citep{2019ApJ...886L...4Z}, and G339.88 \citep{2019ApJ...873...73Z}. The color description is the same as Figure~\ref{fig.12CO21-cos0525-allmass-4color-casa-alma-dis2}. {We also present the 5$\sigma$ contour of each component for reference. The grey dashed line represents the direction of outflows identified through both \co\ and continuum emission \citep{2019ApJ...873...73Z,2019ApJ...886L...4Z,2022ApJ...936...68Z}.}}
\label{fig.12CO21-realobs-alma-4color}
\end{figure*} 

In order to compare the simulated CO emissions with observational data, we show \co~(2-1) emission maps of outflows from three massive protostellar objects, G35.20$-$0.74N (hereafter G35.20; \citealt{2013A&A...552L..10S,2022ApJ...936...68Z}), G45.47+0.05 (hereafter G45.47; \citealt{2019ApJ...886L...4Z}), and G339.88$-$1.26 (hereafter G339.88; \citealt{2019ApJ...873...73Z}). The distances to G35.20, G45.47, and G339.88 are 2.2 kpc, 8.4 kpc, and 2.1 kpc, respectively.


The presented \co~(2-1) data were obtained with ALMA in the C36-3 configurations
on September 8 (G35.20), April 24 (G45.47), and April 4 (G339.88) of 2016 (ALMA project ID: 2015.1.01454.S)
with baselines ranging from 15~m to 463~m.
The integration time for each source was 3.5 min.
The \co~(2-1) data of G35.20 and G339.88 were previously presented by 
\citet{2022ApJ...936...68Z,2019ApJ...873...73Z}, and the continuum data of G45.47 obtained in the same observation
were presented by \citet{2019ApJ...886L...4Z}. {The major axis sizes of the synthesized beams for G35.20, G45.47, and G339.88 are 0.87\arcsec, 0.90\arcsec, and 0.93\arcsec, respectively. The minor axis sizes of the synthesized beams for the same sources are 0.83\arcsec, 0.84\arcsec, and 0.74\arcsec, respectively. The position angles of the synthesized beams for G35.20, G45.47, and G339.88 are -88.8\deg, 14.3\deg, and -51.8\deg, respectively.} We refer the reader to these papers for more details of the observations.

The \co~(2-1) data were calibrated and imaged in CASA. After pipeline calibration, self-calibration using the continuum data was performed and applied to the CO line data. The CASA {\it tclean} task
was used to image the data, using Briggs weighting with the robust parameter set to 0.5. It is important to highlight that the adoption of different weightings, such as natural and Briggs, could introduce {an uncertainty} 10\%. However, this is expected to have a modest impact on the ultimate calculation of the physical properties of the outflows. {From investigation,} it introduces a potential uncertainty of up to 20\% in the final estimation of outflow momentum and energy. For G35.20, G45.47, and G339.88, the synthetic beams of the \co\ images are Gaussian with a size of 0.\arcsec 25. Figure~\ref{fig.12CO21-realobs-alma-4color} shows the integrated \co\ emissions of these sources in a similar way as the simulated emission maps presented above.

\subsection{Comparison of Observational and Synthetic Outflow \co\ (2-1) Spectra }
\label{Comparison of Observational and Synthetic Outflow Spectra}

\begin{figure*}[hbt!]
\centering
\includegraphics[width=0.99\linewidth]{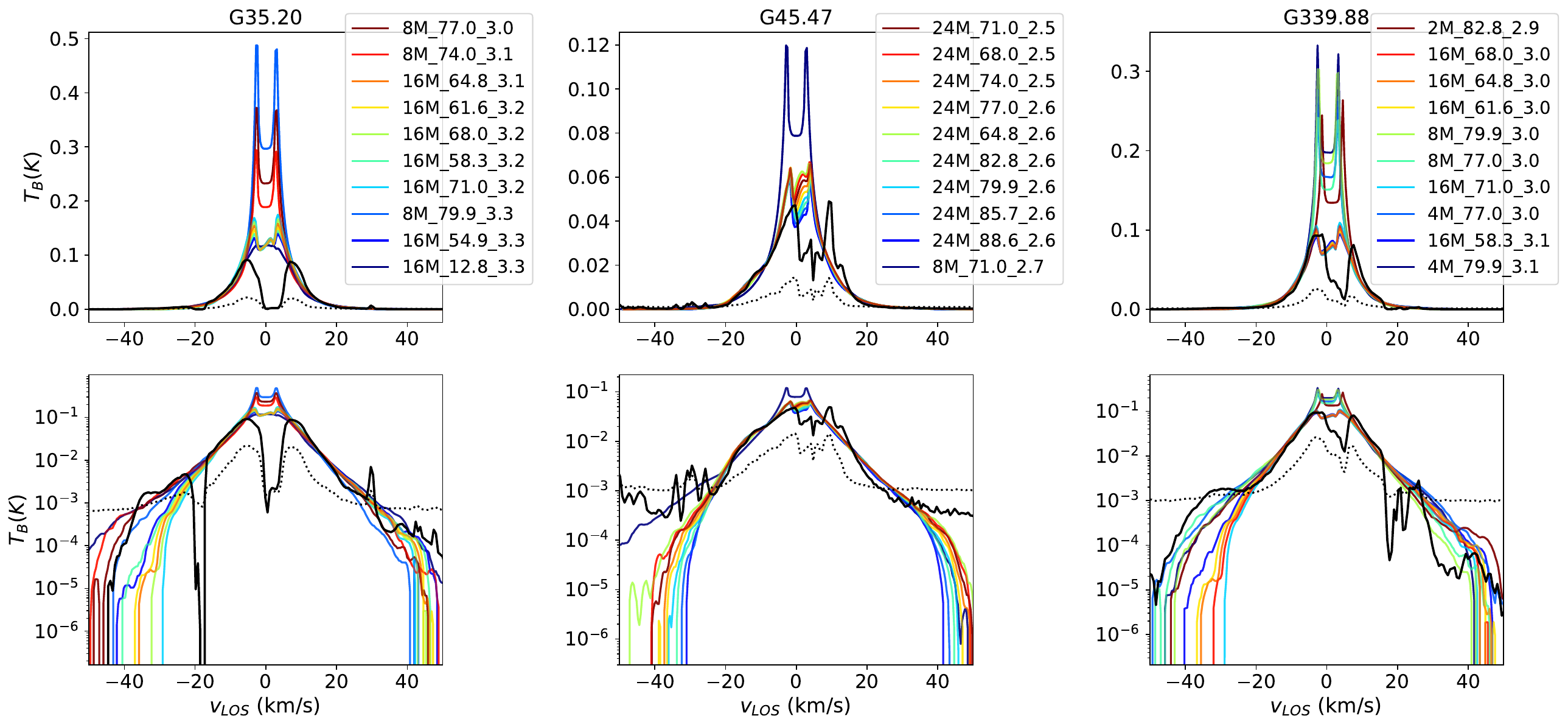}
\caption{Fitting between the observed \co\ (2-1) outflows and the synthetic ALMA \co\ (2-1) spectra post-processed with CASA/\texttt{simalma}. The observed outflow \co\ (2-1) spectra are depicted using black lines. {The 5$\sigma$ noise level of the spectrum is depicted by the dotted line.} The legends present the top ten best fitting results, where the first number corresponds to the protostellar mass, the middle number represents {the inclination angle}, and the last number denotes the {reduced} $\chi^{2}$ value. }
\label{fig.12CO21-realobs-spec-alma-fitting}
\end{figure*}


\begin{figure}[hbt!]
\centering
\includegraphics[width=0.99\linewidth]{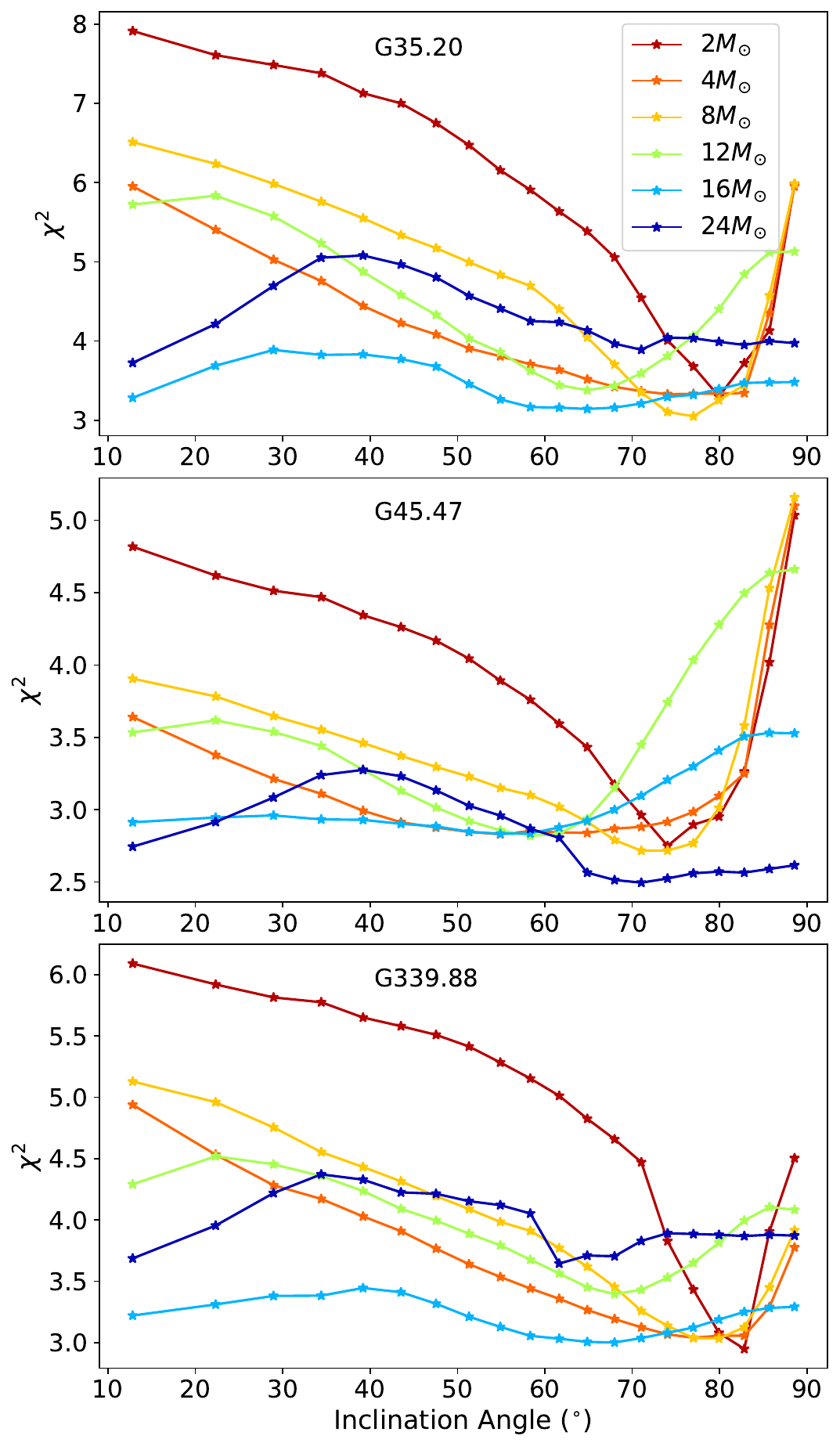}
\caption{Reduced $\chi^{2}$ values resulting from the fitting process between the observed \co\ (2-1) outflows and the synthetic ALMA \co\ (2-1) spectra post-processed with CASA/\texttt{simalma}, plotted against the protostellar mass and the {inclination angle}.}
\label{fig.12CO21_realobs_spec_alma_fitting_chi_theta}
\end{figure}

In this section, we conduct $\chi^{2}$ fittings of the observed outflow \co~(2-1) spectra with our synthetic ALMA \co~(2-1) spectra. To avoid the missing flux issue near the central velocity, we restrict our fitting to spectra with an absolute LOS velocity exceeding 6 \kms. {It is important to emphasize that our model's synthetic ALMA \co~(2-1) spectra, post-processed with CASA/simalma, are idealized, lacking ambient cloud structures and devoid of foreground and background molecular clouds. In actual observations, outflows are likely embedded in molecular clouds exhibiting diverse structures over various velocity ranges. This complexity can result in a more intricate drop in emission near the central velocity.} Additionally, since accurately determining the central velocity is challenging, we introduce a free parameter $v_{\rm offset}$ to the fitting process. This parameter is restricted to within $\pm$6 km/s. Another free parameter $f_{\rm scale}$ is used to set the emission scale of the synthetic spectrum. {This free parameter, $f_{\rm scale}$, makes allowance for variation in source distance and intrinsic protostellar properties, i.e., accommodating cores that are either more massive or less massive than the fiducial 60~\msun\ core we have simulated. It is essential to emphasize that caution is required when interpreting the fitted protostellar mass results, especially when the initial core mass deviates significantly from the 60~\msun\ core used in our single sequence of simulations. } Thus, we only have two free parameters in the $\chi^{2}$ fitting process for each protostellar mass and inclination angle. We obtain the minimum $\chi^{2}$ for each protostellar mass and inclination angle and subsequently arrange the $\chi^{2}$ values to identify the best-fitting parameters. {It is important to mention that in the $\chi^{2}$ fitting process, we mask out emissions below the $5\sigma$ noise level in the real observed spectra. Therefore, only emissions above the $5\sigma$ noise level are considered in the $\chi^{2}$ fitting.} The results of fitting the observed outflows and our synthetic spectra are presented in Figure~\ref{fig.12CO21-realobs-spec-alma-fitting}. The ten best fitting results, along with their respective {reduced} $\chi^{2}$ values, are listed in the figure legend. 
Additionally, we present the reduced $\chi^{2}$ values resulting from the fitting procedure in Figure~\ref{fig.12CO21_realobs_spec_alma_fitting_chi_theta}, providing insights into the constraints on the fitted parameters. It is important to mention that the initial core mass in the simulation is fixed at 60 \msun, while in reality, initial core masses may vary. {The fitted protostellar mass may not necessarily reflect the true protostellar mass in observational data. Rather, the obtained protostellar mass from the fitting process might serve as an indicator of the outflow's evolutionary stage. } 

{

Our fitting analysis suggests that the protostellar mass of G35.20 likely falls within the range of 8~\msun\ to 16 \msun. However, it is crucial to note that the constraint on the protostellar mass for G35.20 is not very tight, as indicated by the reduced $\chi^{2}$ plot in Figure~\ref{fig.12CO21_realobs_spec_alma_fitting_chi_theta}. Meanwhile, the inclination angle is also likely to be less constrained. Although there are more good-fitting results suggesting the cosine value of the inclination angle between 0.1 to 0.6, with $\theta_{\rm view}$ between 51\deg and 82\deg, there are a few good-fitting results suggesting an extremely face-on case, i.e., with $\theta_{\rm view}$ of 12.8\deg. For comparison, the study by \citet{2023ApJ...942....7F} determined the protostellar mass of G35.20 to be within the range of 13~\msun\ to 28 \msun, with an inclination angle of $\theta_{\rm view}=62^{\circ} \pm 17^{\circ}$. Notably, the SED fitting conducted by \citet{2023ApJ...942....7F} constrained the initial core mass of G35.20 to be between 79~\msun\ and 189 \msun, which is larger than that in our simulations. This difference might explain why the \co~(2-1) spectra-fitted protostellar mass of G35.20 is slightly smaller than that obtained from the SED fitting results. The inclination angle constrained by SED fitting seems to be somewhat consistent with the most abundant good-fitting results, which are between 51\deg and 82\deg.


Based on our fitting analysis, the protostellar mass of G45.47 is estimated to be around 24 \msun. The reduced $\chi^{2}$ of the synthetic spectra with a protostellar mass of 24~\msun\ in fitting the G45.47 spectra is significantly smaller than that of other protostellar masses. Furthermore, the inclination angle is likely to be oriented in a more edge-on configuration, with the cosine value of the inclination angle smaller than 0.4, corresponding to $\theta_{\rm view}$ exceeding 65 degrees. The SED-fitting study by \citet{2023ApJ...942....7F} determined the protostellar mass of G45.47 to be within the range of 23~\msun\ to 53~\msun, with an inclination angle of $\theta_{\rm view}=61^{\circ} \pm 18^{\circ}$. \citet{2023ApJ...942....7F} also constrained the initial core mass of G45.47 to be between 228~\msun\ and 444~\msun. The \co~(2-1) CO-spectra-fitted protostellar mass of G45.47 is consistent, although slightly smaller, than that obtained from the SED fitting results. This slight underestimation is likely caused by the much larger initial core mass in reality. The inclination angles obtained from \co~(2-1) spectra fitting are consistent with the results of the SED fitting in \citet{2023ApJ...942....7F} for G45.47. {It is important to consider that G45.47 is four times farther away than the other two sources and the synthetic ALMA observations. Consequently, the actual physical scale of G45.47 may be substantially larger than the scale captured in our synthetic observations. Therefore, caution is warranted in interpreting the fitting results for G45.47, as it could potentially represent a scaled-up version of our fiducial model with an initial core mass of 60 \msun. }


Concerning G339.88, our fitting analysis suggests that the protostellar mass is poorly constrained, ranging from 2~\msun\ to 16~\msun. The inclination angle is likely to be more edge-on, with $\theta_{\rm view}$ exceeding 60 degrees. In contrast, the study by \citet{2023ApJ...942....7F} determined the protostellar mass of G339.88 to be within the range of 11~\msun\ to 42~\msun, with an inclination angle of $\theta_{\rm view}=65^{\circ} \pm 16^{\circ}$. The SED fitting in \citet{2023ApJ...942....7F} constrained the initial core mass of G339.88 to be between 112~\msun\ and 288~\msun. As our protostellar mass fitting result is poorly constrained, a fair comparison with the results obtained from SED fitting is challenging. However, the inclination angles obtained from \co~(2-1) spectra fitting are consistent with the results of the SED fitting in \citet{2023ApJ...942....7F} for G339.88.

In summary, the fitting of \co~(2-1) spectra for the three sources, G35.20, G45.47, and G339.88, provides a general constraint on the inclination angles of the outflows, aligning with the results constrained by SED fitting in \citet{2023ApJ...942....7F}. However, accurately constraining the protostellar mass remains challenging, except for the case of G45.47.

}

\subsection{Estimating the Mass Outflow Rate and
Momentum Outflow Rate in Observations}
\label{Estimating the Mass Outflow Rate and
Momentum Outflow Rate in Observations}

\begin{table*}[htbp]
  \centering
  \caption{Summary of outflow properties for G35.20, G45.47 and G339.88$^{a}$}
    \begin{tabular}{ccccccccc}
      \hline
    \hline
          & $m_{\rm w,out}$     & $p_{\rm w,out}$     & $E_{\rm w,out}$     & $\overline{v}$     & $L_{\rm out,half}$ & $\tau_{\rm dyn}$  & $\dot{m}_{\rm w,out}$  & $\dot{p}_{\rm w,out}$ \\
        & ($M_{\odot}$)     & ($M_{\odot}\, {\rm km\, s^{-1}}$)     & ($10^{45}$ ergs)     & (\kms)    & ($10^4$ AU) & ($10^{4}$ yr)  & ($10^{-5}\, M_{\odot} \,{\rm yr}^{-1}$)   & ($10^{-4}\, M_{\odot} \,{\rm km\, s^{-1}\, yr}^{-1}$) \\\hline
    G35.20 & 0.37$\pm$0.06  & 3.97$\pm$0.61  & 1.27$\pm$0.20  & 10.72 & 3.30  & 1.46  & 2.53$\pm$0.39  & 2.72$\pm$0.42 \\\hline
    G45.47 & 2.53$\pm$0.39  & 29.95$\pm$4.63 & 11.94$\pm$1.84 & 11.85 & 12.60 & 5.04  & 5.01$\pm$0.77  & 5.94$\pm$0.92 \\\hline
    G339.88 & 0.26$\pm$0.04  & 2.64$\pm$0.41  & 0.92$\pm$0.14  & 10.05 & 3.15  & 1.49  & 1.76$\pm$0.27  & 1.77$\pm$0.27 \\\hline
    \multicolumn{9}{p{0.99\linewidth}}{Notes:}\\
\multicolumn{9}{p{0.99\linewidth}}{$^a$ The uncertainty in the estimates represents the standard deviation of the values obtained when using different excitation temperatures ranging from 10 to 50 K. }  
    \end{tabular}%
  \label{tab.Summary_outflow_obs_3sources}%
\end{table*}%

\begin{figure*}[hbt!]
\centering
\includegraphics[width=0.48\linewidth]{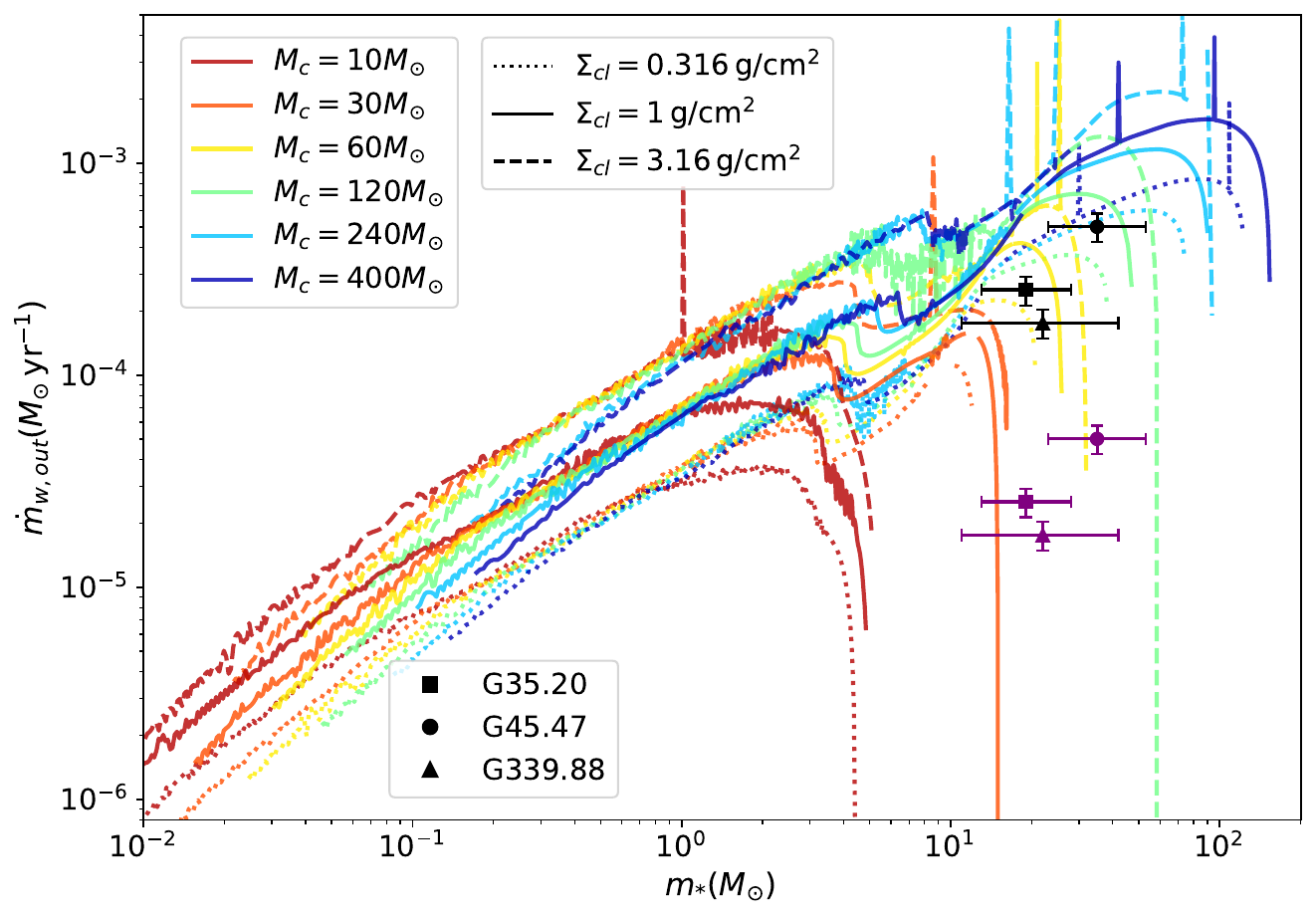}
\includegraphics[width=0.48\linewidth]{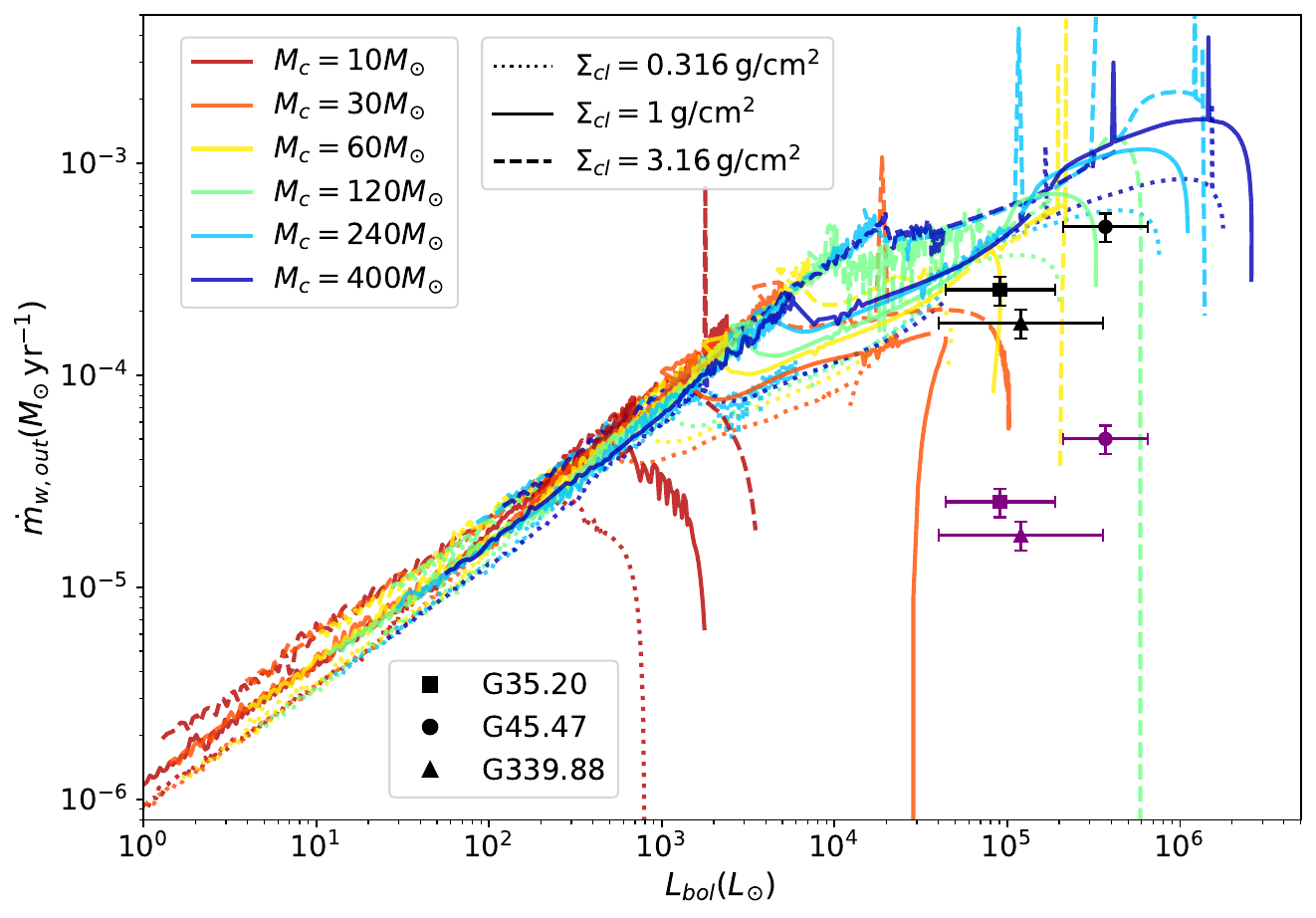}
\caption{Evolution of mass outflow rate as a function of protostellar mass $m_{*}$ ({\it left}) and bolometric luminosity $L_{\rm bol}$ ({\it right}) along various protostellar evolutionary sequences. These sequences are characterized by different clump mass surface densities $\Sigma_{\rm cl}$ and initial core masses $M_{c}$ \citep{2018ApJ...853...18Z}.
The square, circle, and triangle symbols represent the data points for G35.20, G45.47, and G339.88, respectively. The purple symbols represent the raw values of the mass outflow rates, while the black symbols represent the corrected values after applying a correction factor of 10 to the mass outflow rates. }
\label{fig.mdot_plots}
\end{figure*}

\begin{figure*}[hbt!]
\centering
\includegraphics[width=0.48\linewidth]{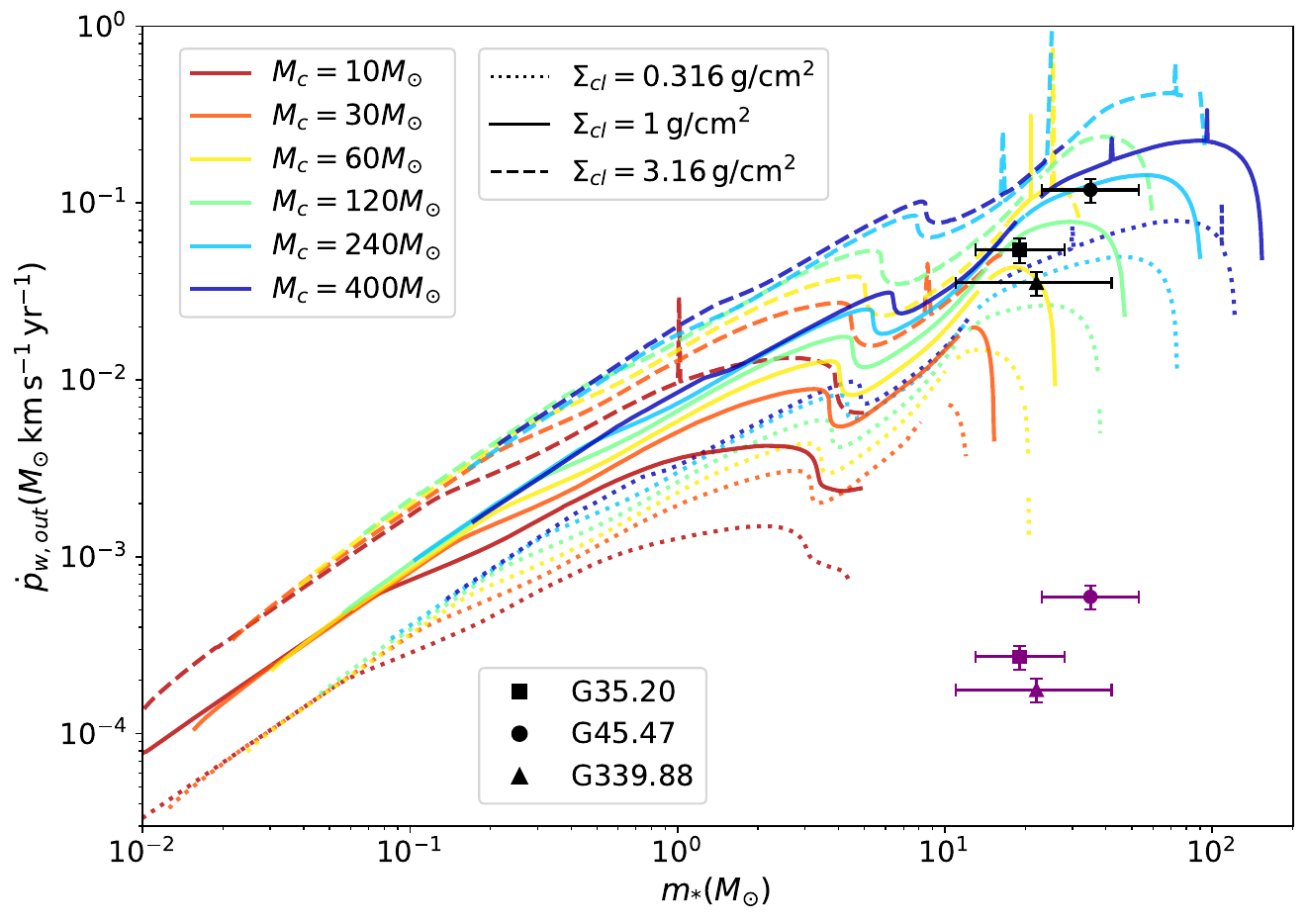}
\includegraphics[width=0.48\linewidth]{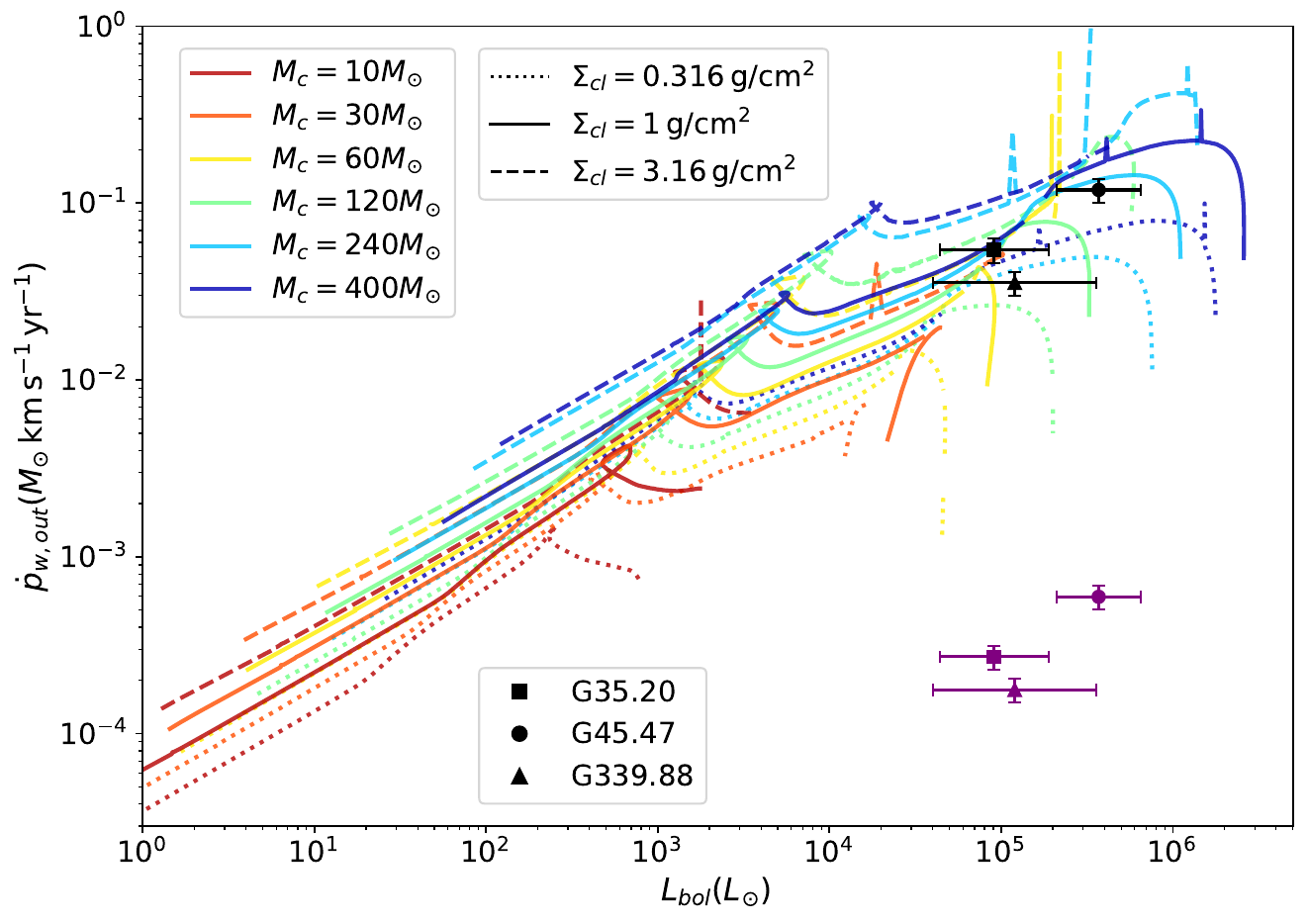}
\caption{Evolution of momentum outflow rate as a function of protostellar mass $m_{*}$ ({\it left}) and bolometric luminosity $L_{\rm bol}$ ({\it right}) along various protostellar evolutionary sequences. These sequences are characterized by different clump mass surface densities $\Sigma_{\rm cl}$ and initial core masses $M_{c}$ \citep{2018ApJ...853...18Z}. 
The square, circle, and triangle symbols represent the data points for G35.20, G45.47, and G339.88, respectively. The purple symbols represent the raw values of the momentum outflow rates, while the black symbols represent the corrected values after applying a correction factor of 200 to the momentum outflow rates. }
\label{fig.pdot_plots}
\end{figure*}

In this section, we calculate the mass outflow rate and momentum outflow rate for the three observed outflows. The method used for these calculations follows the approach described in \S~\ref{Estimating the Mass, Momentum and Energy of Synthetic Outflows}. Additionally, we determine the mass-weighted mean LOS velocity and dynamical timescale of these outflows, using the procedure explained in \S~\ref{Estimating the Outflow Mass Injection Rate and Momentum Injection Rate}. For estimating the typical outflow length, we adopt the assumption of a one-sided lobe of the outflow corresponding to a half-length of approximately 15\arcsec, which is roughly half the size of the field of view from \citet{2019ApJ...873...73Z}. We acknowledge that this could potentially represent a lower limit for the size of the outflow, as it is plausible that outflowing gas may extend beyond the boundaries of the field of view. {It is worth noting that the calculated observed outflow dynamical timescale is around 12 kyr \citep{2019ApJ...873...73Z}, aligning with the range of our time scale estimates for synthetic outflows in Section \ref{Estimating the Outflow Mass Injection Rate and Momentum Injection Rate}. The variability in our time scale estimates is, in part, attributed to the fact that the outflow length used in synthetic observations is individually calculated for each case, contrasting with the assumption of a fixed value in actual observations.} 
The obtained mass outflow rates and momentum outflow rates for the three outflows are summarized in Table~\ref{tab.Summary_outflow_obs_3sources}.


In \S\ref{Comparison of Observational and Synthetic Outflow Spectra}, we determined the inclination angle of the three outflows, and all of them are likely to have an inclination angle exceeding 60 degrees. When the inclination angle falls within the range of 60 to 70 degrees, the conversion factor between the true mass outflow rate and the observed mass outflow rate is approximately 10, while the conversion factor between the true momentum outflow rate and the observed momentum outflow rate is around 200. Based on this, we derived the final estimates for the mass outflow rates of G35.20, G45.47, and G339.88 as $2.53\times10^{-4}M_{\odot}\, {\rm yr}^{-1}$, $5.01\times10^{-4}M_{\odot}\, {\rm yr}^{-1}$, and $1.76\times10^{-4}M_{\odot}\, {\rm yr}^{-1}$, respectively. Additionally, our final estimations for the momentum outflow rates of G35.20, G45.47, and G339.88 are $5.43\times10^{-2}M_{\odot} {\rm km\, s^{-1}\, yr}^{-1}$, $1.19\times10^{-1}M_{\odot} {\rm km\, s^{-1}\, yr}^{-1}$, and $3.55\times10^{-2}M_{\odot} {\rm km\, s^{-1}\, yr}^{-1}$, respectively.

We then place the values for the three sources on the protostellar evolutionary tracks obtained from \citet{2018ApJ...853...18Z}.
The protostellar mass $m_{*}$ and bolometric luminosity $L_{\rm bol}$ are determined from SED fitting as reported in \citet{2023ApJ...942....7F}. Figure~\ref{fig.mdot_plots} illustrates the evolution of mass outflow rate as a function of $m_{*}$ and $L_{\rm bol}$ along various protostellar evolutionary sequences. After applying a correction factor of 10 to the observed mass outflow rate, the final mass outflow rate aligns more closely with the theoretical protostellar evolutionary tracks, though some potential underestimation is still apparent. Figure~\ref{fig.pdot_plots} presents the evolution of momentum outflow rate as a function of $m_{*}$ and $L_{\rm bol}$ along various protostellar evolutionary sequences. After correcting the observed momentum outflow rate by a factor of 200, the final momentum outflow rate {agrees with} the theoretical protostellar evolutionary tracks. {This implies that the mass outflow rate and momentum outflow rate derived from the \co~(2-1) spectrum align with the theoretical protostellar evolutionary tracks. While this agreement is to be expected given that the MHD simulation in this study is built upon the theoretical protostellar evolutionary tracks proposed by \citet{2018ApJ...853...18Z}, the fact that the observed sources have both SED-derived and CO-spectra derived properties that are consistent with the models is important for giving new validation for assumptions made in the model, especially with respect to outflow launching. In summary, measurements of CO outflow properties help to constrain protostellar properties, especially those related to outflow launching, orientation of outflow axis to the line of sight, and protostellar evolutionary stage, complementing SED fitting methods, which typically suffer from significant degeneracies \citep[e.g.,][]{2023ApJ...942....7F}.}

\section{Conclusions}
\label{Conclusions}

In this study, we have used \radmc\ and CASA/\texttt{simalma} to perform radiative transfer and generate synthetic molecular line emission images from 3D MHD simulations of a disk wind outflow. The main results are summarized as follows:

\begin{enumerate}

\item We have presented synthetic observations of the outflows for multiple transitions of \co, \13co, and \c18o. {The outflow morphology is generally uniform across various \co\ transitions, with the exception of the \co~(14-13) transition, which displays a more centralized morphology, emphasizing denser regions of the outflow. Additionally, \13co\ and \c18o\ transitions effectively capture both the overall outflow morphology and the dense portions of the outflow regions.} Our analysis shows that the opening angle of the outflow, {as traced by CO}, increases as the protostellar mass increases. Moreover, the high-velocity components of the outflow are located closer to the outflow launching axis than the low-velocity components. As the inclination angle decreases, more of the outflow forward velocity falls along the line-of-sight, and the high-velocity components become more prominent and adopt a larger spatial distribution.

\item Synthetic interferometric observations of \co~(2-1) outflows were simulated using CASA/\texttt{simalma}, and these synthetic ALMA \co~(2-1) spectra were then compared with three observed ALMA \co~(2-1) spectra. {The fitting outcomes offer a broad constraint on the inclination angles of the outflows, consistent with the findings from SED fitting. While there is some sensitivity of the CO spectra to protostellar mass and evolutionary stage, achieving precise constraints on the protostellar mass via fitting of CO outflows remains a challenging task, which will require synthetic emission to be computed from a more comprehensive grid of simulations.}

\item {We have quantified the total outflow mass and compared it with the outflow mass that is potentially observable in a given velocity range (first for the case ignoring the effects of radiative transfer and imperfect flux recovery from interferometric observations). The main finding is that a significant amount of mass, about ten times or more, is missed if excluding velocities near the rest frame velocity ($v_{\rm LOS}<3$ \kms). However, exclusion of very high velocity material, i.e., $>50\:$\kms, has a negligible impact on the estimate of the total outflow mass.}


\item {We have also analyzed the conversion factor between the total 3D momentum and kinetic energy and the 1D LOS momentum and kinetic energy (first for the case ignoring the effects of radiative transfer and imperfect flux recovery from interferometric observations).}
Although the high-velocity gas ($v_{\rm LOS}>50$ \kms) has negligible mass, it contributes significantly to the overall momentum and energy. Consequently, when estimating the total 3D momentum and energy from the LOS momentum with an absolute LOS velocity ranging from 3 to 50 \kms, there is a probable underestimation of the total 3D momentum and 3D energy by a factor of 4 and 50, respectively.

\item The estimated outflow masses derived from the raw \co~(2-1), \13co~(1-0), and synthetic ALMA \co~(2-1) spectra show {a range of values differing by a factor of} 4, 2, and 8, respectively, when compared to the true values obtained from the simulation.

\item The estimated outflow momentum and energy derived from the raw \co~(2-1), \13co~(1-0), and synthetic ALMA \co~(2-1) spectra show a scatter within a factor of 4 when compared to the true values obtained from the simulation.

\item We have quantified the mass outflow rate and momentum outflow rate from the synthetic ALMA \co~(2-1) spectra and compared them with the true values derived from simulations. Generally, a conversion factor of 10 is needed to obtain the true mass outflow rate, and a conversion factor of 200 is required for the true momentum outflow rate from the synthetic ALMA \co~(2-1) spectra. Furthermore, we have examined the variation of these conversion factors as a function of inclination angle of outflow axis to line-of-sight.

\item We have computed the mass outflow rate and momentum outflow rate for the three massive protostars, G35.20, G45.47, and G339.88. After applying correction factors of 10 and 200 to the observed mass outflow rate and momentum outflow rate, respectively, both rates {agree} with the expected theoretical protostellar evolutionary tracks. {Thus these measurements of CO outflows help to constrain protostellar properties, especially parameters related to outflow launching, inclination and evolutionary stage, as a complement to SED fitting.}

\end{enumerate}

{We express our sincere gratitude to the anonymous referee for their valuable comments, which have significantly enhanced the manuscript.} D.X. and J.P.R. acknowledge support from the Virginia Initiative on Cosmic Origins (VICO). J.C.T. acknowledges support from NSF grant AST-2009674 and ERC Advanced Grant MSTAR. JPR is supported in part by NSF grants AST-1910106, AST-1910675, and NASA grant 80NSSC20K0533. The authors acknowledge Research Computing at The University of Virginia for providing computational resources and technical support that have contributed to the results reported within this publication. The authors further acknowledge the use of NASA High-End Computing (HEC) resources through the NASA Advanced Supercomputing (NAS) division at Ames Research Center to support this work. 

\appendix

{
\section{Structure of Outflow Density and Velocity}
\label{Structure of Outflow Density and Velocity}

In this section, we present figures and conduct statistical studies on the structure of the outflows. Figure~\ref{fig.density_velocity_outflow_slice} illustrates example slices of the density and velocity structures of the outflows with different protostar masses (see also Paper II).

\begin{figure*}[hbt!]
\centering
\includegraphics[width=0.91\linewidth]{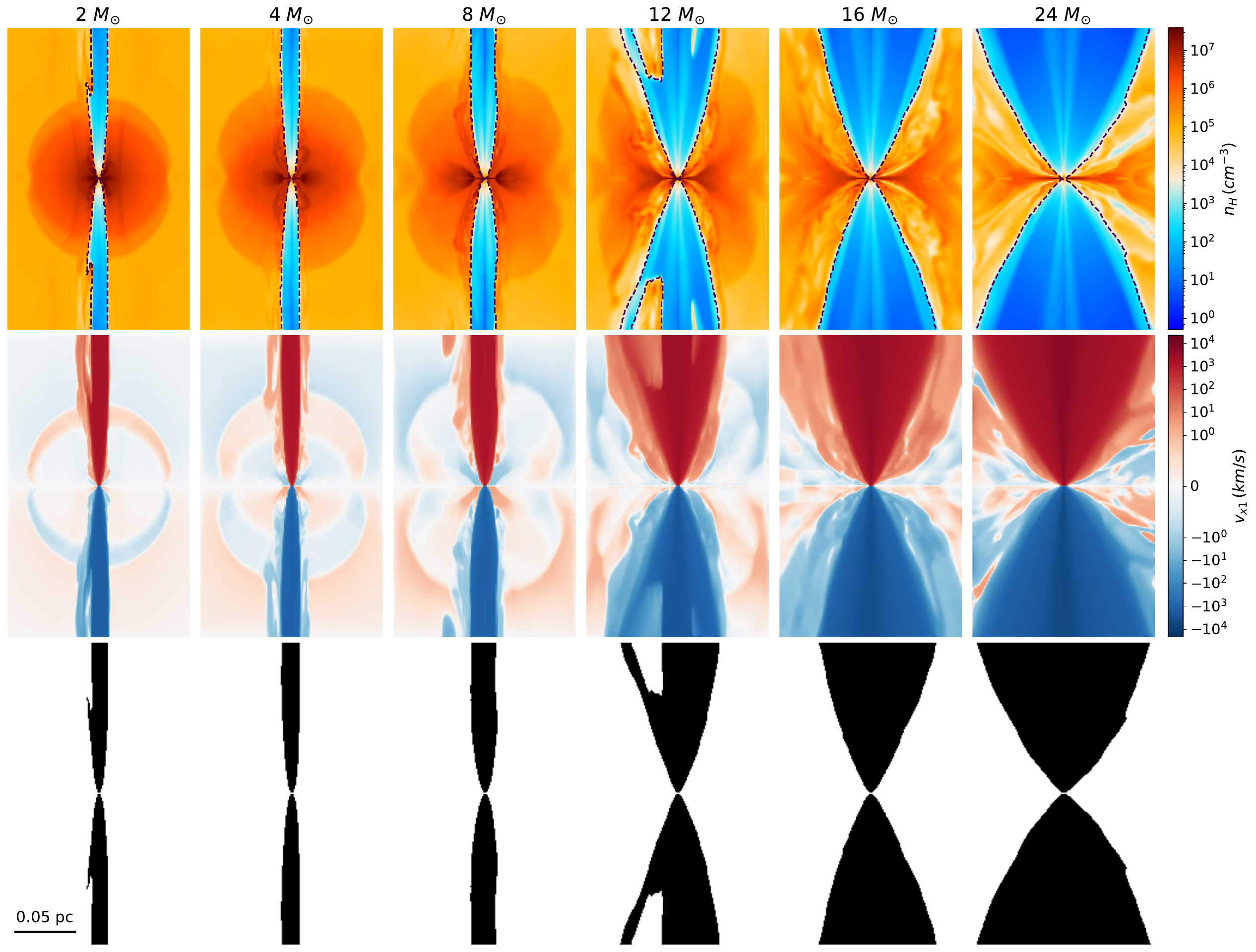}
\caption{Density ({\it $1^{st}$ row}), $x_1$-direction (i.e., outflow launching axis) velocity structures ({\it $2^{nd}$ row}), and outflow mask indicating velocities above 1 km/s outward ({\it $3^{rd}$ row}) at the central slice of outflows with different protostar masses. Dashed contours in the first row indicate the outflow mask.}
\label{fig.density_velocity_outflow_slice}
\end{figure*}

\begin{figure*}[hbt!]
\centering
\includegraphics[width=0.69\linewidth]{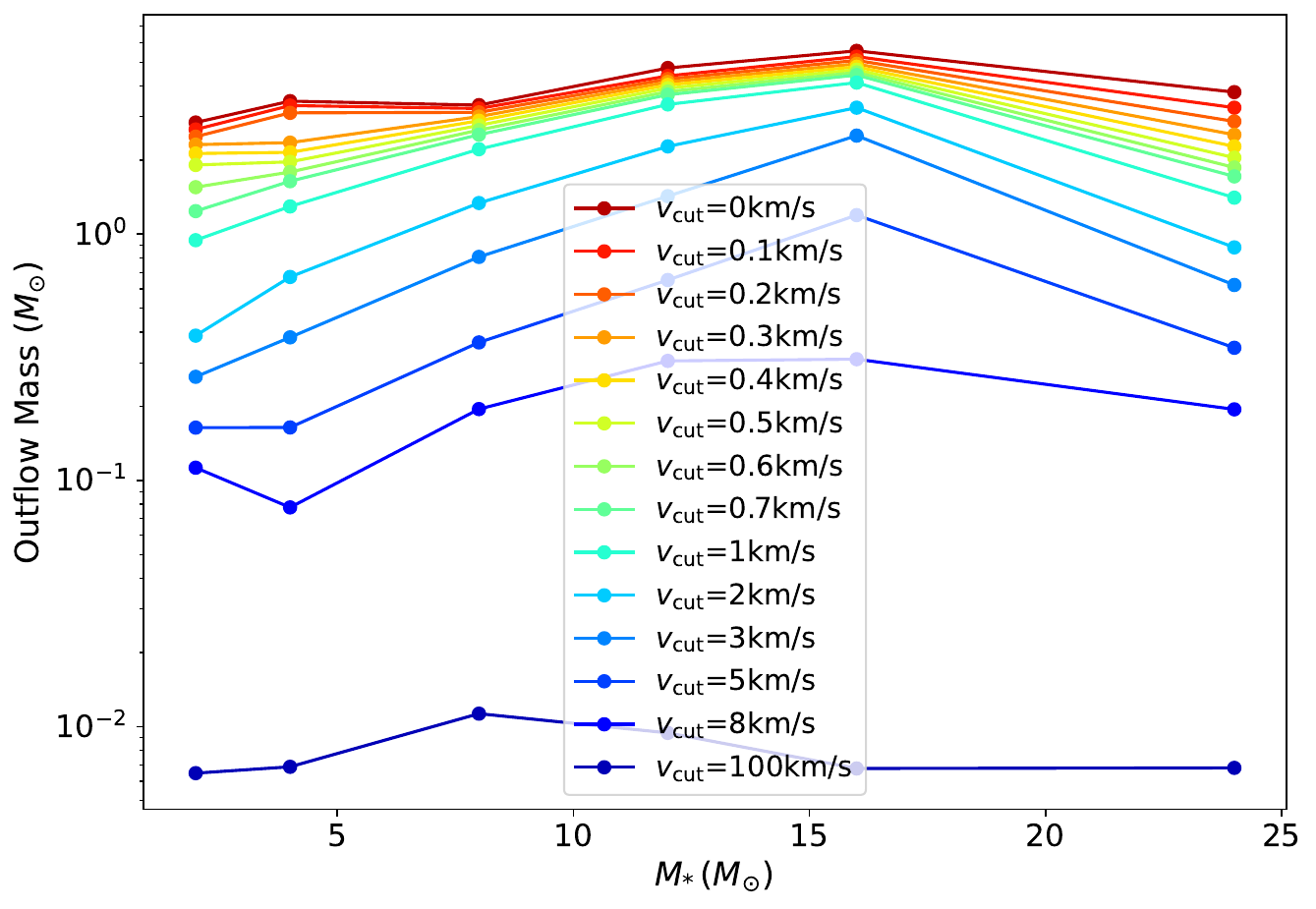}
\caption{Outflow mass defined by varying velocity thresholds across different evolutionary stages, corresponding to different protostellar masses. }
\label{fig.outflow_mass_mstar_vcut_log}
\end{figure*} 

{It is imperative to emphasize that when calculating outflow mass, momentum, and energy, we employ a different definition for outflow material compared to the approach involving a velocity cutoff of 100 km/s applied to different dust components in the context of dust radiative transfer, as discussed in Section~\ref{Dust Temperature}. Here, outflow material is defined as gas exhibiting an outward direction and a velocity surpassing the velocity dispersion of the turbulent core, corresponding to velocities above 1 km/s. We investigated various velocity cutoffs for defining outflow material in Figure~\ref{fig.outflow_mass_mstar_vcut_log}. Notably, when considering only gas with velocities above 100 km/s, the calculated outflow mass is negligible, amounting to less than 0.01 \msun, failing to account for the significant contribution from entrained gas, which also moves with the outflowing gas. In reality, actual outflow gas is often mixed with ambient gas, and each gas pixel exhibiting some outgoing velocity from the star is influenced by the outflowing gas. Consequently, all of this gas, including entrained gas, collectively contributes to the kinetic impact on the host cloud. Therefore, in our practical calculations, we assess the mass, energy, and momentum of outflow material, encompassing both the outflow gas and the entrained gas, rather than focusing solely on the low-density outflow gas. In earlier investigations, such as in the work of \citet{2017ApJ...847..104O}, the mass loading factor of the entrained gas was reported to be 300\%. Additionally, both observational and simulation studies demonstrate that outflows remove dense material, reducing the available material for accretion \citep{2014ApJ...783...29D,2016ApJ...832..158Z}. Numerical simulations suggest that outflows reduce the efficiency of dense gas by 30\%-40\%, implying that outflow material constitutes 30\%-40\% of the accreted mass on the star \citep{2013MNRAS.431.1719M,2014ApJ...784...61O}. Therefore, we argue that a 100 km/s velocity cutoff is inappropriate. Instead, we adopt a 1 km/s velocity cutoff, where the outflow mass linearly increases with protostellar mass for masses below 16 \msun. The ratio between outflow mass and protostellar mass falls within the range of 10\%-40\%, consistent with simulation studies. We acknowledge that the choice of velocity cutoff is somewhat arbitrary, but even with a cutoff of 0.1 km/s, the outflow mass is on average twice as large as that with a 1 km/s cutoff. To visualize the position of outflow material with a 1 km/s velocity cutoff, we provide the outflow mask in Figure~\ref{fig.density_velocity_outflow_slice}. This mask accurately delineates the outflow cavity.
}

Figure~\ref{fig.density_hist_outflow_all} displays the distribution of the volume-weighted outflow gas density and the mass-weighted outflow gas density. Notably, a substantial volume exhibits relatively low density ($<10^{4}$ \cmc), yet the mass within these low-density regions is small, constituting less than 0.7\% of the total outflow gas mass. {Given that the critical density of CO is on the order of $10^{3-4}$ \cmc, this indicates that the LTE assumption is reasonable when calculating outflow properties from synthetic CO line emission, despite the fact that the synthetic CO observations are generated by \radmc\ under the non-LTE assumption. }


\begin{figure*}[hbt!]
\centering
\includegraphics[width=0.91\linewidth]{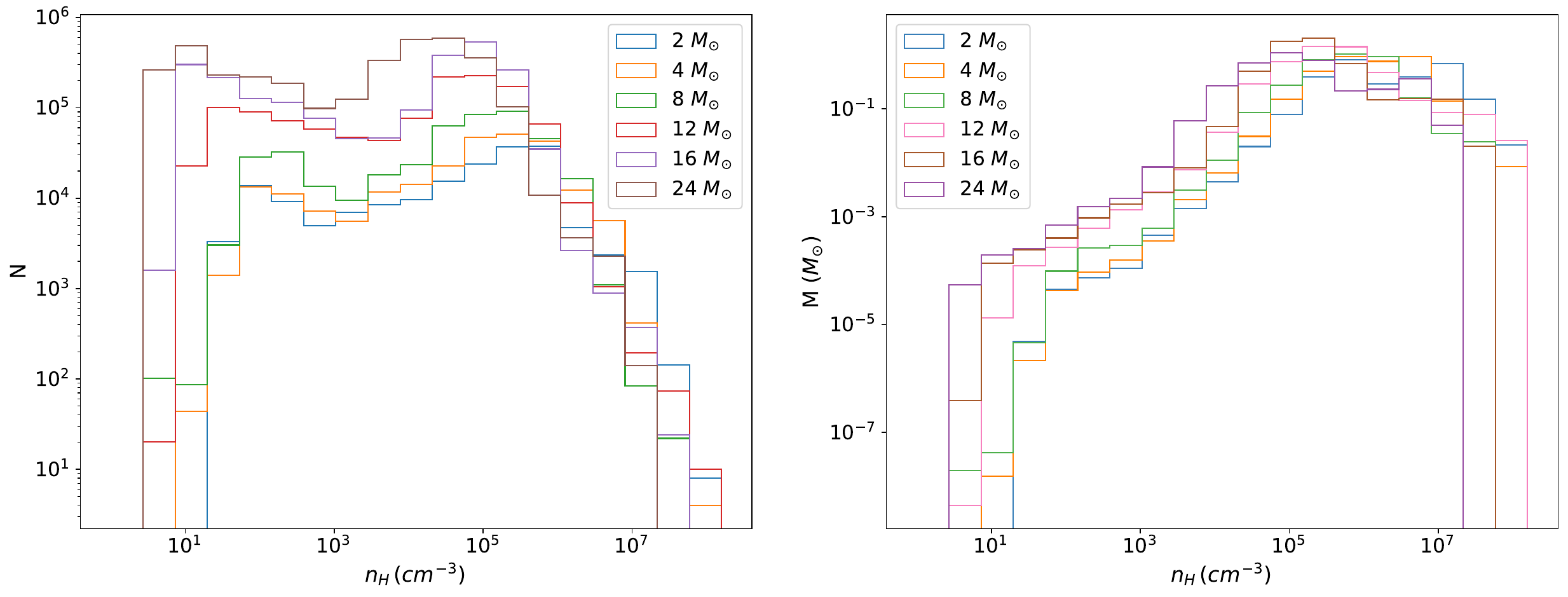}
\caption{Distribution of volume-weighted outflow gas density ({\it left}) and mass-weighted outflow gas density distribution ({\it right}). }
\label{fig.density_hist_outflow_all}
\end{figure*}

\begin{deluxetable}{ccccc}
  \tablewidth{0.8\columnwidth}
  \tabletypesize{\footnotesize}
  \tablecaption{Protostellar input parameters (i.e., protostellar mass, bolometric luminosity, effective photospheric temperature, and radius) for dust temperature calculations models (Sect.\ \ref{Dust Temperature}) adopted from \cite{2014ApJ...788..166Z}.\label{tab:Table_protostellar_properties}}
  \tablehead{\colhead{$m_*$ ($M_\odot$)} & \colhead{$L_{\rm bol}$\tablenotemark{a} ($L_\odot$)} & \colhead{$T_{\rm eff}$\ (K)} & \colhead{$R_*$\ ($R_\odot$)}}
  \startdata
  4  & 1301.15 & 7955.72 & 19.007 \\
  8  & 12499.0 & 10640.8 & 32.930  \\
  12 & 44323.6 & 23303.7 & 12.929 \\
  16 & 65461.5 & 36650.5 & 6.3524 \\
  24 & 84459.8 & 39021.1 & 6.3655 \\
  \enddata
  \tablenotetext{a}{Combined protostellar + accretion luminosity.}
\end{deluxetable}

}

\section{Moment Maps of Synthetic Outflows}
\label{Moment Maps of Synthetic Outflows}

\begin{figure*}[hbt!]
\centering
\includegraphics[width=0.91\linewidth]{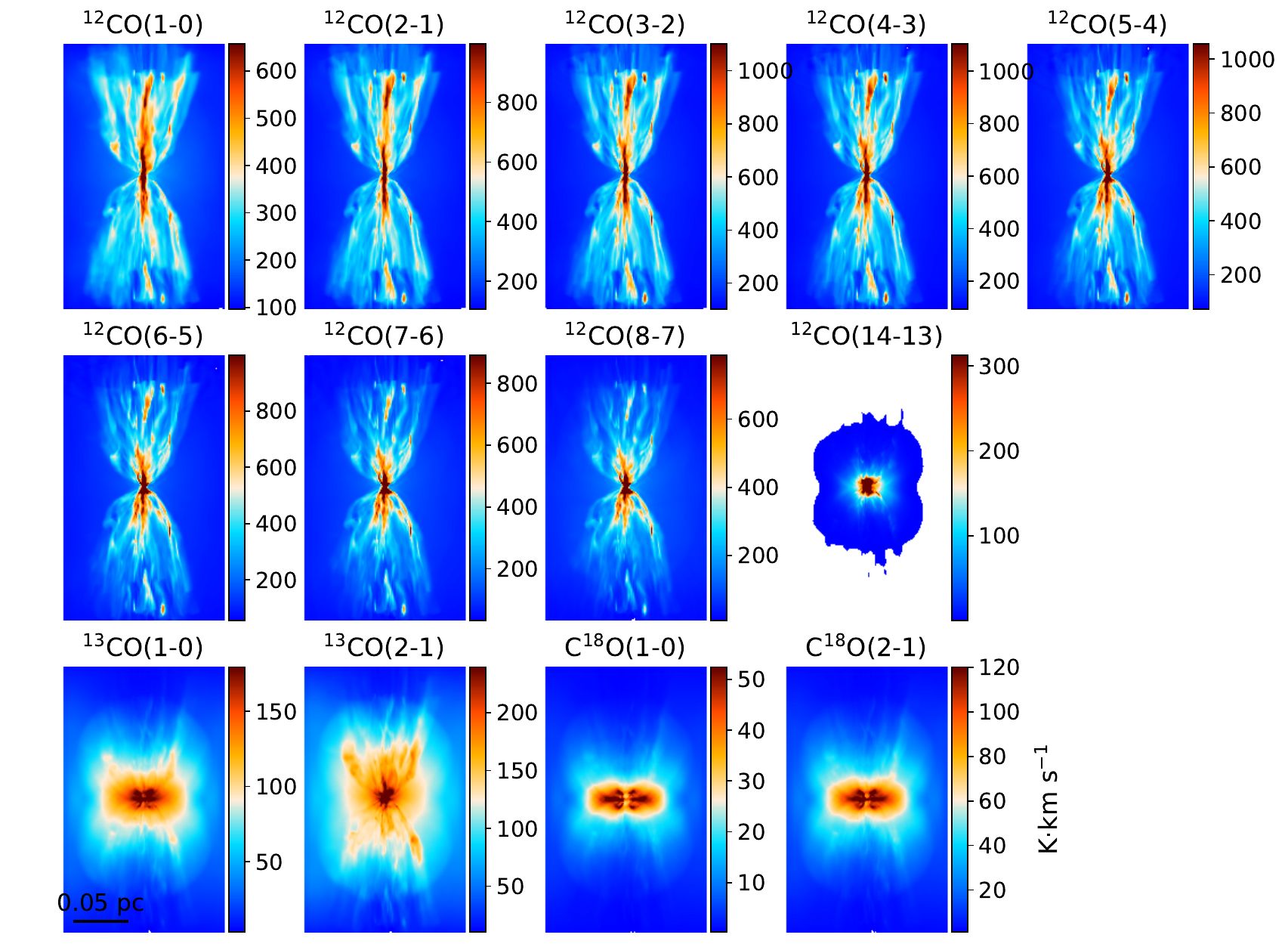}
\caption{Moment zero maps (integrated intensity) of multiple transitions of \co, \13co\, and \c18o\, of the outflow from a 12~\msun\ star at an inclination angle of 58\deg.}
\label{fig.co21-mass12-cos0525-allco-m0}
\end{figure*}

\begin{figure*}[hbt!]
\centering
\includegraphics[width=0.91\linewidth]{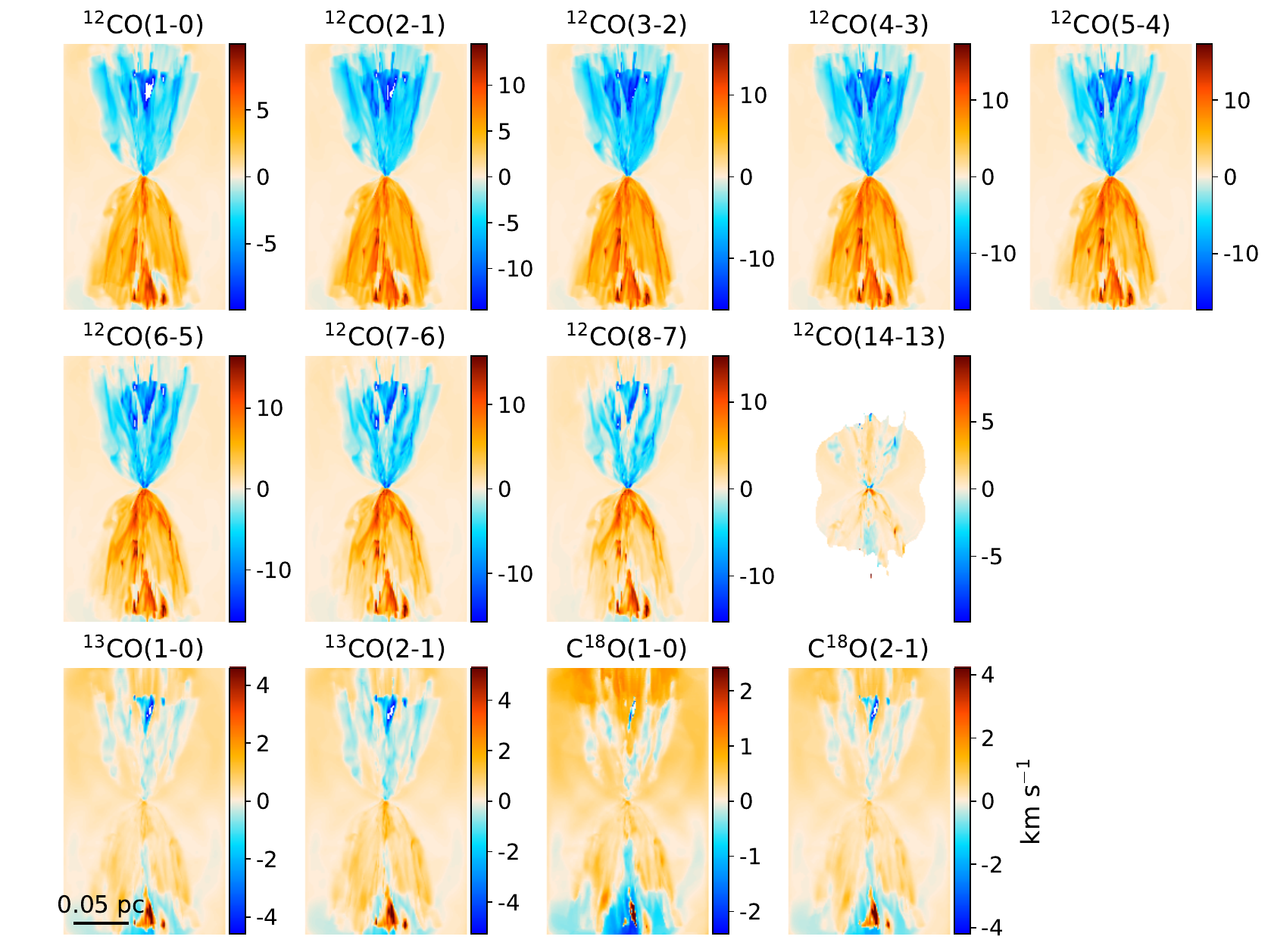}
\caption{Moment one maps (velocity field) of multiple transitions of \co, \13co\, and \c18o\, of the outflow from a 12~\msun\ star at an inclination angle of 58\deg.}
\label{fig.co21-mass12-cos0525-allco-m1}
\end{figure*}

\begin{figure*}[hbt!]
\centering
\includegraphics[width=0.91\linewidth]{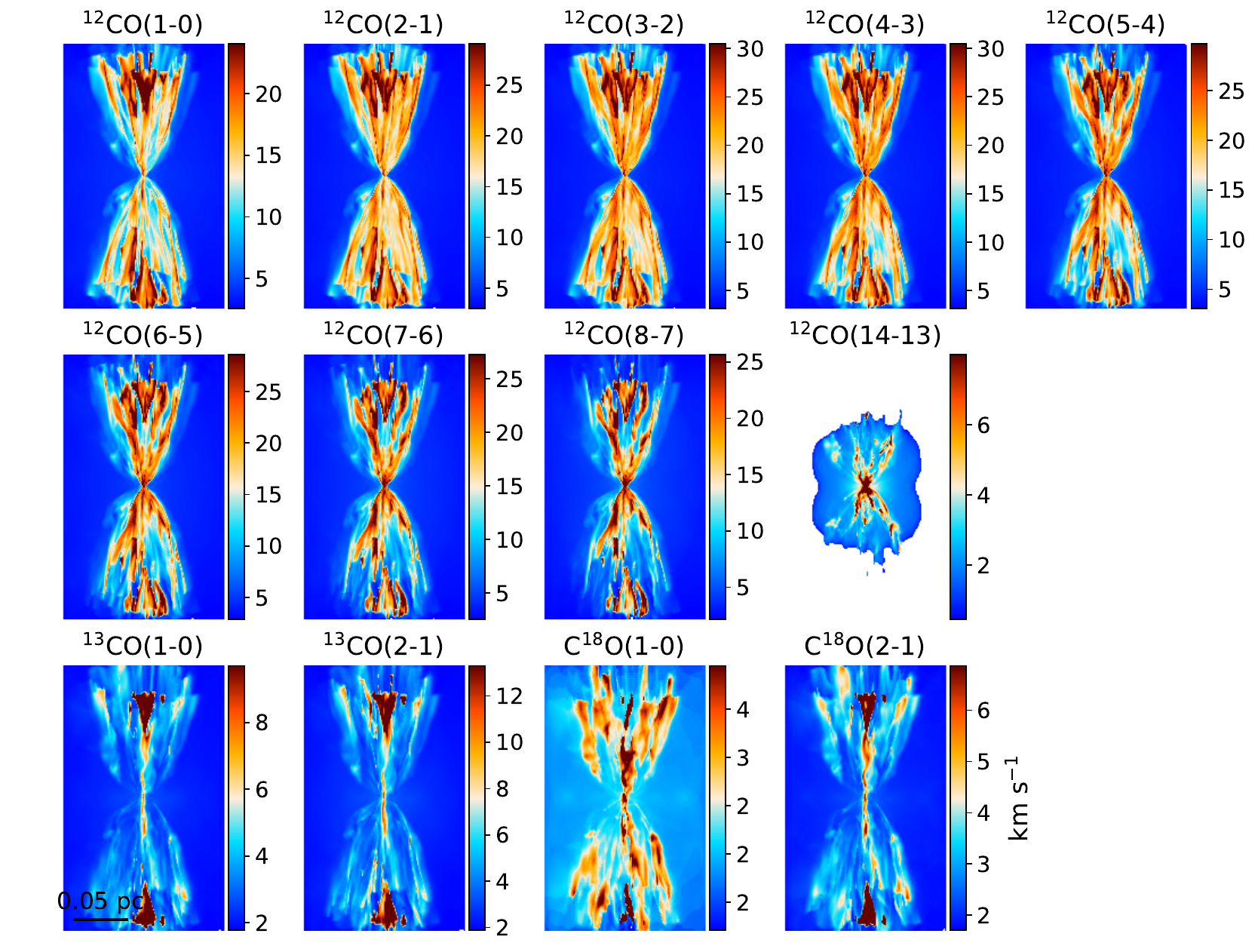}
\caption{Full width at half maximum (FWHM) maps of multiple transitions of \co, \13co\, and \c18o\, of the outflow from a 12~\msun\ star at an inclination angle of 58\deg.}
\label{fig.co21-mass12-cos0525-allco-fwhm}
\end{figure*}

In this section, we present a gallery of moment maps of synthetic outflows as shown in Figure~\ref{fig.co21-mass12-cos0525-allco-m0}-\ref{fig.co21-mass12-cos0525-allco-fwhm}.

\section{Estimates of Mass, Momentum, and Energy of Synthetic Outflows using \co~(1-0) and $^{13}$CO~(2-1)}
\label{Estimates of Mass, Momentum, and Energy of Synthetic Outflows using co 1-0}

\begin{figure*}[hbt!]
\centering
\includegraphics[width=0.49\linewidth]{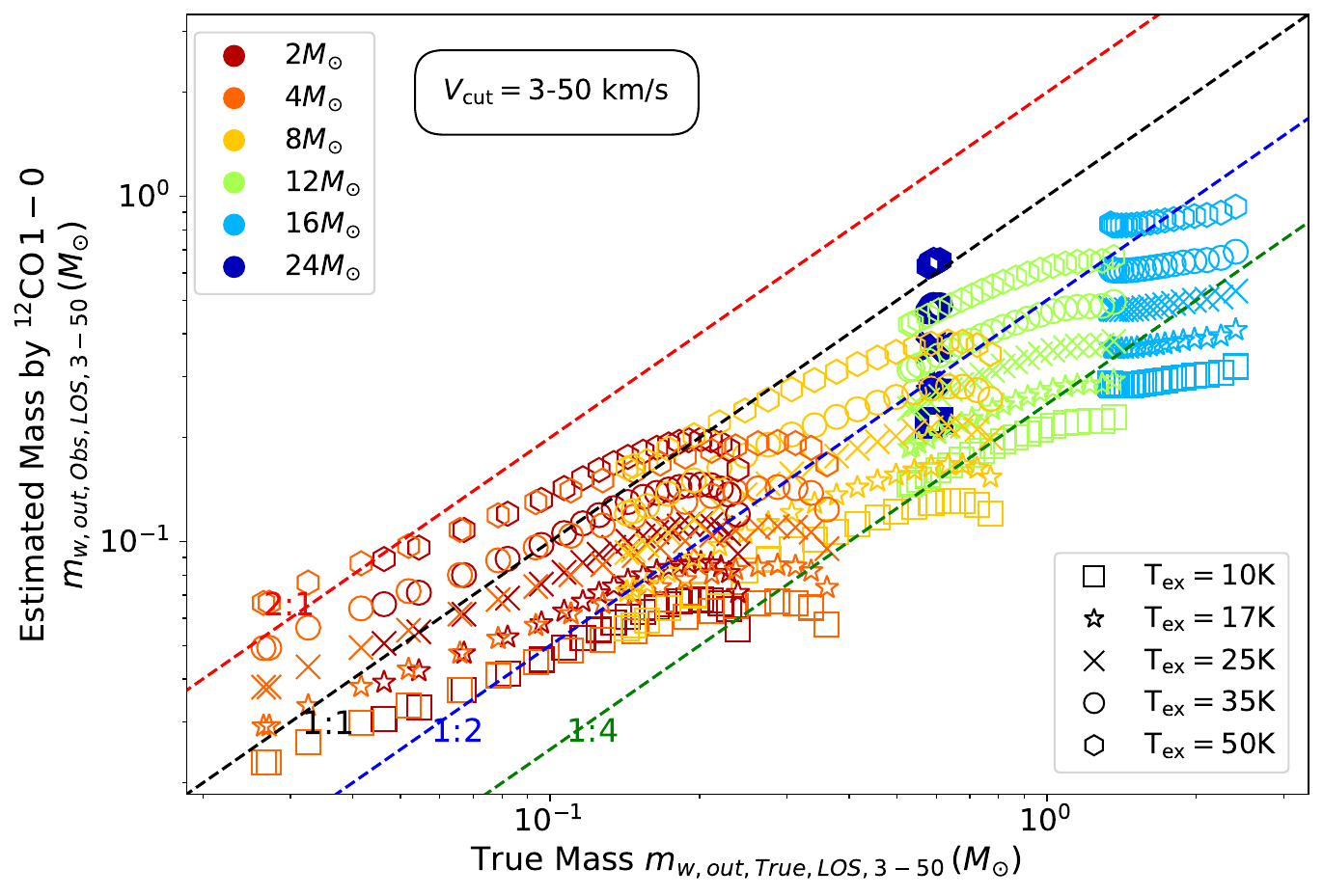}
\includegraphics[width=0.47\linewidth]{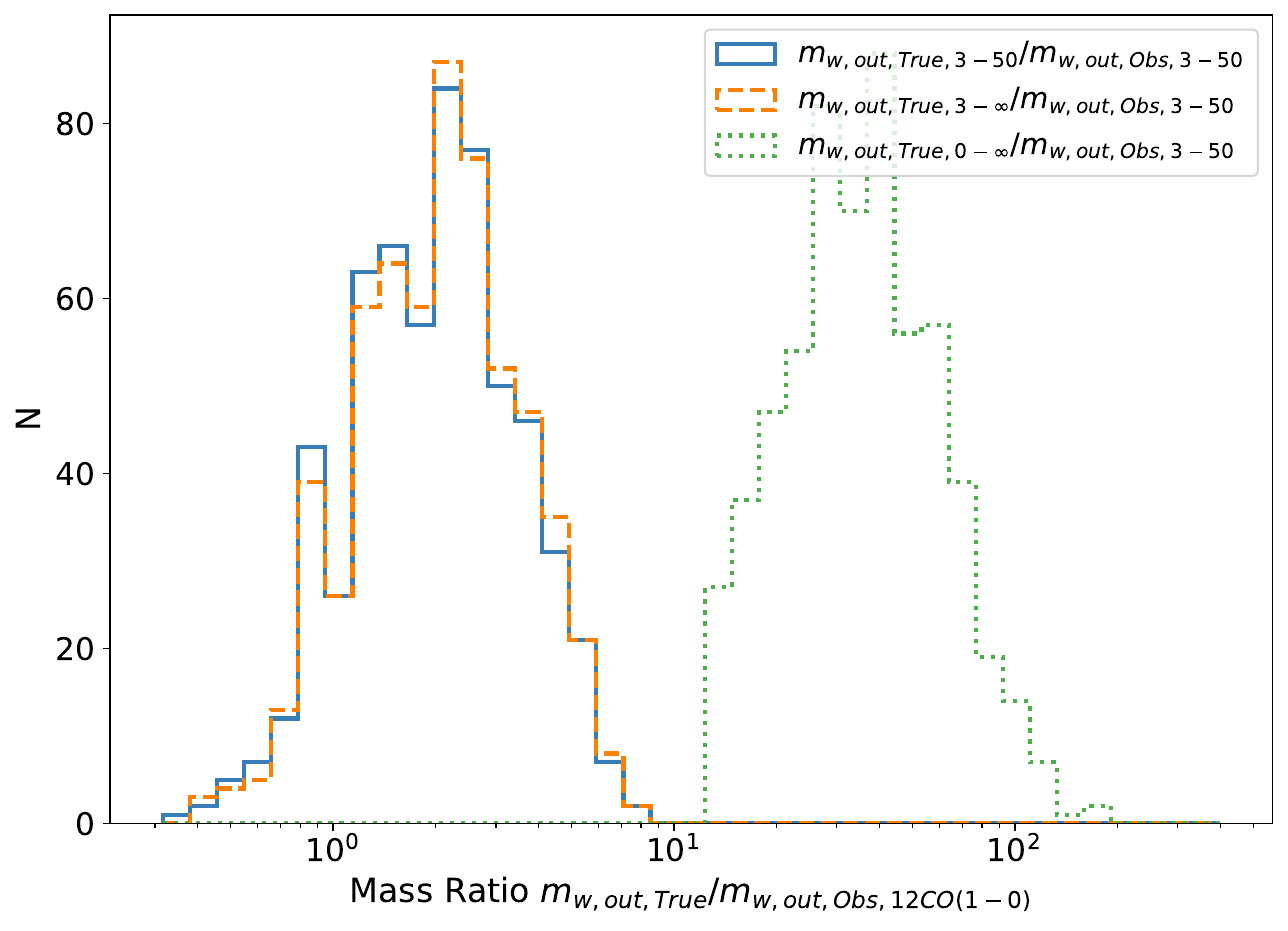}
\includegraphics[width=0.49\linewidth]{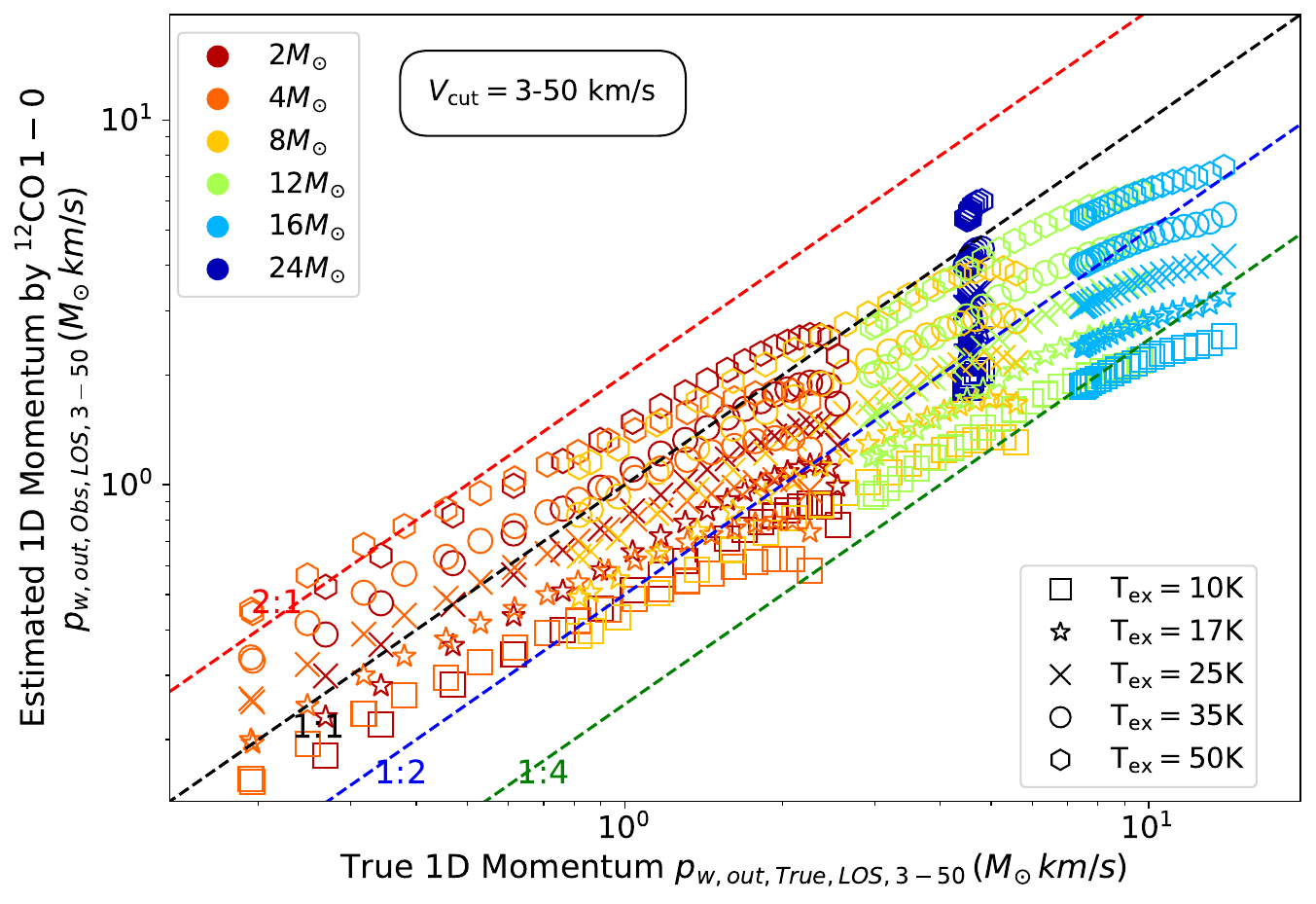}
\includegraphics[width=0.47\linewidth]{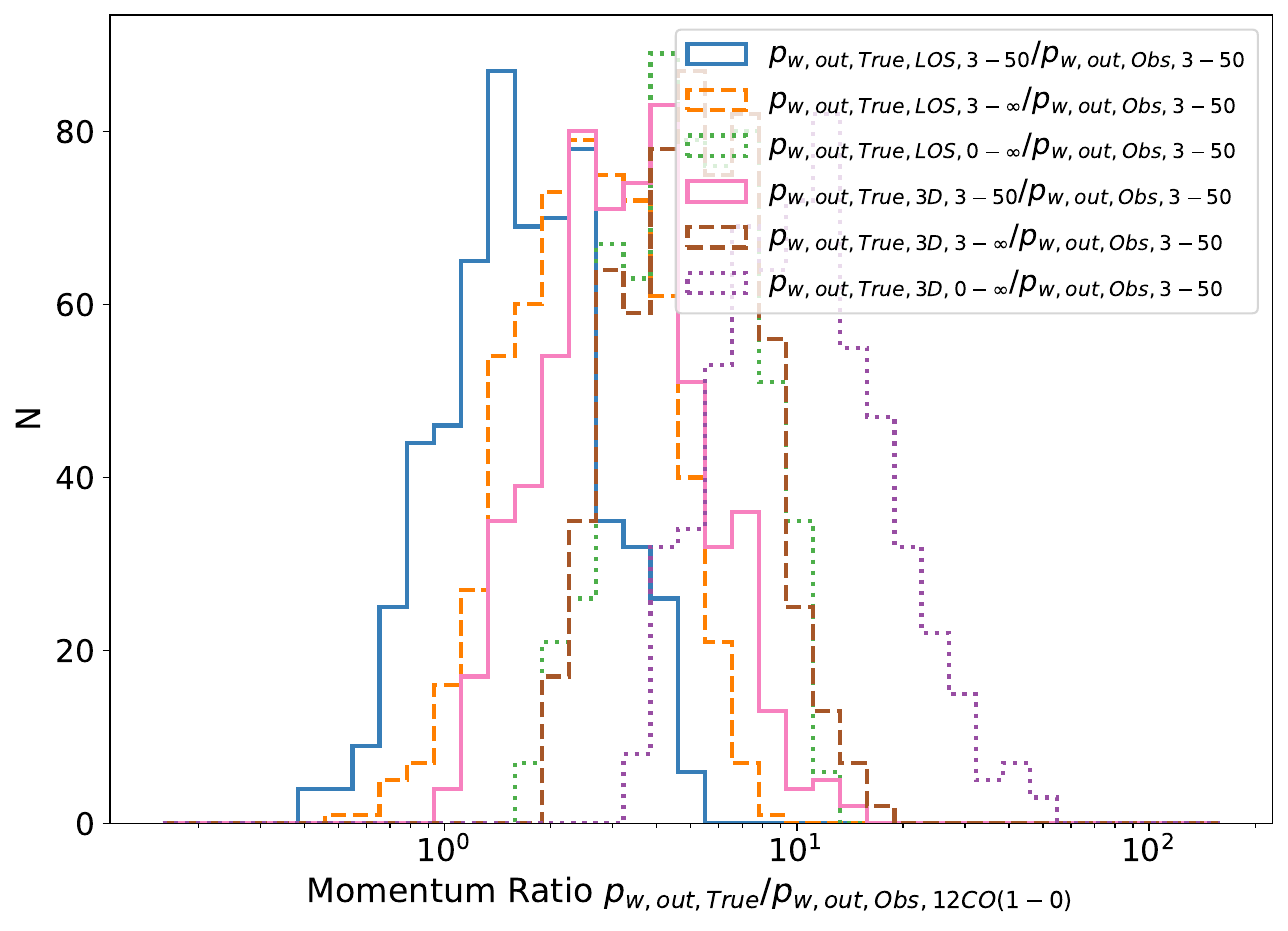}
\includegraphics[width=0.49\linewidth]{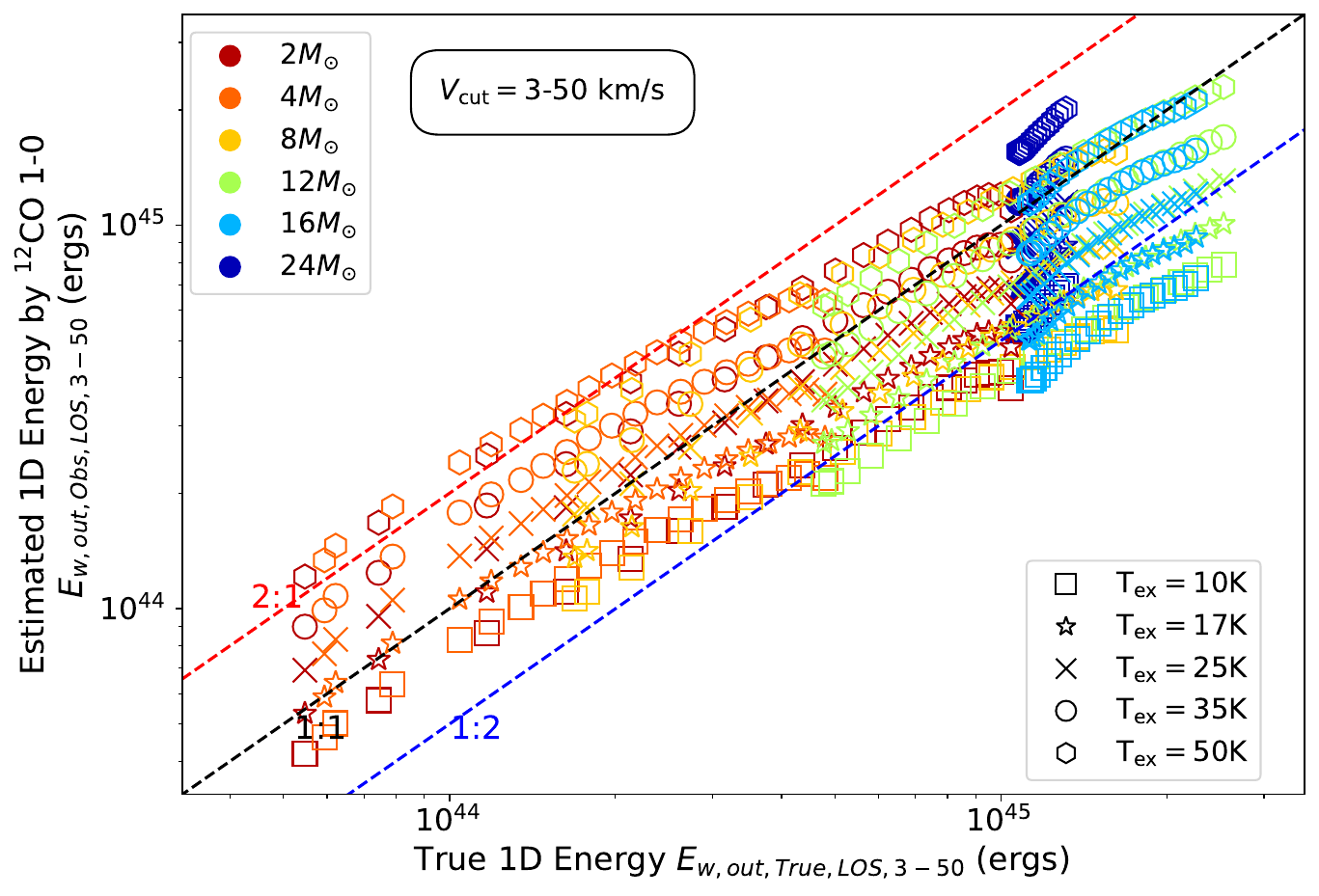}
\includegraphics[width=0.47\linewidth]{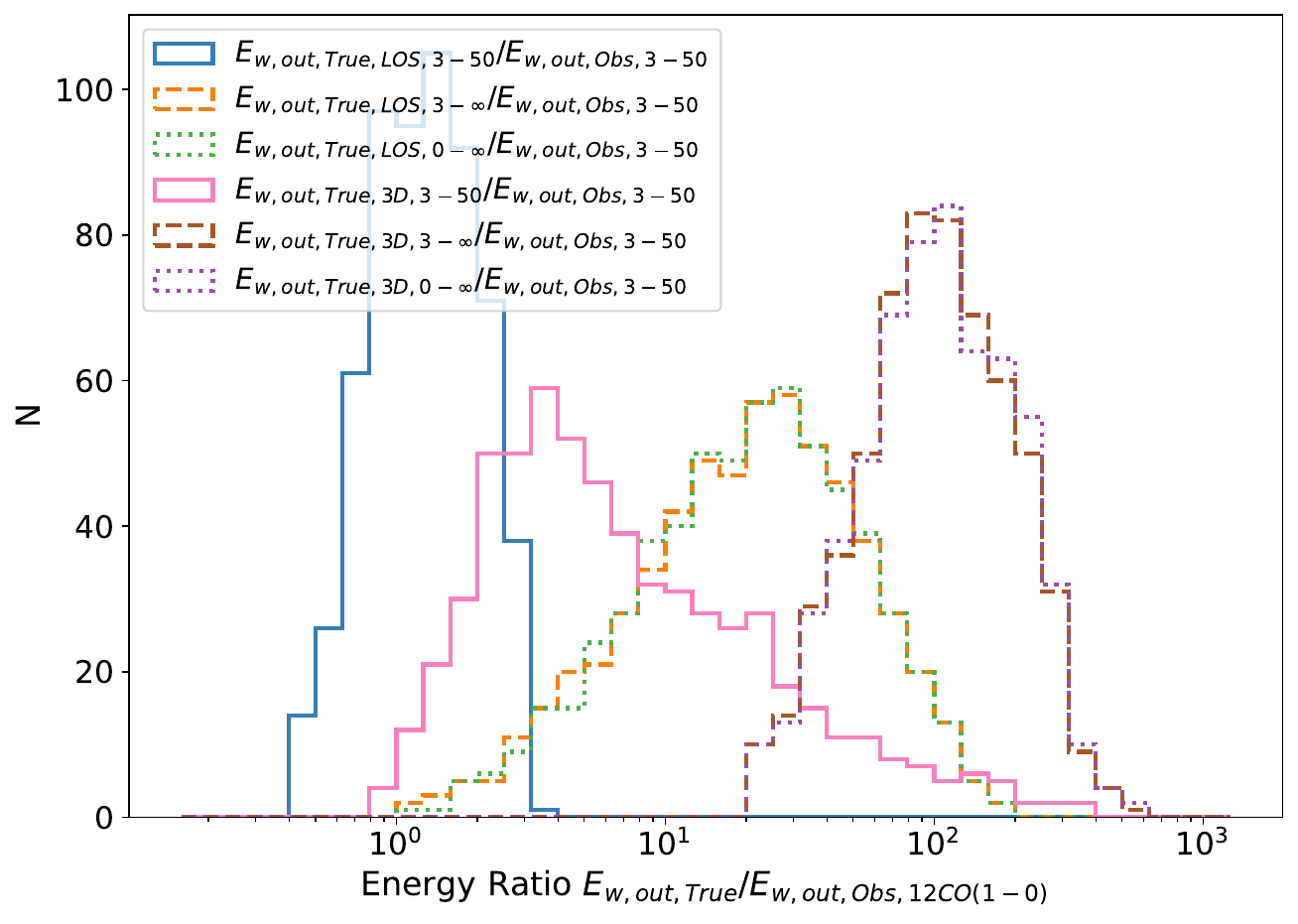}
\caption{{\it Left column:} Comparison of the mass ($1^{st}$ row), momentum ($2^{nd}$ row) and energy ($3^{rd}$ row) estimates obtained for different inclination angles using \co~(1-0) with a LOS velocity cutoff between 3 to 50 \kms, and the corresponding true outflow values with the same velocity cutoff. {\it Right column:}  the ratios between \co~(1-0) calculated mass ($1^{st}$ row), momentum ($2^{nd}$ row) and energy ($3^{rd}$ row) and the true simulation-derived values for different LOS velocity cuts.}
\label{fig.scatter_outflow_mass_P_E_comp_true_12co10_vcut}
\end{figure*}

\begin{figure*}[hbt!]
\centering
\includegraphics[width=0.49\linewidth]{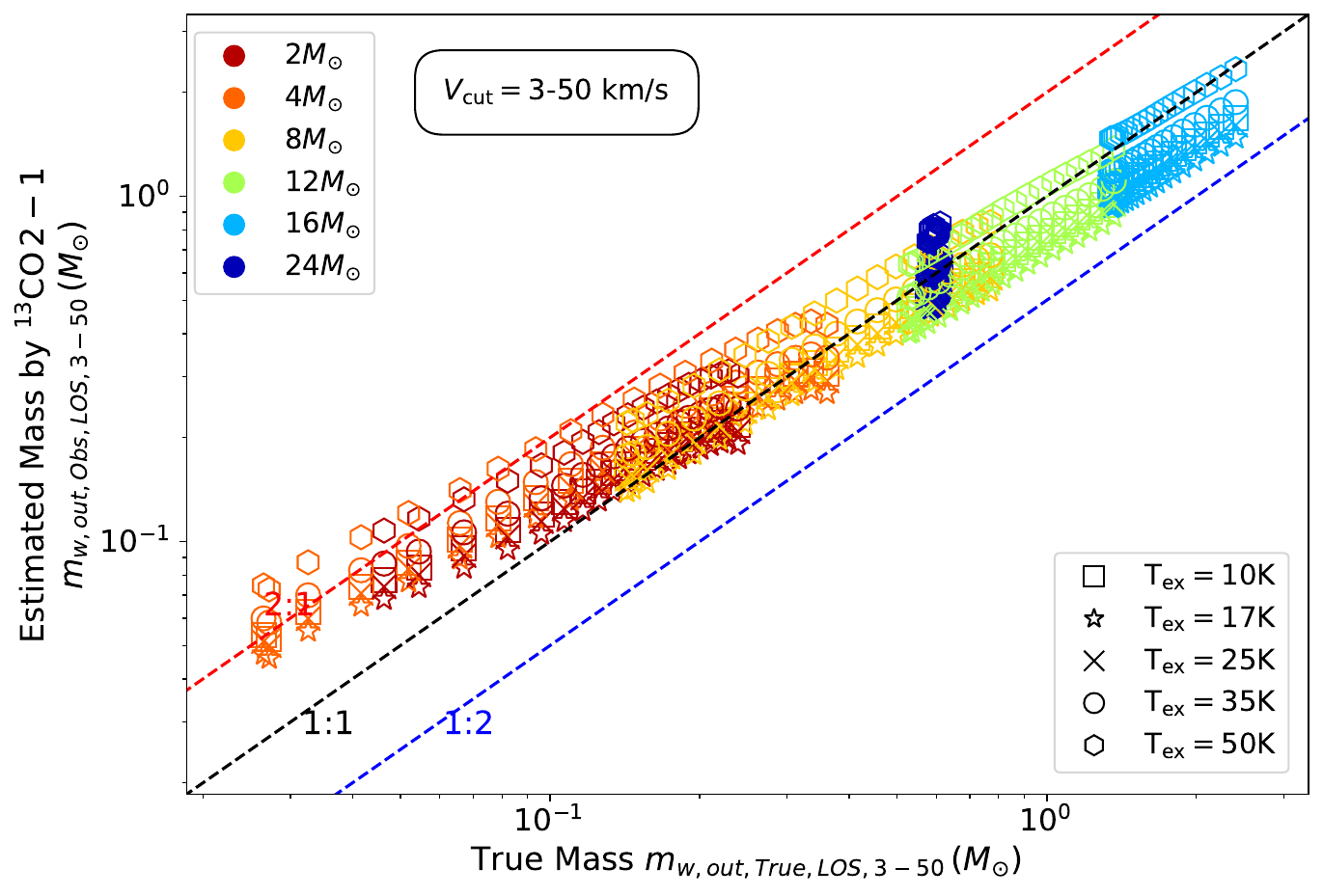}
\includegraphics[width=0.47\linewidth]{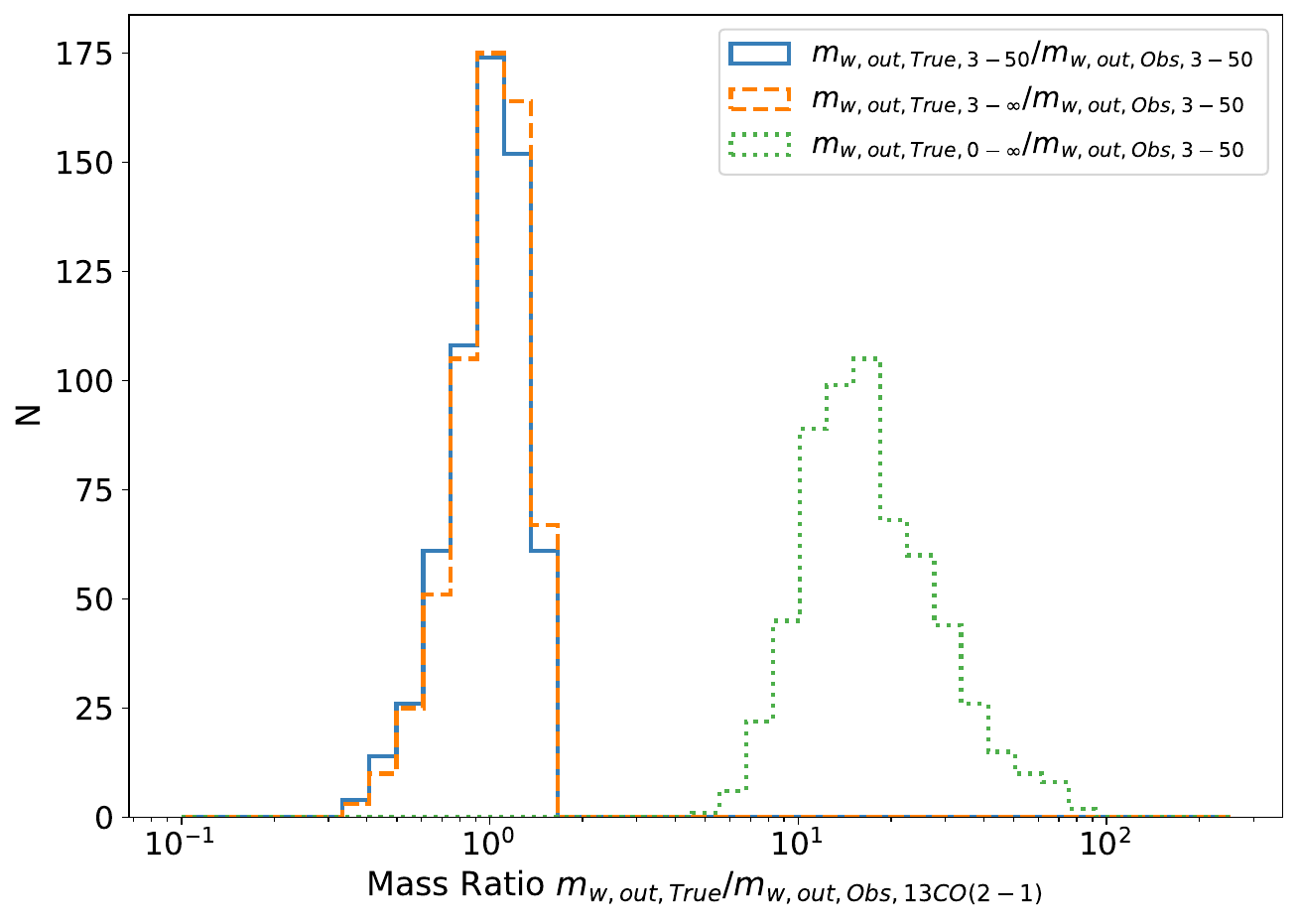}
\includegraphics[width=0.49\linewidth]{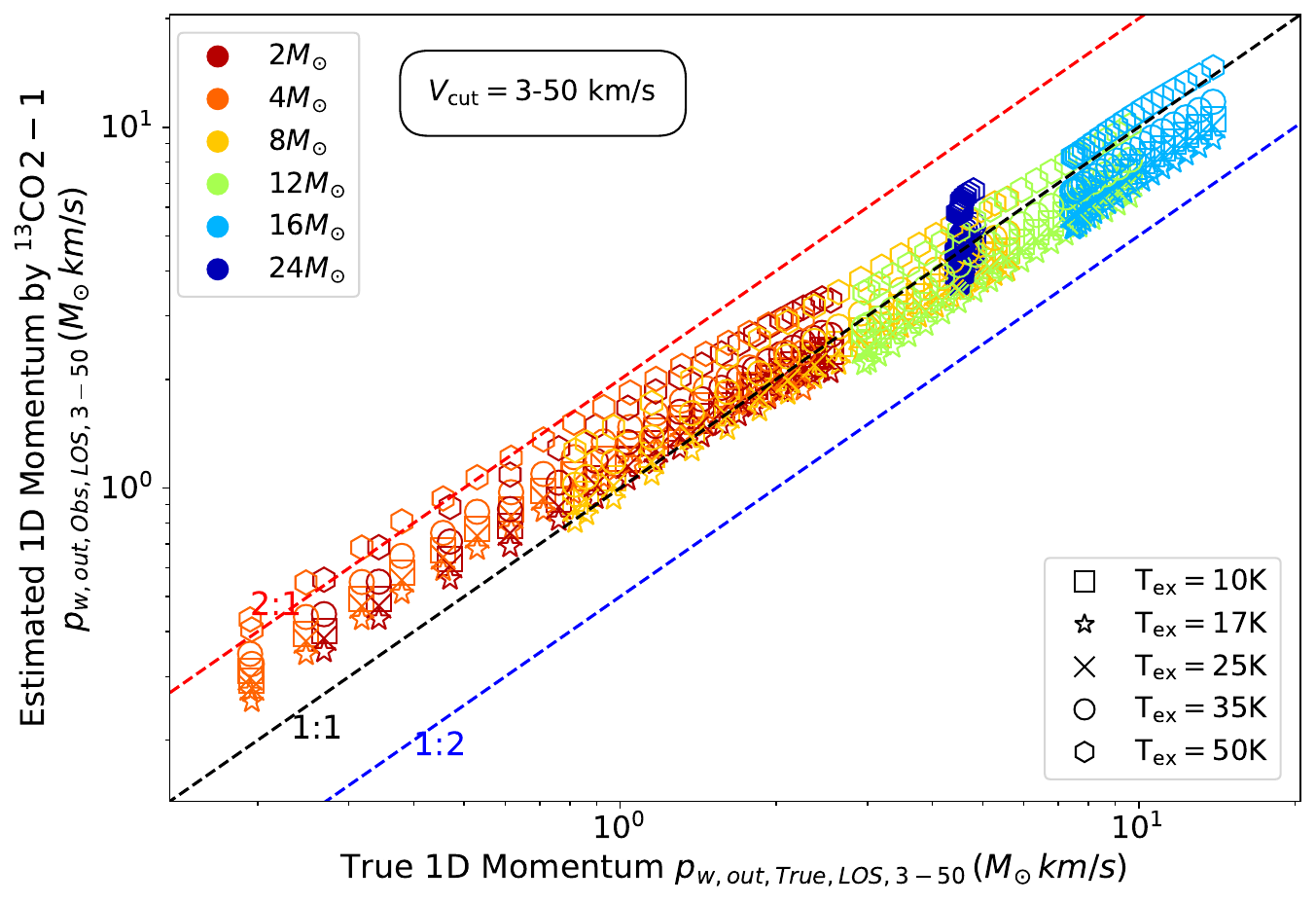}
\includegraphics[width=0.47\linewidth]{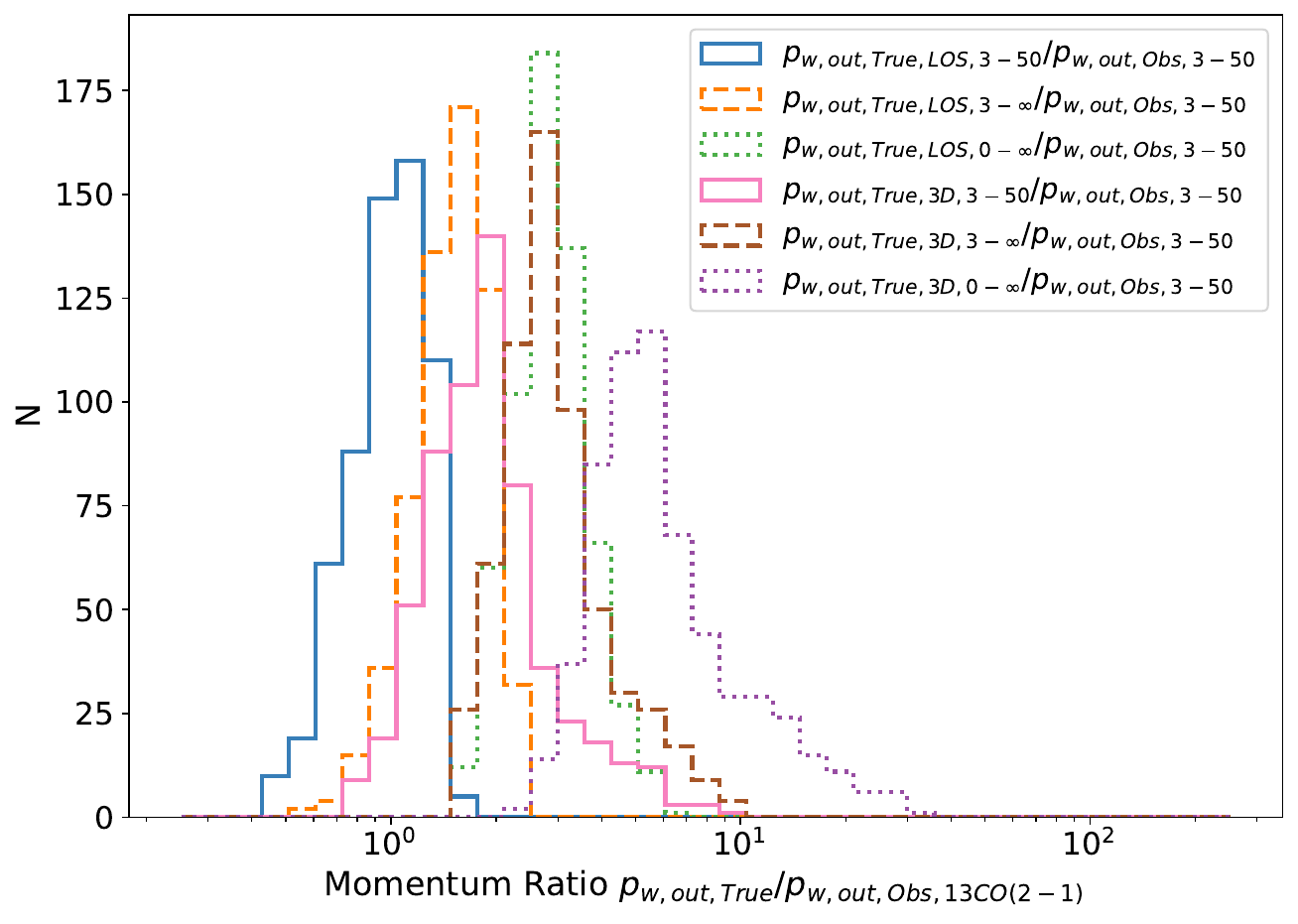}
\includegraphics[width=0.49\linewidth]{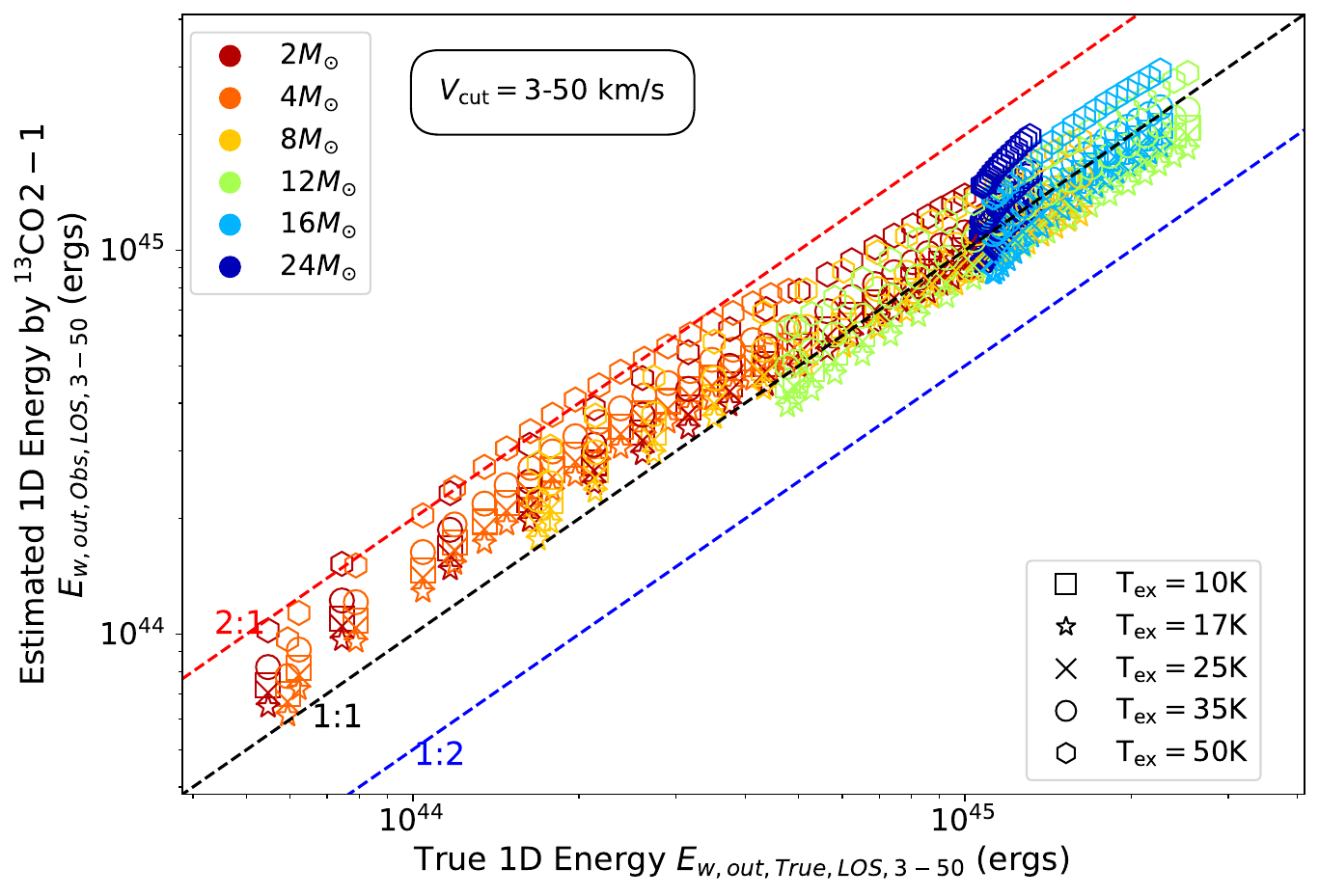}
\includegraphics[width=0.47\linewidth]{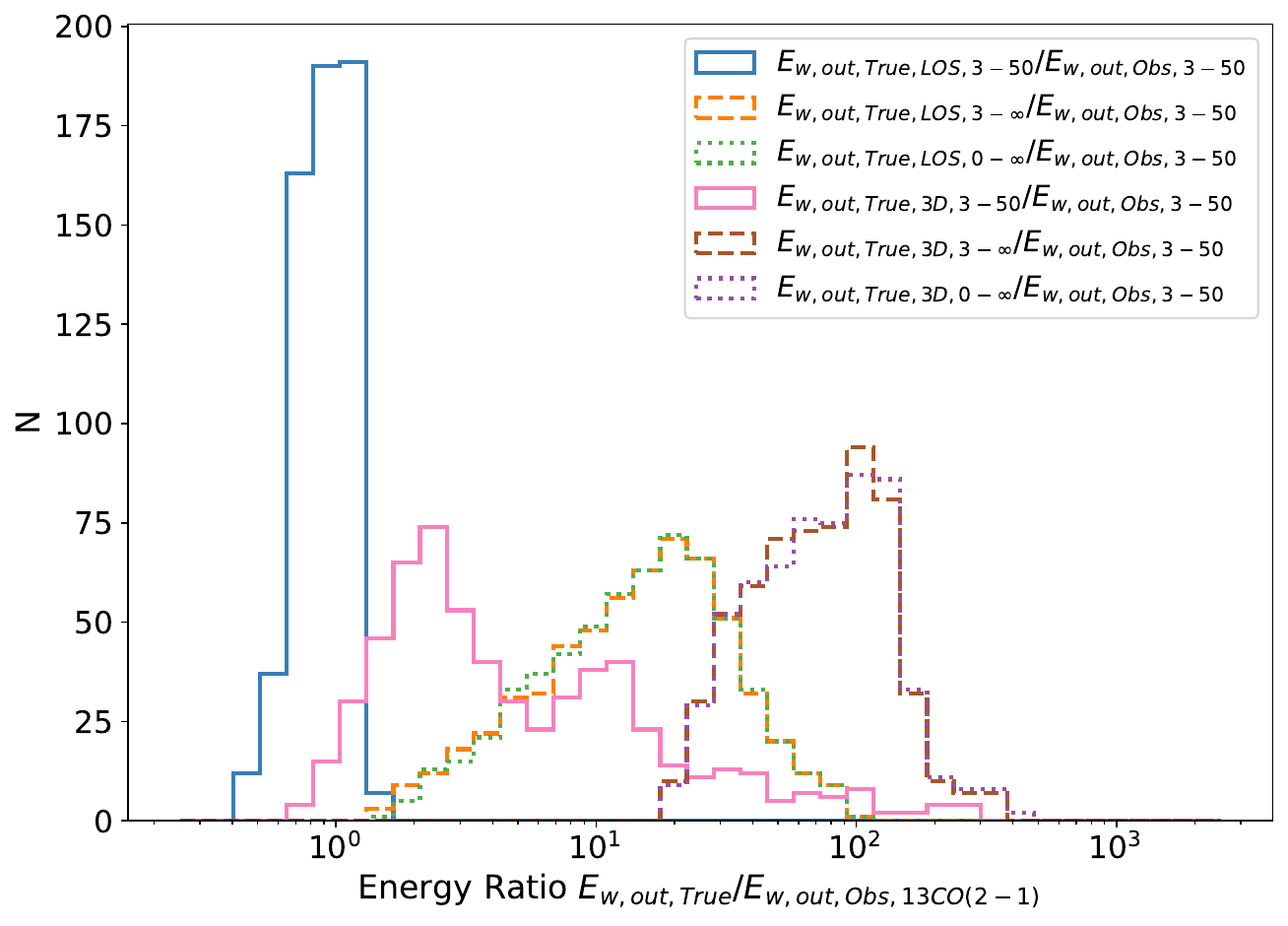}
\caption{{\it Left column:} Comparison of the mass ($1^{st}$ row), momentum ($2^{nd}$ row) and energy ($3^{rd}$ row) estimates obtained for different inclination angles using \13co~(2-1) with a LOS velocity cutoff between 3 to 50 \kms, and the corresponding true outflow values with the same velocity cutoff. {\it Right column:}  the ratios between \13co~(2-1) calculated mass ($1^{st}$ row), momentum ($2^{nd}$ row) and energy ($3^{rd}$ row) and the true simulation-derived values for different LOS velocity cuts.}
\label{fig.scatter_outflow_mass_P_E_comp_true_13co21_vcut}
\end{figure*}

In this section, we report the mass, momentum, and energy estimates of synthetic outflows based on the \co~(1-0) {and \13co~(2-1) line emission, respectively. We then evaluate their accuracy against the actual values obtained directly from the simulations. {It is important to highlight that we examined impacts I and II in these tracers without incorporating synthetic ALMA observations.} The comparison between the \co~(1-0) estimates and the actual values is illustrated in Figure~\ref{fig.scatter_outflow_mass_P_E_comp_true_12co10_vcut}. The comparison between the \13co~(2-1) estimates and the actual values is illustrated in Figure~\ref{fig.scatter_outflow_mass_P_E_comp_true_13co21_vcut}. The similarity between \co~(1-0) and \co~(2-1) is evident, with certain regions exhibiting noticeable optical depth, leading to some scatter and a slight underestimation of outflow mass. Conversely, \13co~(2-1) demonstrates a robust correlation between the estimated properties and the true values, establishing its efficacy as a reliable tracer for outflow gas. However, it is essential to note that, owing to the lower abundance of \13co, its emission may fall below the noise level in real observations. Therefore, combining \co~(2-1) to trace mass in regions where \13co~(2-1) emission is sub-threshold for noise provides a more accurate approach for inferring outflow properties.
}

{

\section{Estimates of Mass, Momentum, and Energy of Synthetic Outflows using ALMA 12m+ACA}
\label{Estimates of Mass, Momentum, and Energy of Synthetic Outflows using ALMA+ACA}

In this section, we present the estimates of mass, momentum, and energy for synthetic outflows derived from synthetic ALMA 12m+ACA \co~(2-1) observations. Subsequently, we assess their accuracy by comparing them with the actual values obtained directly from the simulations. It is crucial to note that we address impacts I, II, and III in this evaluation. The comparison between the estimates derived from synthetic ALMA 12m+ACA \co~(2-1) observations and the actual values is depicted in Figure~\ref{fig.scatter_outflow_mass_P_E_comp_true_12co21_casa_aca_vcut}. Comparing this figure with Figure~\ref{fig.scatter_outflow_mass_P_E_comp_true_12co21_casa_vcut}, a slight difference is evident, with the mass, momentum, and energy consistently higher when using 12m+ACA compared to using only the 12m array, by approximately 25\%. {Consequently, e conclude that the 12m array captures the majority of the total flux density but not the entirety of the flux density.}


\begin{figure*}[]
\centering
\includegraphics[width=0.49\linewidth]{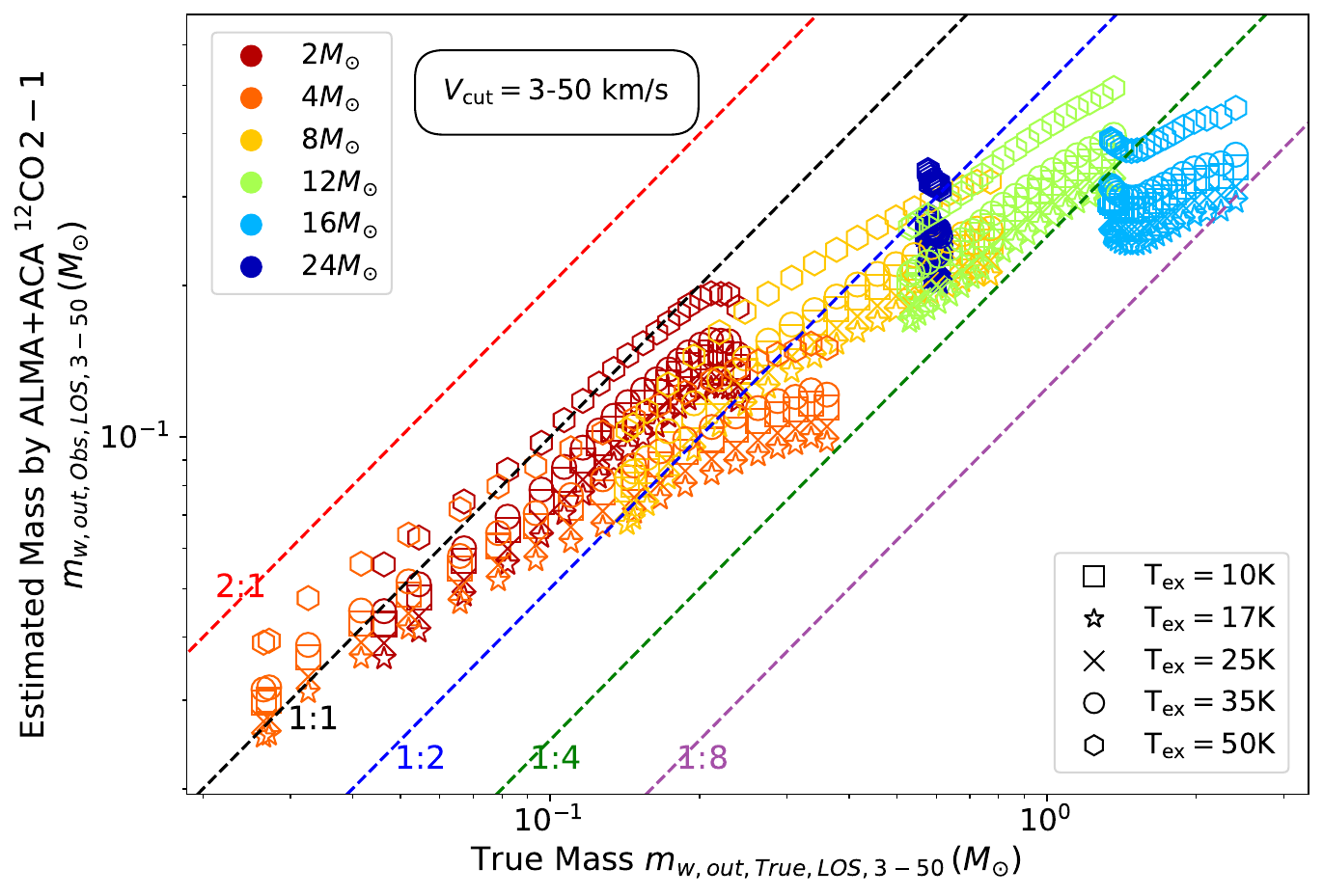}
\includegraphics[width=0.47\linewidth]{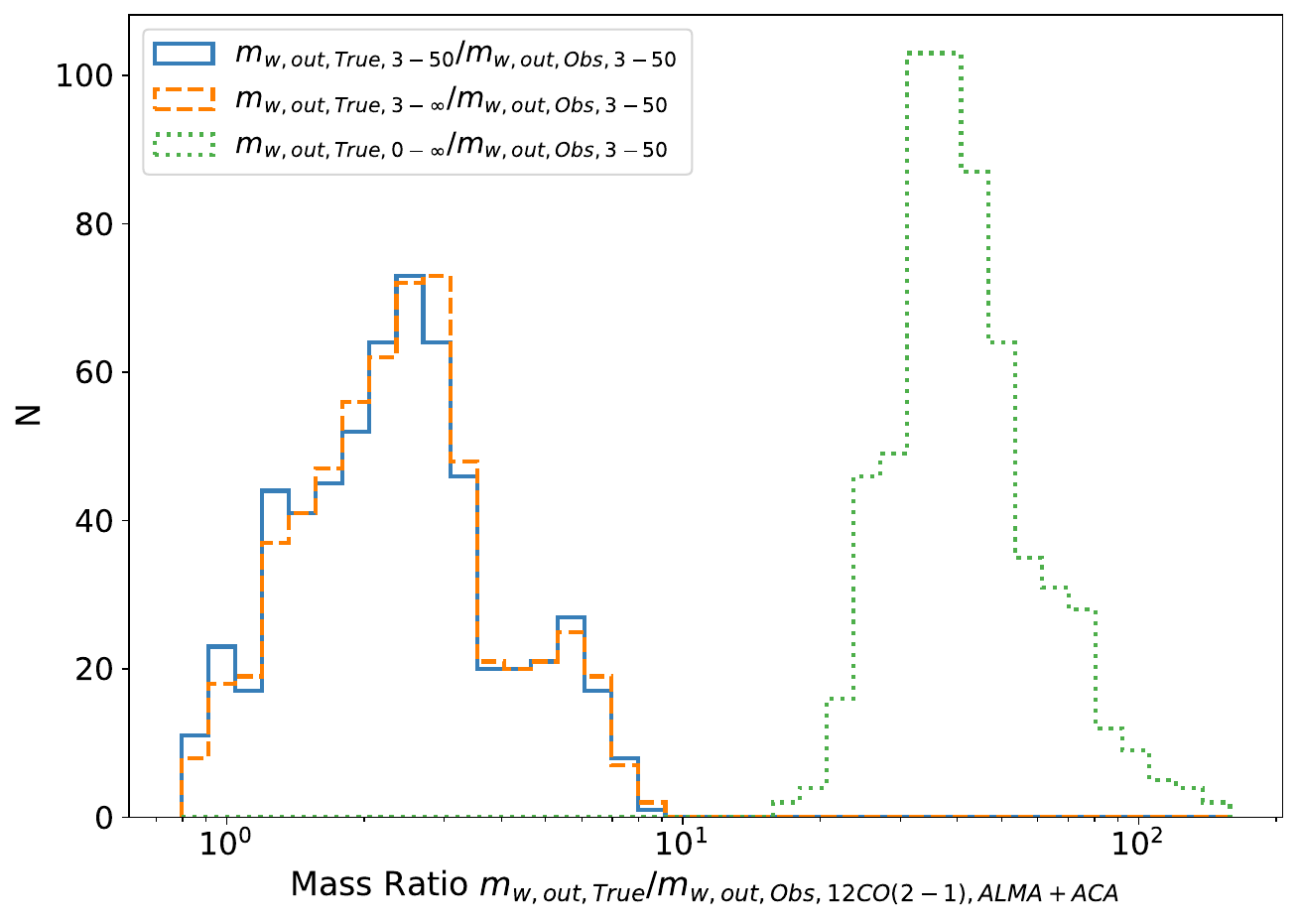}
\includegraphics[width=0.49\linewidth]{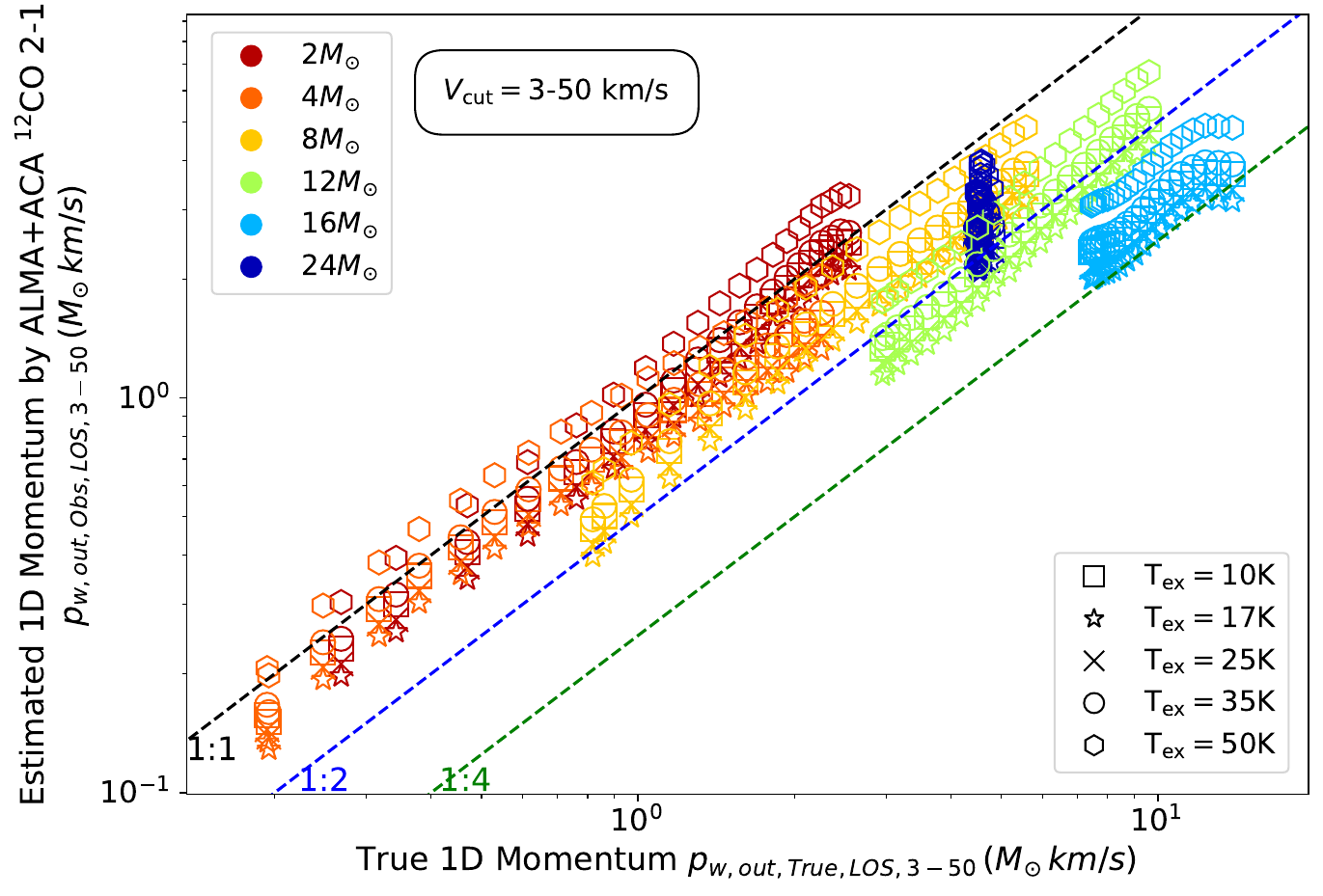}
\includegraphics[width=0.47\linewidth]{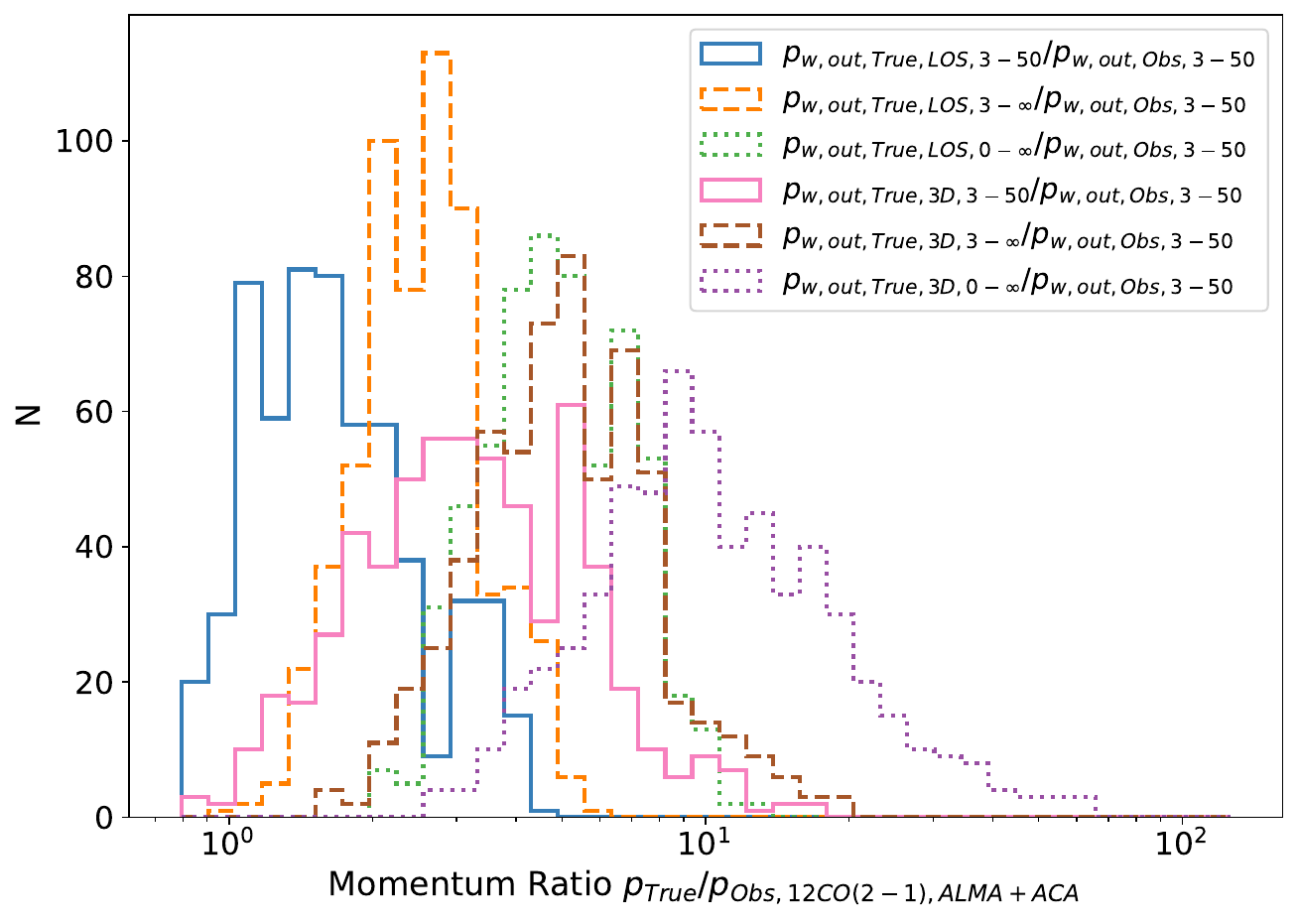}
\includegraphics[width=0.49\linewidth]{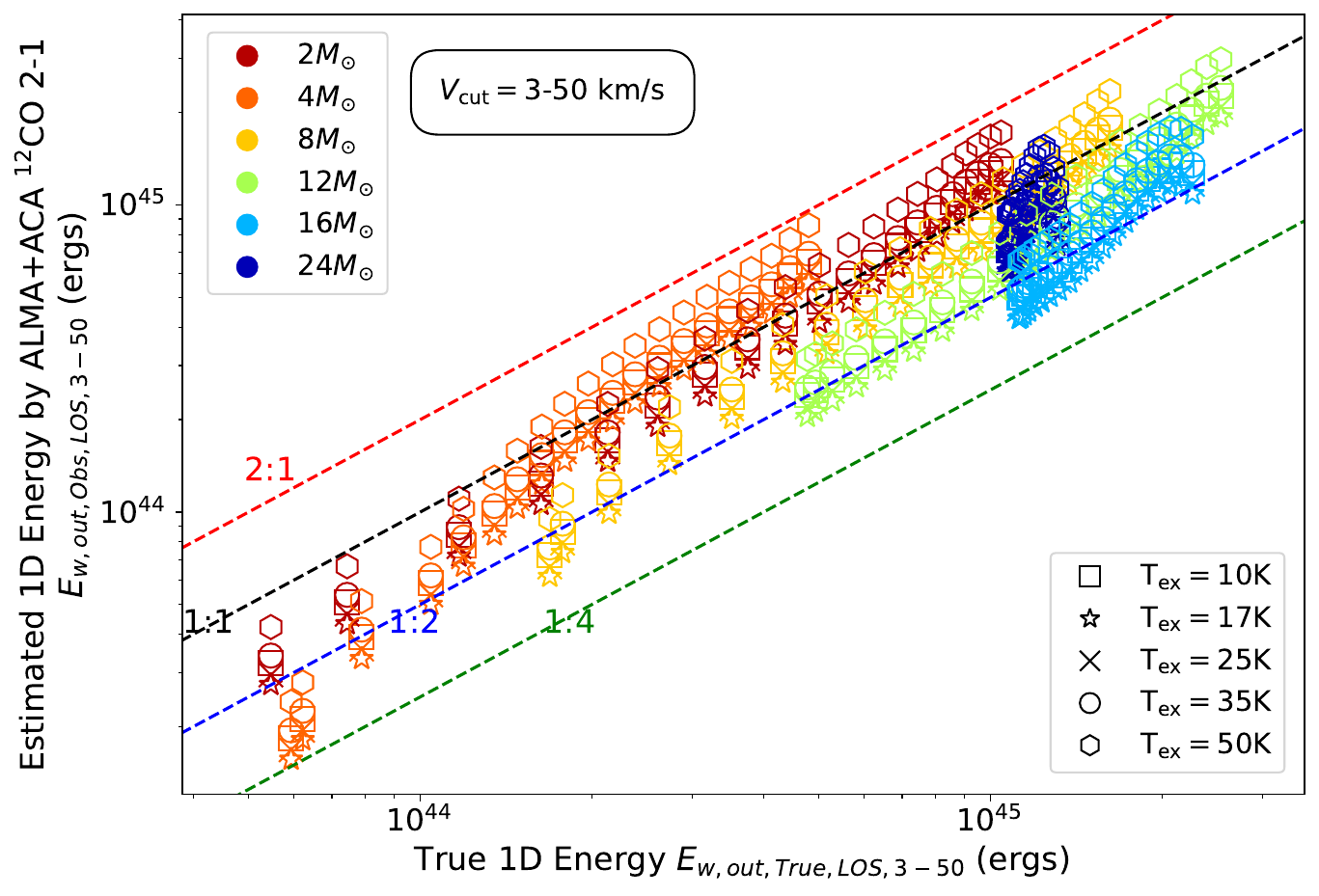}
\includegraphics[width=0.47\linewidth]{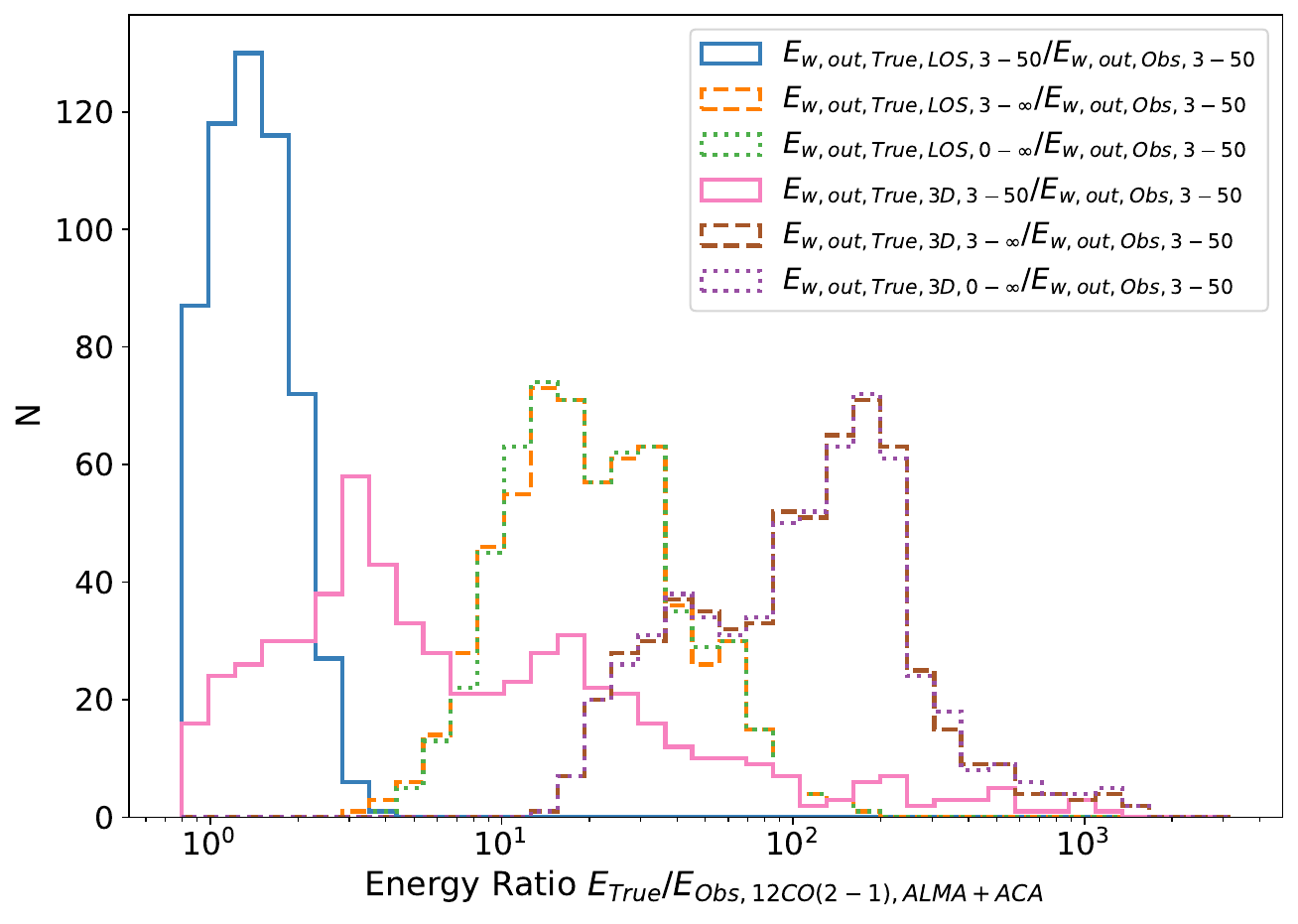}
\caption{{\it Left column:} Comparison of the mass ($1^{st}$ row), momentum ($2^{nd}$ row) and energy ($3^{rd}$ row) estimates obtained for different inclination angles using synthetic ALMA 12m+ACA \co~(2-1) post-processed with CASA/\texttt{simalma} with a LOS velocity cutoff between 3 to 50 \kms, and the corresponding true outflow values with the same velocity cutoff. {\it Right column:} the ratios between synthetic ALMA \co~(2-1) calculated mass ($1^{st}$ row), momentum ($2^{nd}$ row) and energy ($3^{rd}$ row) and the true simulation-derived values for different LOS velocity cuts.}
\label{fig.scatter_outflow_mass_P_E_comp_true_12co21_casa_aca_vcut}
\end{figure*}

}

\bibliographystyle{aasjournal}
\bibliography{references}

\end{CJK*}

\end{document}